\documentclass[10pt,a4paper,twocolumn]{article}
\usepackage{f1000_styles}

\definecolor{paleblue}{HTML}{7892A4}
\definecolor{darkblue}{HTML}{45596E}
\definecolor{midblue}{HTML}{4D637A}

\usepackage{f1000_styles}
\usepackage{graphicx,wrapfig}
\usepackage{enumitem}
\usepackage{hyperref}
\usepackage{multirow}
\usepackage{multicol}
\usepackage{graphics,rotating}%,amssymb,amsmath}
\usepackage{xcolor}
\usepackage{longtable}
\usepackage{booktabs}
\usepackage{subfig}
\usepackage{float,capt-of}
\usepackage[english]{babel}

\usepackage{atlast}

\hypersetup{colorlinks,linktocpage,linkcolor={darkblue}, citecolor={darkblue}, urlcolor={darkblue}}

\begin{document}
\pagestyle{fancy}

\title{Atacama Large Aperture Submillimeter Telescope\\ (AtLAST) Science: Surveying the distant Universe}

%Please provide a concise and specific title that clearly reflects the content of the article.

\author[1]{Eelco van Kampen} 

\author[2]{Tom Bakx}
\author[1]{Carlos De Breuck}
\author[3]{Chian-Chou Chen}
\author[4,5]{Helmut Dannerbauer}
\author[6]{Benjamin Magnelli}
\author[1,7]{Francisco Miguel Montenegro-Montes}
\author[3]{Teppei Okumura}
\author[8,3]{Sy-Yun Pu}
\author[9,10,11]{Matus Rybak} 
\author[12,13]{Am\'elie Saintonge}

\author[14]{Claudia Cicone}
\author[1,4,5]{\\ Evanthia Hatziminaoglou}
\author[10,15]{Juli\"ette Hilhorst}
\author[16]{Pamela Klaassen}
\author[17,18]{Minju Lee}
\author[19]{\\ Christopher C. Lovell}
\author[20]{Andreas Lundgren}
\author[21,22,23,24]{Luca Di Mascolo}
\author[1]{Tony Mroczkowski}
\author[25]{Laura Sommovigo}

% SWG leads and others
\author[16]{Mark Booth}
\author[26]{Martin A. Cordiner}
\author[1,27,28,29]{Rob Ivison}
\author[30,31]{Doug Johnstone}
\author[32,33]{Daizhong Liu}
\author[34]{Thomas J. Maccarone}
\author[35]{Matthew Smith}
\author[36]{Alexander E. Thelen}
\author[14, 37]{Sven Wedemeyer}

%Eelco, Carlos, Evanthia, Tony, Rob
\affil[1]{European Southern Observatory, Karl-Schwarzschild-Str. 2, 85748 Garching bei M\"unchen, Germany}
%(e-mail: evkampen@eso.org)

% first batch of co-authors (contributed significantly)

%Tom
\affil[2]{Department of Space, Earth, \& Environment, Chalmers University of Technology, Chalmersplatsen 4, Gothenburg SE-412 96, Sweden}

%TC & Okumura
\affil[3]{Academia Sinica Institute of Astronomy and Astrophysics (ASIAA), No. 1, Sec. 4, Roosevelt Road, Taipei 10617, Taiwan}

%Helmut
\affil[4]{Instituto de Astrof\'{i}sica de Canarias (IAC), E-38205 La Laguna, Tenerife, Spain}
\affil[5]{Universidad de La Laguna, Dpto. Astrof\'{i}sica, E-38206 La Laguna, Tenerife, Spain}

%Benjamin
\affil[6]{Université Paris-Saclay, Universit\'e Paris-Cit\'e, CEA, CNRS, AIM, 91191 Gif-sur-Yvette, France}

%Francisco Montenegro  (ESO) + Madrid:
\affil[7]{Universidad Complutense de Madrid, Departamento de Física de la Tierra y Astrofísica e Instituto de F\'isica de Part\'iculas y del Cosmos (IPARCOS), Ciudad Universitaria, 28040 Madrid, Spain}

%Sy-Yun
\affil[8]{Graduate Institute of Astronomy, National Tsing Hua University, No. 101, Section 2, Kuang-Fu Road, Hsinchu 30013, Taiwan}

%Matus
\affil[9]{Faculty of Electrical Engineering, Mathematics and Computer Science, Mekelweg 4, 2628 CD Delft, the Netherlands}
\affil[10]{Leiden Observatory, Leiden University, Niels Bohrweg 2, 2333 CA Leiden, the Netherlands}
\affil[11]{SRON - Netherlands Institute for Space Research, Niels Bohrweg 4, 2333 CA Leiden, The Netherlands}

%Amelie
\affil[12]{Department of Physics and Astronomy, University College London, Gower Street, London WC1E 6BT, UK}
\affil[13]{Max-Planck-Institut für Radioastronomie (MPIfR), Auf dem Hügel 69, D-53121 Bonn, Germany}

% second batch (contributed with comments, suggestions, etc.)

%Claudia
\affil[14]{Institute of Theoretical Astrophysics, University of Oslo, P.O. Box 1029, Blindern, 0315 Oslo, Norway}

%Evanthia  (ESO)

%Juliette (Leiden)
% J. Hilhorst
\affil[15]{Department of Astronomy, Yale University, New Haven, CT 06511, USA}

%Pamela (ATC)
\affil[16]{UK Astronomy Technology Centre, Royal Observatory Edinburgh, Blackford Hill, Edinburgh EH9 3HJ, UK}

%Minju
\affil[17]{Cosmic Dawn Center (DAWN), Denmark}
\affil[18]{DTU-Space, Technical University of Denmark, Elektrovej 327, DK2800 Kgs. Lyngby, Denmark}

%Chris Lovell
\affil[19]{Institute of Cosmology and Gravitation, University of Portsmouth, Burnaby Road, Portsmouth, PO1 3FX, UK}

%Andreas
\affil[20]{Aix Marseille Univ, CNRS, CNES, LAM, Marseille, France}

%Luca
\affil[21]{Laboratoire Lagrange, Université Côte d'Azur, Observatoire de la Côte d'Azur, CNRS, Blvd de l'Observatoire, CS 34229, 06304 Nice cedex 4, France}
\affil[22]{Astronomy Unit, Department of Physics, University of Trieste, via Tiepolo 11, Trieste 34131, Italy} 
\affil[23]{INAF -- Osservatorio Astronomico di Trieste, via Tiepolo 11, Trieste 34131, Italy}
\affil[24]{IFPU -- Institute for Fundamental Physics of the Universe, Via Beirut 2, 34014 Trieste, Italy}

%Tony (ESO)

%Okumira (same as TC)

%Laura Sommovigo
\affil[25]{Center for Computational Astrophysics, Flatiron Institute, 162 5th Avenue, New York, NY 10010, USA}

% third batch  (other WG leaders/authors)

%Mark Booth  (ATC)

%Martin Cordiner
\affil[26]{Astrochemistry Laboratory, Code 691, NASA Goddard Space Flight Center, Greenbelt, MD 20771, USA.}

%Rob (ESO) +extra:
\affil[27]{School of Cosmic Physics, Dublin Institute for Advanced Studies, 31 Fitzwilliam Place, Dublin D02 XF86, Ireland}
\affil[28]{Institute for Astronomy, University of Edinburgh, Royal Observatory, Blackford Hill, Edinburgh EH9 3HJ, UK}
\affil[29]{ARC Centre of Excellence for All Sky Astrophysics in 3 Dimensions (ASTRO 3D)}

%Doug
\affil[30]{NRC Herzberg Astronomy and Astrophysics, 5071 West Saanich Rd, Victoria, BC, V9E 2E7, Canada}
\affil[31]{Department of Physics and Astronomy, University of Victoria, Victoria, BC, V8P 5C2, Canada}

%Daizhong
\affil[32]{Max-Planck-Institut f\"{u}r extraterrestrische Physik, Giessenbachstrasse 1, Garching, Bayern, D-85748, Germany}
\affil[33]{Purple Mountain Observatory, Chinese Academy of Sciences, 10 Yuanhua Road, Nanjing 210023, China}

%Thomas Maccarone
\affil[34]{Department of Physics \& Astronomy, Texas Tech University, Box 41051, Lubbock TX, 79409-1051, USA }

%Matthew Smith
\affil[35]{School of Physics \& Astronomy, Cardiff University, The Parade, Cardiff CF24 3AA, UK}

%Alexander Thelen
\affil[36]{Division of Geological and Planetary Sciences, California Institute of Technology, Pasadena, CA 91125, USA.}

%Sven
\affil[37]{Rosseland Centre for Solar Physics, University of Oslo, Postboks 1029 Blindern, N-0315 Oslo, Norway}

\maketitle
\thispagestyle{fancy}

\clearpage
\thispagestyle{fancy}

\begin{abstract}

During the most active period of star formation in galaxies, which occurs in the redshift range $1<z<3$, strong bursts of star formation result in significant quantities of dust, which obscures new stars being formed as their UV/optical light is absorbed and then re-emitted in the infrared, which redshifts into the mm/sub-mm bands for these early times. To get a complete picture of the high-$z$ galaxy population, we need to survey a large patch of the sky in the sub-mm with sufficient angular resolution to resolve all galaxies, but we also need the depth to fully sample their cosmic evolution, and therefore obtain their redshifts using direct mm spectroscopy with a very wide frequency coverage. 

This requires a large single-dish sub-mm telescope with fast mapping speeds at high sensitivity and angular resolution, a large bandwidth with good spectral resolution and multiplex spectroscopic capabilities. The proposed 50-m Atacama Large Aperture Submillimeter Telescope (AtLAST) will deliver these specifications. We discuss how AtLAST allows us to study the whole population of high-z galaxies, including the dusty star-forming ones which can only be detected and studied in the sub-mm, and obtain a wealth of information for each of these up to $z\sim7$: gas content, cooling budget, star formation rate, dust mass, and dust temperature.

We present worked examples of surveys that AtLAST can perform, both deep and wide, and also focused on galaxies in proto-clusters. In addition we show how such surveys with AtLAST can measure the growth rate f$\sigma_8$ and the Hubble constant with high accuracy, and demonstrate the power of the line-intensity mapping method in the mm/sub-mm wavebands to constrain the cosmic expansion history at high redshifts, as good examples of what can uniquely be done by AtLAST in this research field. 

\end{abstract}

\section*{\color{OREblue}Keywords}

cosmology --- galaxy surveys --- galaxy formation --- \\ sub-mm galaxies --- cluster galaxies

\clearpage
\pagestyle{fancy}

%\newtext{
%\textbf{To-do list (\today):
%\begin{itemize}[leftmargin=*,topsep=2pt,itemsep=2pt,parsep=2pt]
%    \item synergy with other AtLAST cases
%    \item acknowledgements (please add)
%\end{itemize}}
%}

\section*{Plain language summary}
\medskip
Galaxies come in a wide variety of shapes, sizes, and colours, despite all of them having originated from similar initial conditions in the early Universe.  Understanding this diversity by tracing back the evolutionary pathways of different types of galaxies is a major endeavour in modern astrophysics. Galaxies build their stellar mass over time by converting gas into stars through various episodes of star formation.  Understanding exactly when, where, and how this star formation process is triggered or suppressed is therefore a crucial question to answer.  

Current observations reveal that the Universe was at its most active (in terms of star formation rate per unit volume) in the distant past, about 10 billion years ago. By measuring the amount of gas and dust in galaxies at that epoch, we also know that the reason for this very high star formation activity is large reservoirs of gas (the fuel for star formation) and the higher efficiency of galaxies at converting their gas into stars. However, recent work also reveals that we are missing significant numbers of distant actively star-forming galaxies in current samples because these are obscured by dust, and therefore our picture is still very incomplete.  

In this paper, we explore how a new proposed telescope, the Atacama Large Aperture Submillimeter Telescope (AtLAST)\footnote{\url{http://atlast-telescope.org/}}, can provide us with the very important missing pieces of this puzzle. AtLAST will allow us to map large areas of the sky at unprecedented depth, resolution and multiplex spectroscopic capabilities. This telecope would  provide us with a complete, homogeneous and unbiased picture of the star-forming galaxy population in the early Universe.  Not only will we be able to discover these galaxies, but also measure their distances, the composition of their gas and dust content, and the rate at which they convert gas into stars.  

\medskip

%Starting with a more general multi-band imaging survey, then either spectroscopic or photometric redshift surveys in which the galaxies are used for various types of measurements, including redshift distributions and clustering estimates. At high redshifts one can constrain cosmological models through line intensity mapping, where the individual galaxies and their surroundings do not need to be resolved, allowing one to target very large cosmological volumes, especially with a telescope that allows large survey speeds due a large field-of-view. The latter also makes it possible to map a large statistical sample of high-$z$ galaxy clusters, the most dense areas in the early large-scale matter distribution.

\section{Introduction}\label{sec:introduction}

\medskip

%\eh{Add something here on the discovery of SMGs, their physical characteristics and their role in galaxy evolution, e.g. in terms of fraction of star formation in the early Universe.}

In the distant Universe, the star formation rate density in galaxies is highest in the redshift range $1<z<3$ (eg.\citealt{Madau2014}), which results in a fair amount of astrophysical dust and gas in these galaxies. 
A similar trend is observed in the cold-gas content of galaxies - as traced by emission lines and cold dust, which is seen to peak at a similar redshift range (e.g. \citealt{Tacconi2020}).
The dust obscures new stars being formed, because their UV/optical light is absorbed and then re-emitted in the infrared (IR, eg. \citealt{SalimNarayanan2020} for an excellent review on the physical mechanisms), contributing to the so-called cosmic far-IR background (CIB). This accounts for about half of the energy density from star formation, integrated over the history of the Universe \citep{Dole06}. The infrared photons emitted in the rest-frame of these galaxies get redshifted into the submillimeter and millimeter ((sub-)mm) observed frame, and the negative K-correction at this wavelength regime enables galaxies to appear roughly constant in the observed flux densities at $z\sim1-10$ \citep{Blain2002}. This means that the (sub-)mm is an essential wavelength range for studying high-$z$ galaxies in order to understand their star formation and growth. Since high-$z$ star forming galaxies are best observed in the (sub-)mm, they are often called 'sub-mm galaxies' (SMGs), although lately the physically-motivated denomination of `dusty star forming galaxies' (DSFGs) is preferred. The study of DSFGs now forms a rich research field:  for reviews on its history see \cite{CarilliWalter2013, Casey2014, Combes2018}. In the 25 years since the discovery of DSFGs a number of surveys have targeted the (dust) continuum and spectral line emission from these high-redshift galaxies. 

\medskip
Wide-field deep continuum surveys with many-pixel bolometer detectors on single-dish telescopes have been efficient in discovering DSFGs out to the epoch of re-ionisation ($z\approx6$). From the ground, the South Pole Telescope (SPT) conducted a $\sim$2500~deg$^2$ shallow survey at 1.4~mm and 2~mm, uncovering almost a hundred (mostly gravitationally lensed) dusty galaxies (\citealt{Reuter2020,Everett2020} and references therein). A similar survey has been completed by the 6-m Atacama Cosmology Telescope (ACT) \citep{Gralla2020}. At 850~$\mu$m, several deg$^2$ have been surveyed by the LABOCA camera on APEX \citep{Weiss2009} and the SCUBA-2 camera on JCMT \citep{Geach2017}. Due to their low angular resolution, these surveys are confusion-limited at the mJy level, i.e., faint sources start overlapping. Current continuum survey facilities include the NIKA-2 camera on the 30-m IRAM telescope, A-MKID 350/850-$\mu$m camera on APEX, and the TolTEC camera on the 50-m LMT.

\medskip
In space, \textit{Herschel} mapped up to $1270$~deg$^2$ at 250 - 500~$\mu$m, revealing 1.7 million dusty galaxies (with multiple detections), as collected in the {\it Herschel} Extragalactic Legacy Project (HELP, \citealt{Shirley2021}), which notably includes
the {\it Herschel} Multi-tiered Extragalactic Survey (HerMES, \citealt{Oliver2012}) and the {\it Herschel} Atlas
survey (H-ATLAS, \citealt{Eales2010}). 
About a dozen extremely bright / highly magnified high-$z$ galaxies were detected in the all-sky (but relatively shallow) continuum imaging with the \textit{Planck} telescope \citep{Lammers2022}.
The redshift distribution depends on the selection wavelength (e.g. \citealt{Bethermin2015}), so there is a clear need for a wide frequency range in observations to enable a complete census of dusty galaxies from cosmic noon to cosmic dawn.

\medskip
ALMA has carried out several ``wide-field'' continuum surveys, starting with the 1.3-mm survey of the HUDF \citep[ $\approx$4.5 arcmin$^2$]{Dunlop2017}. The currently most extensive interferometric survey is the 2-mm MORA survey \citep[184~arcmin$^2$]{Casey2021}, with an extension currently being produced.
In contrast to the continuum mapping, blind spectral-line surveys have been limited to interferometers - e.g., ASPECS \citep{Walter2016},  ALMACAL \citep{Klitsch2019}, and ALCS \citep{Fujimoto2023} on ALMA, and HDF-N survey with NOEMA \citep[8 arcmin$^2$]{Boogaard2023}. %This is due to the lack of IFU / multi-object spectroscopy capabilities at sub-mm wavelengths; interferometers effectively act as an ``IFU'' over their limited field-of-view.  
This is because current spectroscopic instruments on single-dish telescopes are often limited to single-pixel designs, and the few exceptions have a maximum of up to ten spatial elements. Instead, the interferometers act as an ``integral field unit'' (IFU) within their limited field-of-view of interferometers, that cover an area roughly equal to a single pixel element of a single-dish telescope. Multi-object spectroscopy - the ability to obtain spectra of multiple objects in the field-of-view of the telescope, common at optical / near-IR wavelengths - is virtually non-existent in the sub-mm wavebands. As such, these surveys are inherently restricted to pencil-beam observations due to the low mapping speeds. ALMA Large Programs such as ALPINE \citep{Faisst2020, Bethermin2020} and REBELS \citep{Bouwens2022, Inami2022} have provided the first statistical sample of $4<z<7$ (up to 7) dust continuum emitting galaxies (albeit UV-selected, not DSFGs), but these were targeted surveys, not blind ones. The contribution to the star-formation rate density of these more 'normal' (but still massive: $M_*>10^9,1^{10} M_\odot$ for ALPINE/REBELS) dusty galaxies seems to be far from negligible ($>30$\% at $z$=7, see e.g. \citealt{Algera2023}, \citealt{Barrufet2023}). There are many uncertainties due to selection bias, and lack of multiple ALMA observations for most of the targets, but AtLAST will have the sensitivity to detect these sources.

\begin{figure*}
    \centering
    \includegraphics[width=0.49\textwidth]{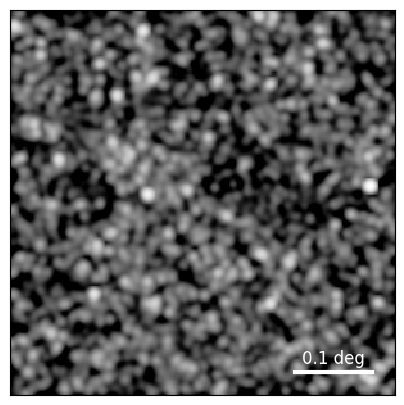}
    \includegraphics[width=0.49\textwidth]{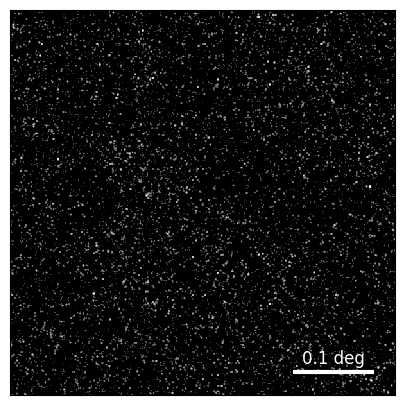}
    \caption{A simulated 277-GHz continuum map with the 6-m ACT telescope (left) and AtLAST (right) \emph{without} instrument noise. The ACT map is limited by confusion noise %at X~mJy level
    (notice the seemingly overlapping faint sources); thanks to its 50-m dish, AtLAST will be able to resolve much fainter (up to four orders of magnitude) individual sources. These mock images are based on simulated galaxy catalogues from \citet{Lagos2020}.}
    \label{fig:ACT_AtLAST}
\end{figure*}

\medskip
To date, the science driven by sub-mm observations has focused on a combination of large-area continuum observations from both ground- and space-based observations, while spectroscopic observations with a sufficient spatial resolution to resolve these galaxies are limited to small areas.
This bimodal approach has left a discovery space that can only be addressed by a telescope with a large primary mirror, exploiting recent advances in instrumentation to provide Integral Field Unit (IFU) capabilities on the scale of individual galaxies. AtLAST is conceived to realise this by combining a large throughput single dish facility with a powerful instrumentation suite.
How will AtLAST outperform current facilities? High angular resolution (lowering confusion noise); large collecting area and large focal plane - high survey speed \citep[see e.g.][]{Klaassen2020, Ramasawmy2022, Mroczkowski2023, Mroczkowski2024}. For example, the expected sky area covered by AtLAST with a single pointing will be about 200 times larger than the LMT. In addition, the photometric confusion noise at 350~$\mu$m will be a factor of $>10000\times$ lower than that of 6-m telescopes like ACT and CCAT-p ($<0.2~\mu$Jy for AtLAST vs. 2600~$\mu$Jy for ACT and CCAT-p).
The improvement in angular resolution provided by AtLAST is demonstrated clearly in Figure \ref{fig:ACT_AtLAST}, which compares what can be achieved by current 6-m telescopes like ACT to what AtLAST can do. 

\medskip
Here we describe the compelling science that requires a new facility in order to be achieved - specifically AtLAST, a 50m single dish sub-mm telescope with the capabilities listed above (incl. high mapping efficiency).
Such an observatory would enable a large-area spectroscopic survey, expose baryon acoustic oscillations, probe the unresolved power-spectrum of galaxies (i.e., line intensity mapping) and reveal the largest coherent structures in the Universe towards the early Universe.

\section{Science Goals}

\medskip
In the following subsections we outline the high-$z$ science goals for AtLAST, focusing on the overall high-$z$ galaxy population as well on galaxies in proto-clusters. We also discuss how to extract cosmological parameters from the large survey, and the novel line-intensity mapping technique. 
We refer to Lee et al.\ (in prep.) and Di Mascolo et al.\ (in prep.) for companion AtLAST case studies focused on emission line probes of the cold circumgalactic medium (CGM) of galaxies and on probing the Intra-Cluster Medium (ICM) using the full Sunyaev-Zeldovich (SZ) spectrum to understand the thermal history of the Universe, respectively. Additional AtLAST science cases outside the research fields of the distant Universe and cosmology are presented by Booth et al. (in prep), Cordiner et al. (in prep), Klaassen et al. (in prep), Liu et al. (in prep), Orlowski-Scherer et al. (in prep), and Wedemeyer et al. (in prep).
For the purpose of this paper, ``high-$z$'' refers to $z>1$.

\medskip
\subsection{A large homogeneous galaxy survey in the distant Universe}

\medskip
%\newtext{... science questions to go here, arguing for large homogenous galaxy surveys, continuum and line ...}

%Some initial thoughts on scientific motivation of the large surveys. Please check and complement - Francisco%

The integrated spectral energy distribution of the CIB is nowadays relatively well constrained thanks to accumulated observations over the past decades from ground and space facilities, in particular close to the peak of thermal IR emission around 150 $\mu$m in the rest-frame. Still, at sub-mm wavelengths, where the contributions from high-redshift galaxies are expected to dominate, a fair fraction of this emission remains unresolved and only the brightest population of DSFGs has been identified and studied in certain detail. Many of those have been identified in blind large area-surveys, their redshift determined thanks to CO spectroscopy in the mm-regime, and then followed up and studied in detail with interferometers like ALMA, as in   \cite{Reuter2020}.

\medskip

Little is known observationally about the less extreme population of normal dusty galaxies, accounting for the bulk of the objects contributing to the CIB at these wavelengths. Studying this population is crucial to progressing our understanding of numerous open questions like the co-evolution of star formation and black-hole growth, as most high-z star formation occurs in galaxies deeply embedded in dust (see \citealt{Carraro2020, Mountrichas2023}). Or the evolution of the dust properties over cosmic time, which is under intense debate (see e.g. \citealt{Sommovigo2022, Dayal2022, DrewCasey2022, Hirashita2022, DiCesare2023}). Till now, the study of these less extreme sub-mm galaxies has been relying partially on extrapolation of properties of MIR- or radio-selected galaxies with emission in FIR/sub-mm. I Another approach has been studying the physical properties of some of the brightest DSFGs which are associated with lensed systems. Both approaches are severely biased against the brightest end of the population.
\medskip

In order to determine accurately the number counts and the redshift distribution of the population of normal dust star-forming galaxies, it is crucial to conduct deep and unbiased surveys (continuum and spectroscopic) over large areas with high enough spatial resolution, as the ones we propose to do with AtLAST. 
%Another idea could be mentioned: Study the residual CIB after substracting the increased population of sources identified in the surveys%
Such large, homogeneous survey of DSFGs allows for an angular clustering analysis, a determination of the (photometric) redshift distribution, number count estimates, and many other statistical properties. It will provide a high-$z$ counterpart to extensively studied large galaxy samples at low and intermediate redshifts. 

\medskip
Perhaps one of AtLAST's most important contributions to this field will be to study the role of the environment (voids, filaments, groups, clusters) and the way the evolution of DSFGs varies as a function of this environment. This is particularly complementary in the era of Euclid, LSST, Roman and SKA. 

\medskip
The relatively strong negative K-correction in the sub-mm for high-redshift dusty galaxies has been extensively exploited to efficiently: detect DSFGs (either lensed or unlensed) out to high redshifts, and study them in various amounts of detail, depending on the sensitivity and angular resolution of the sub-mm observatory used for the study. Currently, the ALMA interferometer has the best of both, it is not, however, able to study large samples due to its small field of view - ALMA is not a survey instrument. To study statistical populations of high-$z$ galaxies, and environmental effects on their evolution, a complementary, dedicated type of instrument is required, with a high survey speed, low confusion noise limits, and sufficient sensitivity, which is what AtLAST will deliver.

\medskip
We aim for a comprehensive multi-band imaging survey, uniquely mapping large parts of the sky to specifically target high-$z$ galaxies and map their distribution (noting that this could be done in combination with imaging surveys for other science cases). Using a multi-chroic camera, several of the bands can be observed simultaneously, allowing for accurate spectral slope determinations and photometric redshifts (especially in combination with complementary data available at other wavebands), as the observing conditions will be identical for each of the bands. This will provide a rich catalogue of sources to follow-up with ALMA, JWST, or ELT, but more importantly, yield a large homogeneous sample of galaxies in the early Universe.

\medskip
Wide-field ``blind'' spectral line surveys with AtLAST will be crucial for mapping the population of ``normal'' star-forming galaxies across the cosmic history. These are often too faint in dust emission to be detected in continuum surveys. However, as suggested by recent predictions from simulations (e.g., \citealt{Lagos2020}), while the cosmic star-forming activity is dominated by sub-mm bright galaxies ($S_\mathrm{350\,GHz}\geq$ 1~mJy), the gas budget of the Universe is dominated by sub-mm faint galaxies (quantitatively: $\approx$75\% of the gas budget at $z=2$, and $\geq$90\% at $z\geq3$). Deep, wide-bandwidth spectroscopic surveys of CO and [CII] emission with AtLAST will be critical for mapping the evolution of cold-gas content across the cosmic history.
Especially [CII] is a very valuable tracer, physically. It correlates well with the star formation rate \citep{Delooze2014}, is a valid tracer of the bulk of the gas mass (see \citealt{Wolfire2022} for a recent review, and \citealt{Zanella2018} for a high-z empirical study), and it is one of the brightest FIR lines out to high-z, as demonstrated by the detection rate of recent ALMA Large Programs, e.g. ALPINE \citep{Faisst2020,Bethermin2020} and REBELS \citep{Bouwens2022,Inami2022}.

\medskip
In the following we explore two worked examples of surveys that AtLAST will make possible. Often a 'wedding-cake" approach is followed, where several surveys are planned with different angular sizes and sensitivity limits, each forming a layer of an imaginary wedding cake. Our two worked examples form the two extremes: the bottom and top layers of the cake, but we will certainly consider the other layers as well, although these could hit the confusion limit for the longer wavelengths if the survey area is too small. An additional pointed survey of galaxy clusters is discussed in Section 2.4.1. With respect to the continuum survey, we list a likely set of frequency bands with associated sensitivities and beam sizes for AtLAST in the companion high-$z$ paper by Di Mascolo et al.\ (in prep.). These are optimized for a range of AtLAST science goals, including the ones presented in this paper. 

\subsubsection{A wide continuum survey}

\begin{figure}[h]
\centering
    \includegraphics[width=0.48\textwidth]{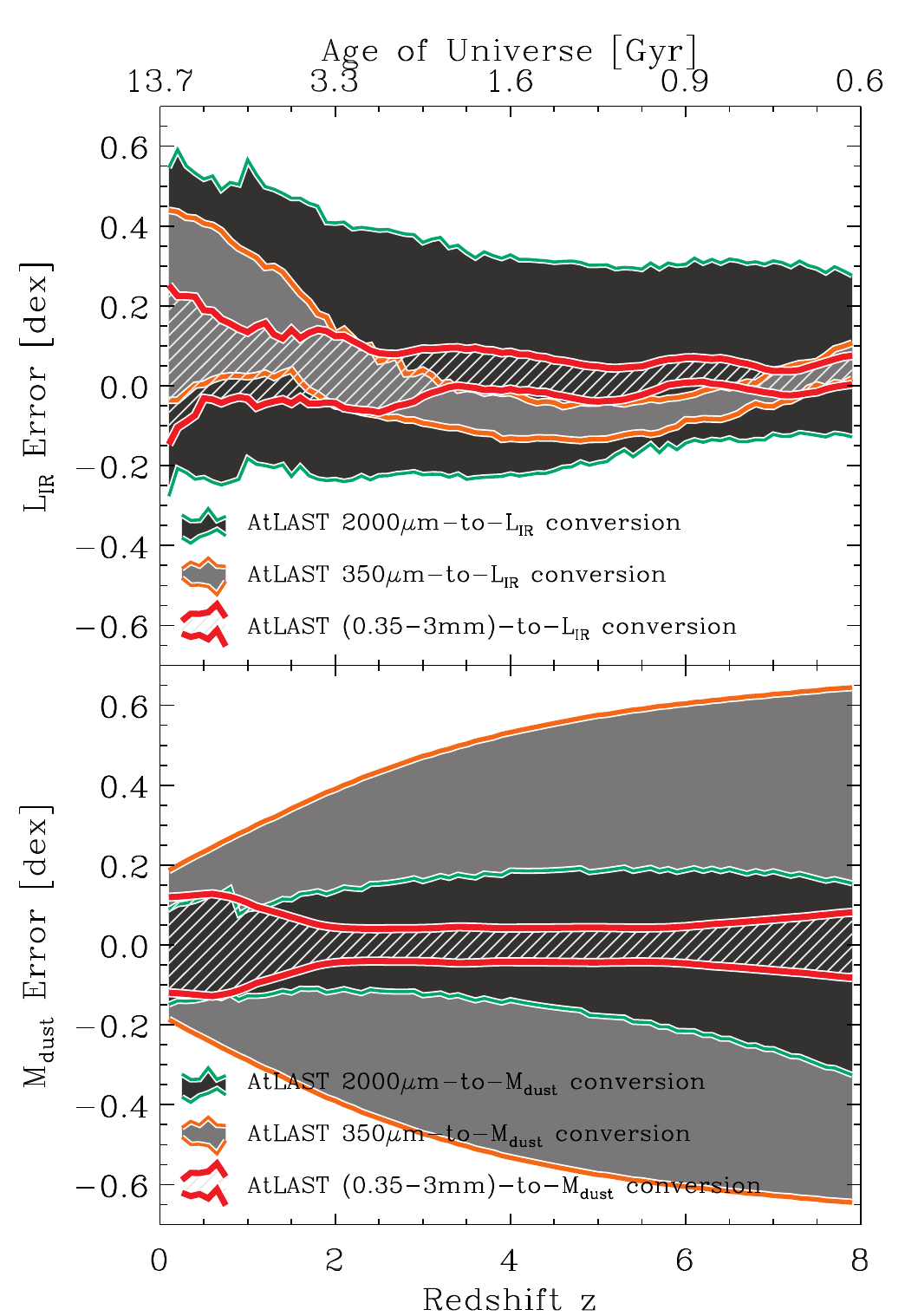}
    \caption{The mean (top panel) and dispersion (bottom panel) of  log$_{10}$(L$_{IR-BB}$ / L$_{IR-True}$) and log$_{10}$(M$_{dust-BB}$ / M$_{dust-true}$) for mock observations of the SED library of \cite{DraineLi2007}.}
    \label{fig:accuracy}
\end{figure}

\medskip
To estimate what can actually be achieved with AtLAST, we consider a 1000 deg$^2$ mock galaxy survey as could be obtained in 1000 hours of observing time in the continuum in two or more bands. Such a continuum survey allows one to infer two important physical properties of galaxies: their infrared luminosities L$_{IR}$ (and thus SFR$_{obscured}$) and dust masses (subsequently M$_{ISM}$, assuming a given gas-to-dust ratio). To demonstrate the advantage of a multi-wavelength approach, we estimated the accuracy one would achieve while deriving L$_{IR}$ and M$_{dust}$ as a function of redshift for different available bands. After exploring almost all possible band combinations, for the purposes of this work we considered the following three cases: (i) galaxies only detected at 2000$\mu$m, (ii) galaxies only detected at 350$\mu$m, and (iii) galaxies detected at (350 or 450) $\mu$m and (750 or 850 or 1100) $\mu$m and (1300 or 2000 or 3000) $\mu$m. To infer these accuracies, we assumed that the diversity of SEDs in the Universe is given by the SED library of \cite{DraineLi2007}, and we fitted mock observations (assuming signal-to-noise ratios of 5) of these templates with a blackbody function (T$_{dust}$ in the range 10-60K). In Fig.~\ref{fig:accuracy} we show the mean and dispersion of the log$_{10}$(L$_{IR-BB}$ / L$_{IR-True}$) and log$_{10}$(M$_{dust-BB}$ / M$_{dust-true}$). This shows that 350$\mu$m is a good L$_{IR}$ proxy from $z\sim$1-8 (i.e. probing the rest-frame 40-120$\mu$m range of the SED), but a poor proxy for the dust mass. 2000 $\mu$m tells the opposite story, i.e., a good dust mass proxy (because of the Rayleigh-Jeans tail) but a poor L$_{IR}$ proxy. Finally, one can do very well for both L$_{IR}$ and the dust mass if data in more than two bands are available, i.e. the third case we considered.
This is quantified in Fig. \ref{fig:accuracy}: the range of $T_{dust}$ explored is that of the \cite{DraineLi2007} SED library, so it does contain SEDs with high temperature. For detection, we used a signal-to-noise limit of 5, and the red lines in \ref{fig:accuracy} show that do better than 0.1 dex (25\%) at $z>1$.

\begin{figure}[h!]
\centering
    \includegraphics[width=0.48\textwidth]{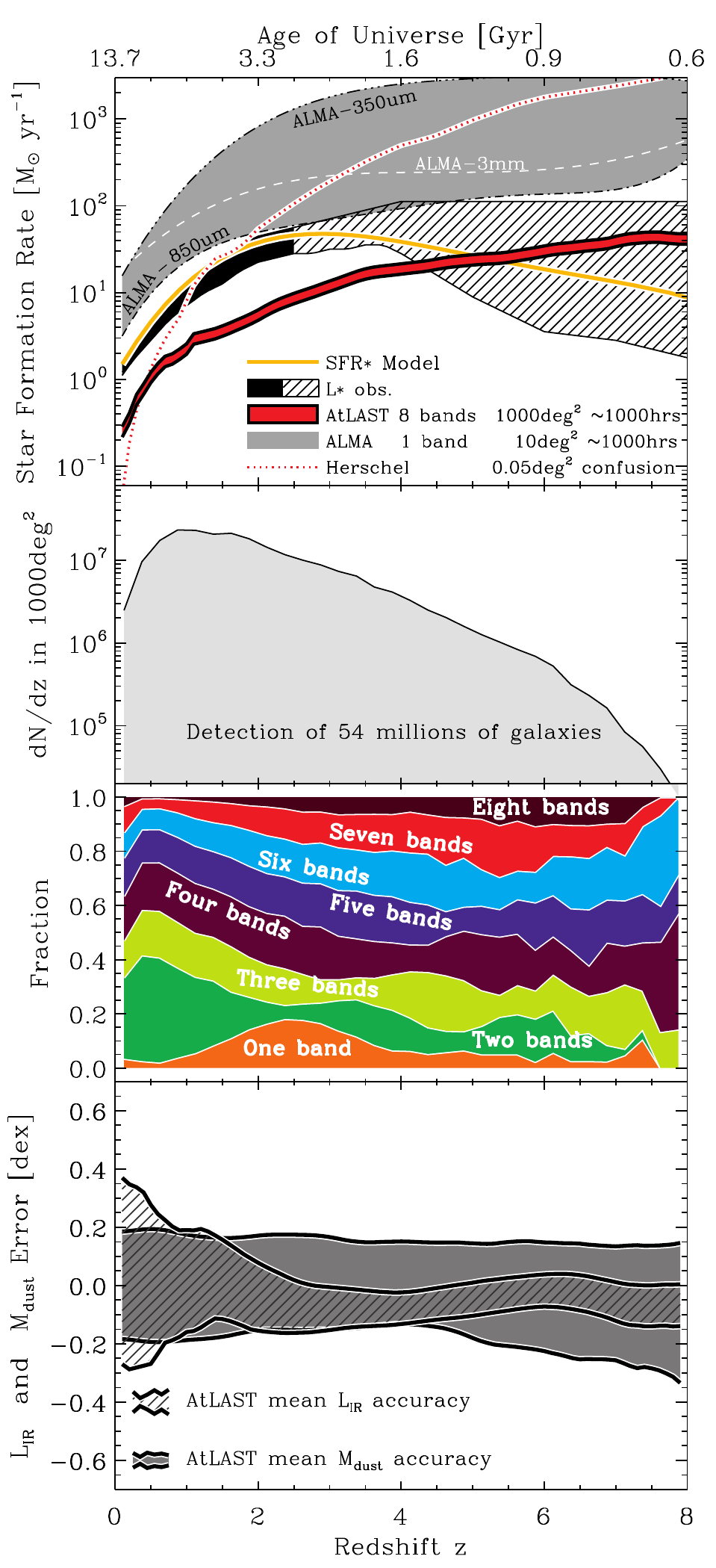}
    \caption{Exploring the parameter space of a 1000 deg$^2$ survey using 1000 hours of AtLAST observing time. Top panel: SFR vs. redshift survey limit. Second panel: $dn/dz$. The third panel shows the fraction of (mock) sources detected on 1, 2, 3, ..., or all 8 bands. 
    The bottom panel shows the expected AtLAST accuracy in inferring $L_{FIR}$ and $M_{dust}$ as a function of redshift (better than 25\% and 50\% at $z>2$, respectively), demonstrating why multi-band detections are important.
    }
    \label{fig:survey1000}
\end{figure}

\medskip
Using the expected mapping speed of AtLAST (fitted with a multi-chroic camera with a million pixels, which is what we expect for our first generation camera), a 1000 deg$^2$ survey (with 1000 hours observing time) results in a 3$\sigma$ sensitivity limit of 570 $\mu$Jy at 350 $\mu$m (at this limit, 82\% of the Cosmic Infrared Background at 350$\mu$m will be resolved into individual sources) and 324 $\mu$Jy at 450 $\mu$m. At lower frequencies we hit the confusion limit (e.g. \citealt{Blain2002}). Note that this implies that going for a much smaller field will only benefit the 350 $\mu$m and and 450 $\mu$m bands, and thus low-redshift galaxies science, where much work has already been done. With these limits we use the model of \cite{Bethermin2017} to explore the parameter space probed by this survey, which is shown in Fig.~\ref{fig:survey1000}.

\medskip
The first panel of Fig.~\ref{fig:survey1000} shows the classic star formation rate (SFR) versus redshift survey limit using the models of \cite{Bethermin2017} and classical scaling relations. AtLAST will be able to detect sources below SFR$_*$ up to z$\sim$5, where SFR$_*$ is the characteristic SFR defined as the value at the 'knee' of the classical Schechter function. We compare this to what Herschel and ALMA can do: for ALMA we considered a hypothetical 1000 hours survey in one ALMA band only (either 350 $\mu$m, 450 $\mu$m or 3mm), over 10 deg$^2$ (requiring tens of thousands of pointings at 0.85-1.3mm), assuming that the ALMA sensitivity will be twice as good as it is now (bandwidth increased by a factor 4, taking the ALMA WSU upgrade into account: \citealt{Carpenter2023}) and no overheads. The shaded region shows the range of star formation rates probed by such a survey. ALMA, with its small field of view, is clearly not efficient. The most optimal survey with ALMA would be at 850um, still barely reaching SFR$_*$ up to $z\sim$4. The second panel of Fig.~\ref{fig:survey1000} displays $dn/dz$ for our 1000 deg$^2$ survey, whereas the third one shows the fraction of sources with detection in only 1 band, only 2 bands, etc., up to all 8 bands. Over the full redshift range, about 70-80\% of our galaxies will have detection in at least 3 bands, and therefore can be used to simultaneously solve for L$_{IR}$ and dust mass, using photometric redshift estimates from AtLAST itself and elsewhere.

\medskip
With a facility like AtLAST, we will, for the first time, have the far-infrared SED (L$_{IR}$, M$_{DUST}$, T$_{DUST}$) of all the relevant star-forming galaxies (SFR larger than halve of SFR$_*$) from $z\sim$0 to $z\sim$5-6. This will allow statistically-sound studies on the star forming and inter-stellar matter content of galaxies even while dividing our sample in many redshift, mass, metallicity, environment, and morphology subsamples. Such a survey will prove extremely valuable to complement already planned large optical/near-infrared surveys from which our photometric redshifts will be drawn (combined with the multi-band AtLAST data). The bottom panel of Fig.~\ref{fig:survey1000} shows the average L$_{IR}$ and M$_{DUST}$ accuracies of our survey. This quantifies why having detections in several bands for most of our galaxies is important (70-80\% of the galaxies will have at least bands, as shown in the third panel).
We will have an accuracy better than $\sim$0.3 dex for both L$_{IR}$ and M$_{DUST}$ for over 78 million galaxies, which will be unprecedented. This survey does need good 350/450$\mu$m conditions, so observing should be spread over four years at least, most likely.
Interestingly, this continuum survey will contain many Virgo/Coma-like structures up to $z\sim$2, and group/poor clusters up to $z\sim$6. We discuss how to obtain a large, complete cluster sample in Section 2.4.

\begin{figure}[h!]
\centering
    \includegraphics[width=0.48\textwidth]{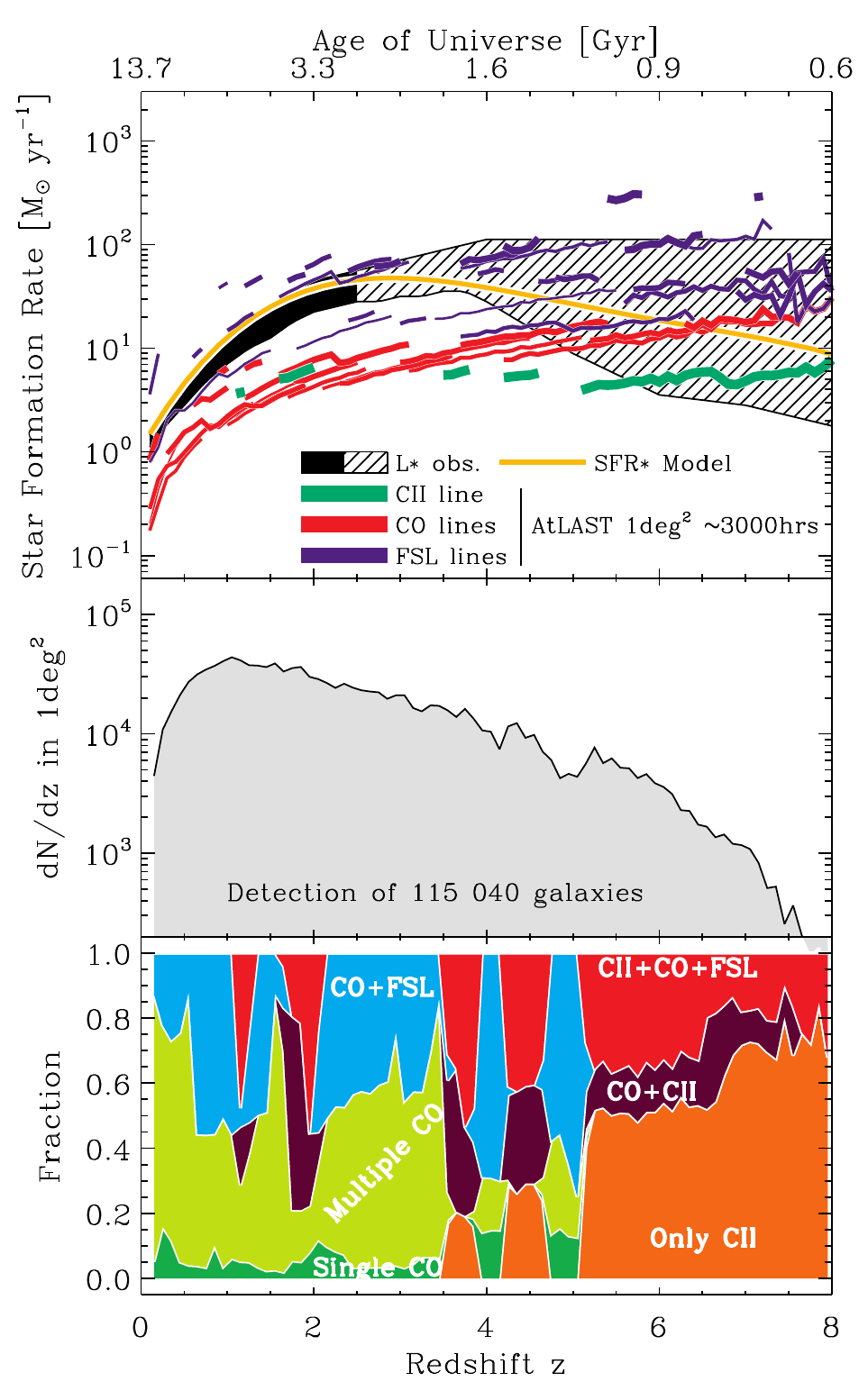}
    \caption{Exploring the parameter space of a 1 deg$^2$ deep line survey using 3000 hours of AtLAST observing time. The top panel shows, for each line considered, the minimum star formation rate (SFR) a galaxy must have for be detected at a given redshift (see main text for more details). The second panel displays $dn/dz$. The third panel shows the fraction of (mock) sources with only CII detection, only a single CO detection, only multiple CO detection (no CII, no FSL). }
    \label{fig:survey_line_1}
\end{figure}

\subsubsection{A deep “blind" spectroscopic survey}

\medskip
The high density of spectral features and large spectroscopic bandwidths of optical spectrographs make the optical regime an excellent wavelength range to determine the redshifts of large samples of galaxies.
% Determining large quantities of spectroscopic redshifts has traditionally been done at optical wavelengths because the high density of spectral features and the large spectroscopic bandwith of optical spectrographs. 
For example, one of the most ambitious new redshift surveys will be done using {\it Euclid}, with a target number of 1.5$\times$10$^8$ galaxy redshifts \citep{Euclid2011}. However, optical redshifts provide a biased view, missing most of the dust obscured objects. This is particularly important for the most highly star-forming objects which tend to be obscured by their large reservoirs of interstellar dust. The redshifts of these dusty star-forming galaxies (DSFGs) have been first attempted with optical spectroscopy \citep[e.g.][]{Chapman2005}, but it has since become clear that direct mm spectroscopy is a much more exact and efficient method \citep[e.g.][]{Weiss2007,Reuter2020,Chen2022,Cox2023}. 

\medskip
Therefore, in addition to a very wide continuum survey, we also estimate what a deep line survey with AtLAST can achieve. We consider a factor of three increase in time (3000 hours), which is realistic as a line survey uses mostly the low-frequency part of the spectrum:  the 350/450$\mu$m bands do not provide many lines but for CII at $z\sim$1.5-2.0. The exercise here is to see what a 3000 hours spectrocopic survey with AtLAST delivers.

\medskip
Again using the expected mapping speed of AtLAST, for a deep 1 deg$^2$ line survey (with 3000 hours observing time) in a 400km/s channel (R=750), we can estimate the sensitivity limits for the various typical bands. Not listing all available bands, we find (assuming no confusion) 3$\sigma$ sensitivity limits ranging from 2330 $\mu$Jy at 350 $\mu$m, 210 $\mu$Jy at 850 $\mu$m, to 27 $\mu$Jy at 3~mm. With these limits we employ the model of \cite{Bethermin2017} to model the galaxy population and predict their CO and fine structure line (FSL) peak flux densities, assuming a line width of $\sim$400 km/s (i.e., matching our channel width). We do this for the CO lines from J$_{up}$=1 to J$_{up}$=7, assuming L’$_{CO(1-0)}$/L$_{IR}$ = 40 and a sub-mm galaxy CO ladder as in Tab.~2 of \cite{CarilliWalter2013}. We take L$_{CII}$/L$_{IR}$ $\sim 3\times10^{-3}$ (not in the deficit part as our survey will be dominated by SFR$_*$ galaxies). Other FSLs are CI$_{610}$, CI$_{370}$, NII$_{205}$, OI$_{146}$, NII$_{122}$, OIII$_{88}$, OI$_{63}$ and OIII$_{51}$, for which we use an FSL$_{line}$/L$_{IR}$ ratio as found in the literature \cite{GraciaCarpio2011, Bothwell2016, Zhao2016, CarilliWalter2013, Schimek24}.

\begin{figure*}[h!]
    \centering
    \includegraphics[width=\textwidth]{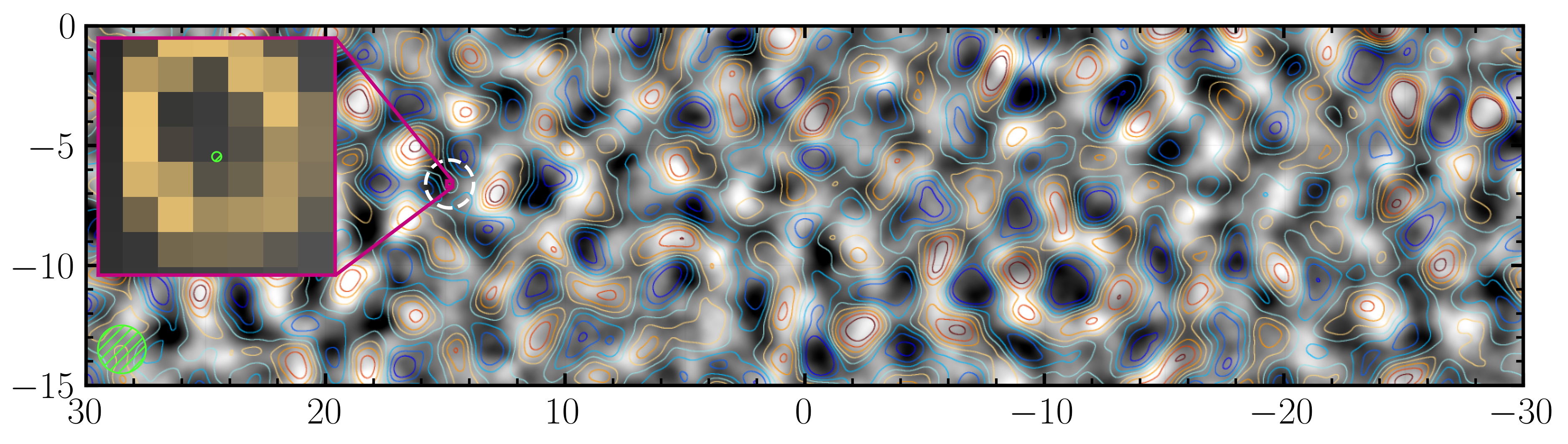}
    \caption{A 900 deg$^{2}$ portion of the map of the gravitational mass distorting the Cosmic Microwave Background obtained using the new Atacama Cosmology Telescope Data Release 6, overlayed with the Cosmic Infrared Background (CIB, sampled at a frequency of 545~GHz by Planck) tracing the distribution of dusty galaxies, peaking at $z\sim0.5-5$. The comparison shows a perfect match between the CMB lensing map and the sub-mm tail of the CIB as seen by Planck, where the over-densities of the latter (orange contours) correspond well with the bright peaks of the CMB lensing map. The white dashed circle shows the 2-deg-diameter field of view of AtLAST, and the inset shows the AtLAST resolution element computed at 545 GHz ($2.4''$, the diffraction-limited beam full width half power). Figure adapted from \cite{Madhavacheril2023}.}
    \label{fig:CIB_ACT}
\end{figure*}

\medskip
For this survey setup we again plot the parameter space probed by this line survey, now in Fig.~\ref{fig:survey_line_1}. The top panel shows, for any given line considered, the minimum SFR a galaxy must have to be detected at a given redshift (for CO, the thin line is for J=1-0, and the thickest line for J=7-6; for FSL, the thin line is for CI$_{610}$, and the thickest line for OIII$_{51}$; i.e. plot line thickness increases with increasing energy). Such a survey will basically provide multiple line detection for galaxies below SFR$_*$ up to $z\sim$7. In particular, SFR$_*$ galaxies at $4<z<7$ should all have one CO detection and one [CII] detection, reminding ourselves that multiple line detection is crucial for redshift determination. Of course, [CII] at high redshift detects sources well below SFR$_*$. One can argue that using energetic considerations, such single line detections could be used on their own to constrain the redshift of these sources.
Note that ALMA will be a factor 
%~sqrt[ (nb_AtLAST_beam/nb_ALMA_beam) * (AtLAST_collecting_area/ALMA_collecting_area) * (AtLAST_bandwidth/ALMA_bandwidth) ] = sqrt[ (500/1) * (1900/4900) *  (395/32) ] = 
$\sim$ 50 less sensitive for a 3000 hrs / 1 deg$^2$ survey. The second panel of Fig.~\ref{fig:survey_line_1} shows $dn/dz$ for our 1 deg$^2$ mock line survey. Finally, the bottom panel of Fig.~\ref{fig:survey_line_1} shows the fraction of sources with only [CII] detection, only a single CO detection, and only multiple CO detection (no [CII], no FSL). About 90\% and 50\% of our galaxies at $z<5$ and $z>5$, respectively, will have multiple line detections, which is excellent. 

\medskip
All this means that an AtLAST 1 deg$^2$/3000 hrs line survey (a 400 km/s channel, i.e. R=750) will basically give us multiple line detection for SFR$_*$ galaxies up to $z\sim$7. Combining this with multiple band continuum detections will allow us to obtain a wealth of information for each of these SFR$_*$ $z<7$ galaxies: the redshift, gas content, cooling budget, star formation rate, dust mass, and dust temperature.
A survey going much wider than 1 deg$^2$ in the same amount of time will loose many of the CO lines, but will still detect [CII].

\subsection{Constraining cosmological parameters via BAO and clustering}

\medskip
One of the most fundamental issues in modern cosmology is the significantly different values of the Hubble constant measured from the CMB (the early universe) and those based on late time observations, colloquially known as 'Hubble tension' (see \citealt{DiValentino2021} for a review). Improving and expanding the state-of-the-art measurement methods is one way to tackle the Hubble tension. Clustering of galaxies measured in large samples with spectroscopic redshifts provides a powerful means in this regard, and has been shown to be one of the standard references since the first significant measurements of the redshift space distortion (RSD; \citealt{Peacock2001}) and the Baryon Acoustic Oscillation (BAO, sound waves from the embryonic Universe; \citealt{Cole2005}). This effective approach has been realized and the cosmological parameters, especially the growth rate, f$\sigma_8$, and the Hubble constant, have been measured to percent level precision up to $z\lesssim1$ \citep{Bautista2021,DES2024}, and the new surveys that will be carried out by Dark Energy Spectroscopic Instrument and the Subaru Prime Focus Spectrograph will push that limit to $z\sim2.4$ \citep{Takada2014,DESI2016}. 

\medskip
Similar measurements beyond $z\sim2.4$ may start to become challenging for optical and near infrared observations, since most of the strong lines move toward longer wavelengths that are hard to access from the ground, and as galaxies become fainter at higher redshifts and so do their lines. Measurements in the FIR, on the other hand, become more easily accessible thanks to the shifting of the bright [CII] and CO lines to the more transparent atmospheric windows. Indeed, detecting these far-infrared fine-structure lines from galaxies at $z>2$ has become a regular practice for extragalactic studies in the early cosmic time (e.g., \citealt{Bethermin2020,Bouwens2022})

\begin{figure*}[h]
    \centering
    \includegraphics[width=\textwidth]{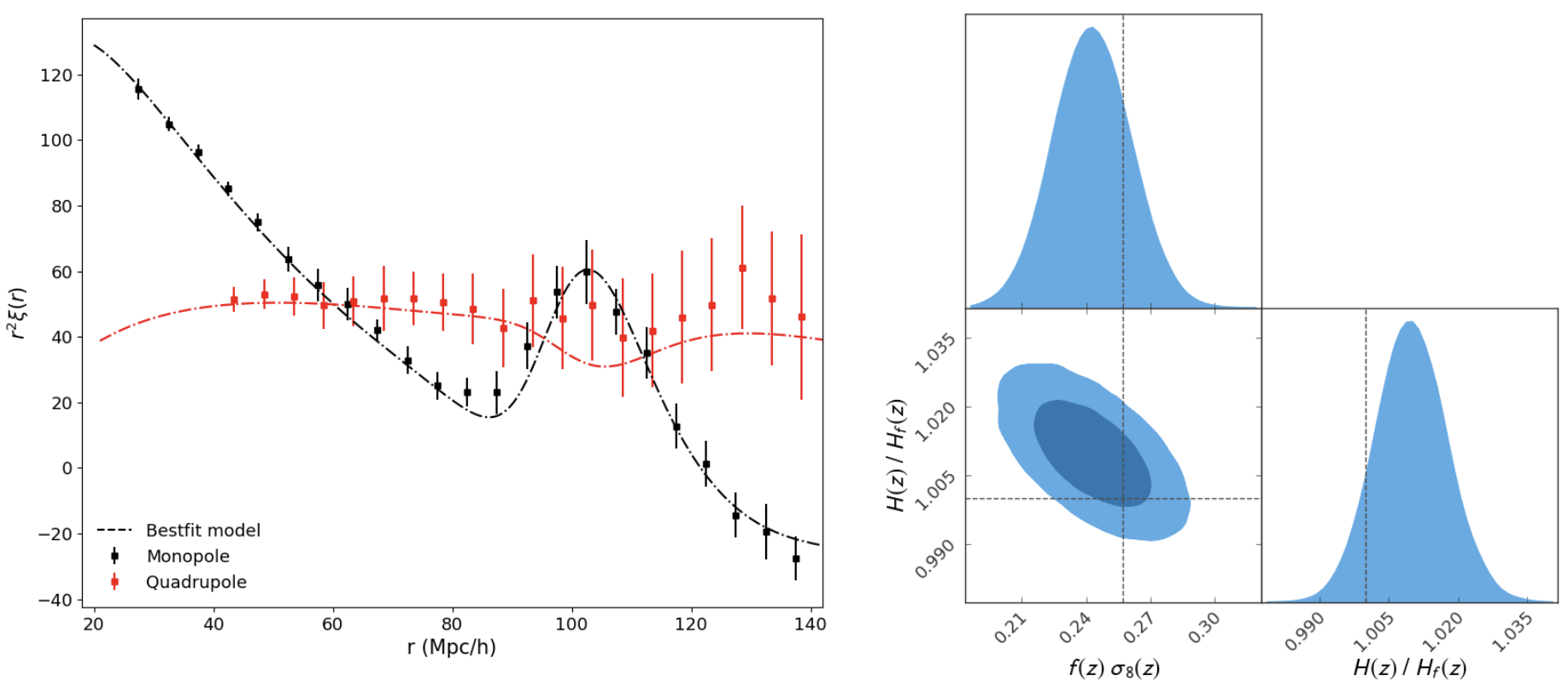}
    \caption{{\it Left:} Data points with errors are monopole and quadrupole moments of the correlation functions measured using dark matter halos in a simulated dark matter only box, and the curves are the best-fit models including four parameters, $f\sigma_8$, $b\sigma_8$, $H(z)$ and $D_A$. Both BAO and RSD are considered in these fittings. Data points are shown within the separation ranges on which the fittings are performed.
    % The dashed vertical line marks the lower boundary of the scale above which the fittings are performed. 
    {\it Right}: The results of the MCMC analyses with one and two sigma confidence regions colour coded by different darkness of blue. Here $b\sigma_8$ and $D_A$ are marginalized over.}
    \label{fig:BAORSD}
\end{figure*}

\medskip
A high-redshift galaxy spectroscopic survey with AtLAST would combine the successful strategy from the optical/infrared surveys that measured galaxy clustering in the last few decades (eg. \citealt{Wilkinson2017,Cochrane2018,Amvrosiadis2019,Johnston2021,Mohammadjavad2023,vanKampen2023}, and many references therein) with the new possibility of measuring strong emission lines from the FIR regime for large samples of galaxies at $z\gtrsim3$. 

\medskip
To test the feasibility and help design such a survey, as a first step, we have conducted a simple forecast study using dark matter only cosmological simulation at $z=3$. We use $N$-body simulation runs that are part of the \texttt{D{\footnotesize ARK} Q{\footnotesize UEST}} project \citep{Nishimichi2019}. The simulation box has 2 Gpc/h on a side with the Planck cosmology as the fiducial cosmological model. Halos are identified using the \texttt{\footnotesize ROCKSTAR} algorithm \citep{Behroozi2013}. The halo mass is defined by a sphere with a radius $R_{\rm 200m}$ within which the enclosed average density is 200 times the mean matter density, as $M_h\equiv M_{\rm 200 m}$. Motivated by recent clustering measurements of dusty star-forming galaxies at $z>1$ \citep{Lim2020,Stach2021}, we apply a selection of halos with masses of $10^{12.5}-10^{13.5}$ solar masses, which yields about 800k halos for the analyses. 

\medskip
% The two-point correlation functions are measured using the Landy-Szalay estimator \citep{Landy1993}. Integral constraint is added to the measured correlation functions in order to address the issue of the finite survey volume \citep{Peebles1976}. Covariance matrix is estimated with the standard bootstrap method \citep{Norberg2009}.

\medskip
The two-point correlation functions are measured using the selected simulated halos, and covariance matrix is estimated with the standard bootstrap method \citep{Norberg2009}. The models considered for the fitting of the correlation functions are the fiducial $\Lambda$-CDM, Baryonic Accustic Occasilations (BAO; \citealt{Eisenstein2005}), linear redshift space distortion (RSD; \citealt{Kaiser1987}), and the Alcock-Paczynski effect \citep{Alcock1979}.  In this model, we have four free parameters in total, bias ($b\sigma_8$), growth rate ($f\sigma_8$), Hubble parameter ($H(z)$) and angular diameter distance ($D_A$). To break the degeneracy between bias and the growth rate, we perform fittings of the monopole and the quadrapole moments of the measured correlation functions, over a pair separation range of 25-140\,Mpc/$h$ and 40-140\,Mpc/$h$, respectively.

\medskip
The results are plotted in Figure \ref{fig:BAORSD}, where the left panel shows the results of the fittings of the correlation functions and the right panel shows the MCMC results of the cosmological parameters, including the Hubble parameter and the growth rate. The other two parameters, bias and angular diameter distance, are marginalized over. 
% bias ($b$), the Hubble constant ($H_0$), matter density ($\Omega_m$), spatial curvature density ($\Omega_k$), and the amplitude of the primordial fluctuations ($A_s$) which is similar in concept as $\sigma_8$ but in the Fourier space. 
In summary, the Hubble parameter can be measured at 0.7\% precision and the growth rate at 7.3\%. 

\medskip
The above analyses demonstrate the potential constraining power of a baseline design of a AtLAST spectroscopic survey; that is, a spectroscopic survey of hundreds of thousands of sources with a footprint of about 1000 square degrees, which can be achieved with a survey of a few thousand hours when employing the bright [CII] line, as shown in the previous sections. 
%\eh{link to the description of the survey in previous sections}. 
% The current, to be one step closer of having a more realistic forecast, similar analyses need to be performed on mock SMG catalogues.

\begin{figure}[h!]
    \includegraphics[width=0.48\textwidth]{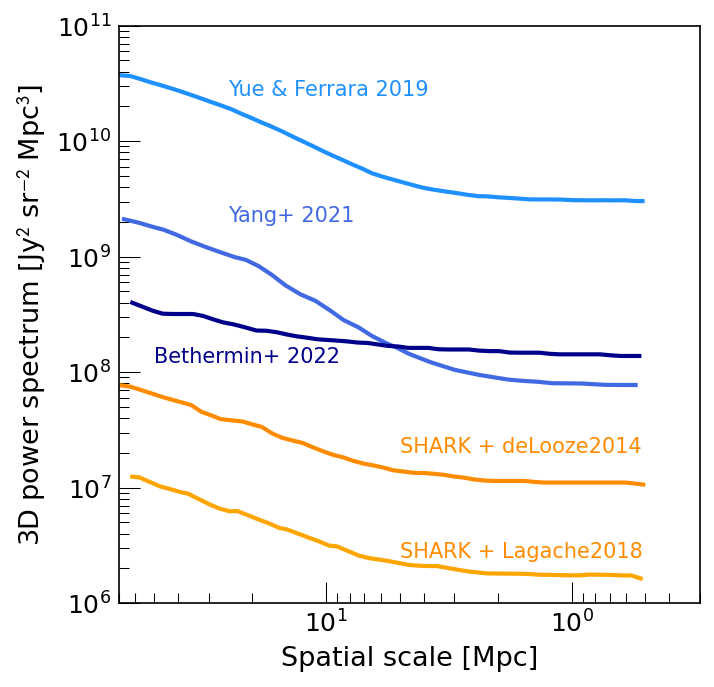}
    \caption{Line-intensity mapping: predicted 3D power spectrum of the [CII] emission at redshift $z=3$. Individual curves correspond to different analytical \citep{Yue2019, Yang2021} and N-body + semi-analytical models \citep{Bethermin2022, Lagos2020}. We also show two realisations of the SHARK semianalytical model coupled with two different prescriptions for [CII] emission \citep{Delooze2014, Lagache2018}. The predictions vary by over 3~dex; the large FoV and sensitivity of AtLAST will be critical for measuring LIM signal across cosmic time.  Adapted from \citet{Bethermin2022}, J. Hilhorst (MSc thesis).}
    \label{fig:LIM}
\end{figure}

\subsection{Line-intensity mapping (tomography)}

\medskip
A novel method for studying the large-scale structure of the Universe is the line-intensity mapping (LIM) of the [CII] and CO emission lines. Specifically, line-intensity mapping measures 2- and 3-dimensional power spectra from spectral cubes, providing a statistical view of the large-scale structure across cosmic time. LIM experiments at sub-mm wavelengths bridge the optical surveys at $z\leq1$ and the upcoming radio surveys of the 21-cm HI emission line in the Epoch of Reionisation ($z\geq6$). Namely, the optical surveys become inefficient at higher redshifts as the bright optical lines move into infrared wavelengths and the dust obscuration increases; conversely, the 21-cm line signal peters out at $z\leq6$ as the neutral hydrogen in the intergalactic medium becomes fully ionised. The [CII] and CO lines remain bright across this redshift range and do not suffer from dust obscuration, making them ideal tracers of the large-scale structure.

\subsubsection{Predictions for LIM signal}

\medskip
The predictions for LIM mapping signal (and the associated foregrounds) have been explored theoretically using different approaches: from simple analytical models (e.g., \citealt{Yue2019}) to dark-matter only simulations combined with semi-analytical models (e.g., \citealt{Bethermin2022}) and cosmological-volume hydrodynamical simulations (e.g., \citealt{Karoumpis2022}). The predictions for the resulting 2D and 3D power spectra vary by \emph{several orders of magnitude} (Fig.~\ref{fig:LIM}), chiefly due to differences in the assumptions on galaxy evolution and predictions of emission line intensity.

\subsubsection{Current observations: lack of constraints}

\medskip
Several teams have conducted early LIM experiments on 10-metre class telescopes. CONCERTO, a scanning Martin-Puplett interferometer with MKID detectors covering 125 -- 310~GHz frequency range, was installed on the APEX telescope in 2021 \citep{Ade2020, Monfardini2022}, mapping $\approx$1.4~deg$^2$. TIME is 16-pixel grating spectrometer observing the 200--300~GHz band, mounted on the 12-m ARO telescope \citep{Crites2014, Li2018}. At lower redshifts, COMAP \citep{Cleary2022, Lamb2022} is a 19-pixel heterodyne spectrometer mapping the CO(1--0)/(2--1) emission lines in the 26--34~GHz band. These experiments are currently limited by the small collecting areas, low pixel count, and small survey areas (few deg$^2$); rather than detecting the LIM signal at high redshift, they will provide upper bounds, potentially ruling out some of the more ``extreme'' models. AtLAST will supersede these facilities by providing a much larger collecting area, large focal plane, and superior site and dish quality.

\medskip
One of the key challenges in measuring the 2/3D power spectra of, e.g., [CII] emission, is the need to remove ``interlopers'': either CO and [CI] lines from lower-redshift galaxies, or [OIII] emitters at higher redshifts. This can be achieved by several techniques, such as masking (known) foreground galaxies (e.g., \citealt{Bethermin2022}) or using the periodicity of CO emission lines to separate the CO and [CII] power spectra (e.g., \citealt{Yue2019}).

\begin{figure}[h!]
    \centering
    \includegraphics[width=0.49\textwidth]{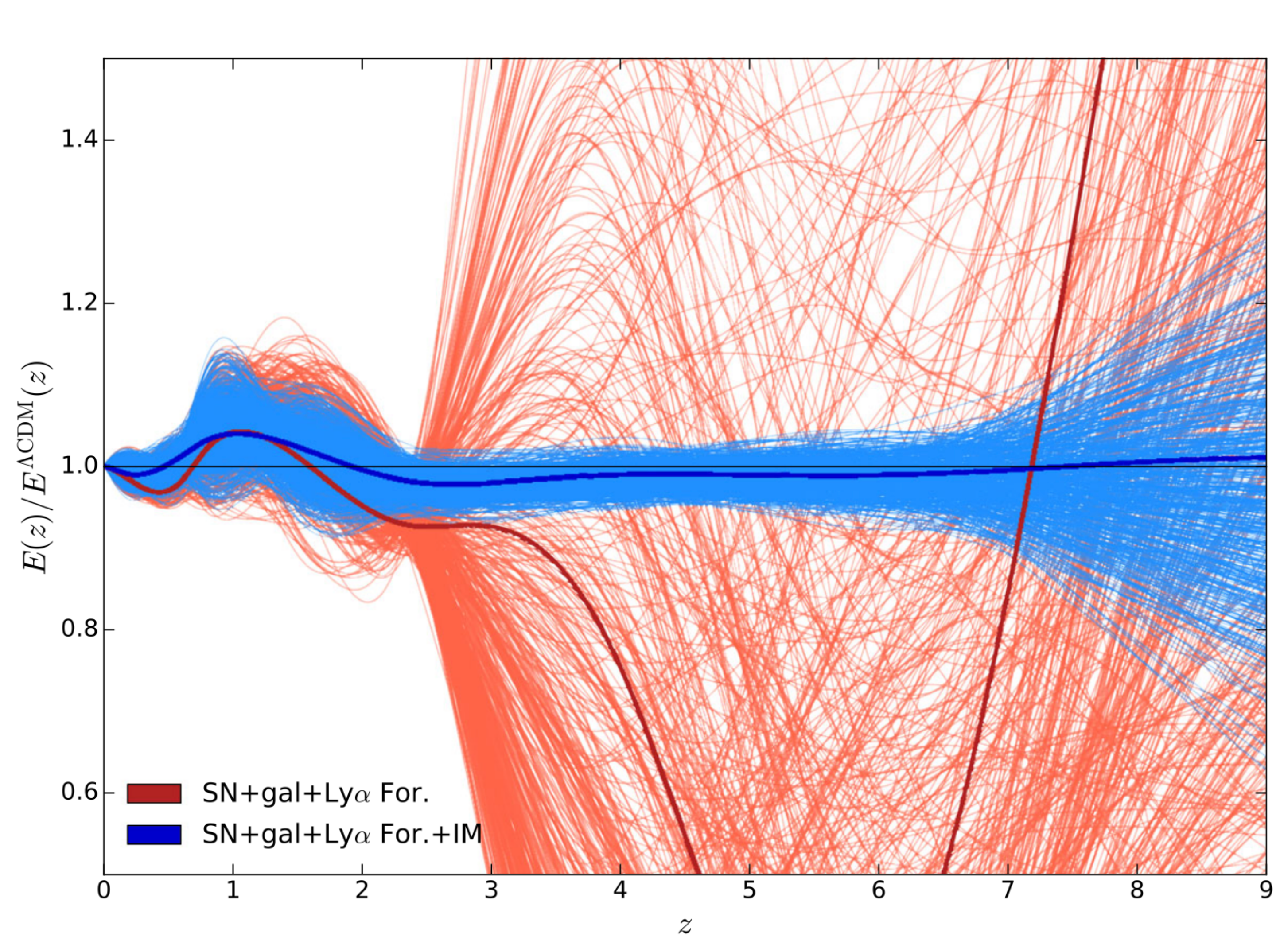}
    \caption{Model-independent constraints on the shape of the cosmic expansion history, $E(z) = H(z)/H_0$ (normalised to $\Lambda$CDM), with (blue lines) and without (red lines) CO(1-0) Line-Intensity Mapping over a 1000 deg$^2$ patch. For details, see \cite{Bernal2019} and \cite{Silva2021}, from which this figure was taken. It emphasizes the importance of LIM in constraining the cosmic expansion history, especially at $z>3$.}
    \label{fig:lim_expansion}
\end{figure}

\medskip 
The power of CO LIM to constrain the cosmic expansion history $H(z)$, i.e the Hubble expansion rate as a function of redshift, is illustrated 
in Fig. \ref{fig:lim_expansion}, especially for $z>3$. This figure, taken from \cite{Silva2021}, and based on the work of \cite{Bernal2019}, compares using only Supernovae (SN), galaxy surveys and the Lyman-$\alpha$
forest (red lines) to a combination of these with a LIM experiment measuring CO(1-0) over a 1000 deg$^2$ area. These calculations were not specifically performed for AtLAST, but do illustrate well the how much LIM with AtLAST can contribute to constraining the cosmic expansion history.

%\tcr{When you put some text please consider answering the following questions.}\\
%\emph{What is the current status from the observational point of view?}
%\emph{Is there any simulation/theoretical forecasts relevant to the proposed science case?}
% Apologies for the stream-of-conciousness write-up. I 've made a small tool that identifies the redshift particularly using CO lines, which have several considerations. If it is too detailed, feel free to remove, but it might be a useful tool to quickly see the effect of different instruments in a sub-mm 3D sky map.

%\medskip
%\emph{What observations are required to answer the remaining open questions?}
%\emph{Why is AtLAST \citep{Ramasawmy2022} the only telescope that will allow to pursue the proposed scientific goal?}
%\emph{What are the potential synergies with other wavelength studies?}

\subsection{Surveying cluster galaxies in the distant Universe}

\begin{figure}[h!]
    \centering
    \includegraphics[width=\linewidth]{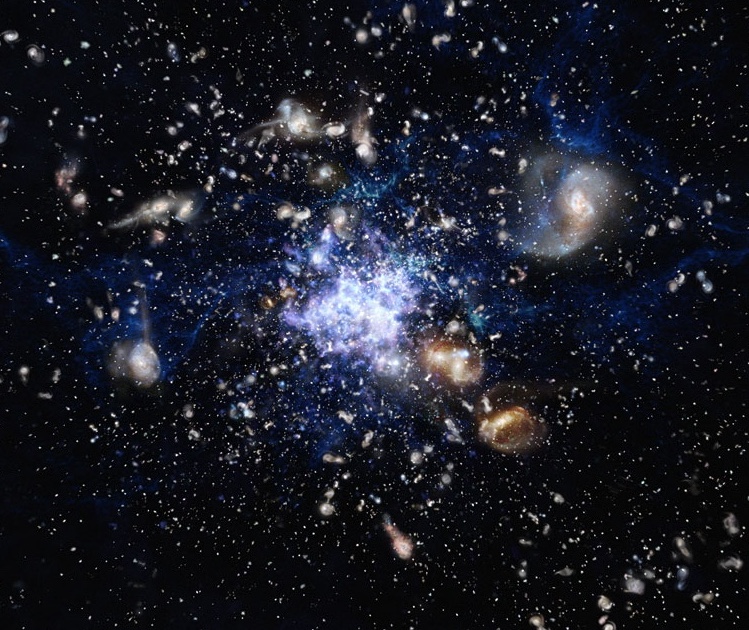}
    \caption{This artist’s impression depicts the formation of a galaxy cluster in the early Universe. The galaxies are vigorously forming new stars and interacting with each other. They are observed as Far-Infrared/Sub-mm Galaxies or Dusty Star-Forming Galaxies. Credit: ESO/M. Kornmesser. Courtesy from ESO Press Release October 2014.}
    \label{fig:clus_f1}
\end{figure}

\medskip
Galaxy clusters are the first large structures to form and eventually evolve into the largest virialised objects in the Universe. They should therefore be seen as the earliest fingerprint of galaxy formation and evolution (e.g. the review of \citealt{Kravtsov2012}). Clusters grow hierarchically through the merging and accretion of smaller units of galaxy halos, which are dominated by (very) young galaxies displaying intense bursts of star-formation --- the dusty star-forming galaxy population (DSFGs; see for a review \citealt{Casey2014}). These are rich in molecular gas but also heavily obscured by dust, which makes them prime targets for far-infrared/submm facilities \citep{AlbertsNoble2022}. These early cluster galaxies are most probably the progenitors of elliptical galaxies (eg. \citealt{Lutz2001, Ivison2013}) which end up dominating local galaxy clusters. Fig.~\ref{fig:clus_f1} (based on work by \citep{Dannerbauer2014} on a $z=2.2$ proto-cluster) shows how violent galaxy clusters could be in the distant Universe.
%These high density regions are remarkable environments for investigating the physical processes responsible for the triggering and suppression of star formation and black hole activity.
In this section we motivate the need of a systematic study of the early galaxy population in proto-clusters in order to make big leaps forward in this emerging research field, which we argue is best done with a large single dish sub-mm telescope at a high site. Please note that we focus here on the cluster galaxy population: a companion AtLAST case study by Di Mascolo et al.\ (in prep.) focuses on probing the ICM using the full SZ spectrum. 

\subsubsection{A systematic mapping survey of distant cluster galaxies}

\medskip
Presently, samples of galaxies in proto-clusters are small and heterogenous, cannot sample the full extent of the cluster infall regions, and take a lot of time to complete using current facilities. For this reason there are still relatively few cold ISM measurements of cluster galaxies. A future systematic (sub-)mm survey of high-redshift (proto-)galaxy clusters will resolve this situation, and allow us to answer the following scientific questions in detail: 1) How do galaxies and clusters co-evolve at early times? 2) How does environment (especially in over-dense regions) affect star formation, enrichment, outflows and feedback processes? and 3) What is the time evolution of each of these processes? When and where do they peak?

\medskip
Currently the number of known spectroscopically confirmed (proto-)clusters beyond $z=2$ is still relatively low (see for a review \citealt{Overzier2016}), where the targets are often sparsely sampled. The heterogeneous datasets collected so far suffer from strong selection biases and projection effects \citep[e.g.][]{Jianhang2023}, which prevent obtaining a complete picture of the build-up of the cluster galaxy population over cosmic time. To remedy this, a large and statistical sample is required. However, each of these clusters and their infall region cover a linear extension of up to 30 Mpc \citep{Muldrew2015, Casey2016, Lovell2018}, which corresponds to about 30$^{\prime}$ at $z\sim1-7$. Therefore, to study and understand the epoch of cluster formation, one really needs to cover areas of up to one square degree, something current sub-mm facilities cannot practically do with adequate sensitivity and survey speed.

\medskip
For example, ALMA allows for a survey speed of at most a few square arcminutes per hour for the brightest CO lines \citep{Popping2016} to yield detections for a sufficient number of cluster galaxies with L$>$ L$_{\ast}$ at $z\approx1$. The survey speed for a given line depends on the sensitivity to that line and the primary beam at its frequency. This determines the number of pointings required for a given desired map size: obtaining an area of a square degree with ALMA would take at least 1000 hours for the line with the highest survey speed, CO(5-4). This line is hard to interpret physically, while CO(3-2) would take around 6000 hours with ALMA, and CO(2-1) would even need three times that \citep{Popping2016}. This renders one degree surveys for the lower transition lines prohibitively expensive with ALMA. Another property of ALMA is the relatively narrow bandwidth (just below 8 GHz) which makes spectral scans slow as well.

\medskip 
The past decade has seen a rise in several hundred detection's of the cold molecular gas supply that fuel the star formation in the distant Universe, albeit focusing mostly on isolated field galaxies \citep{CarilliWalter2013,Tacconi2018}. However, the number of published cold gas measurements of galaxies located in galaxy clusters at $z>1$ is fairly low (of order a hundred). Even though ALMA and ATCA allowed this number to increase significantly  \citep{Coogan2018,Dannerbauer2017,Hayashi2017,Hayashi2018,Noble2017,Noble2019,Rudnick2017,Stach2017,Tadaki2019,Jin2021,Williams2022,Cramer2023,AlbertsNoble2022}, it remains low nonetheless. 
In order to resolve these problems described above, we aim to produce a high-redshift counterpart to local large cluster galaxy surveys (eg. \citealt{ACO1989}).
This will help us understand the contribution of protoclusters to the obscured cosmic star formation rate density evolution \citep{Chiang2017}.

\begin{figure}[h]
\begin{centering} 
\includegraphics[width=1.0\linewidth]{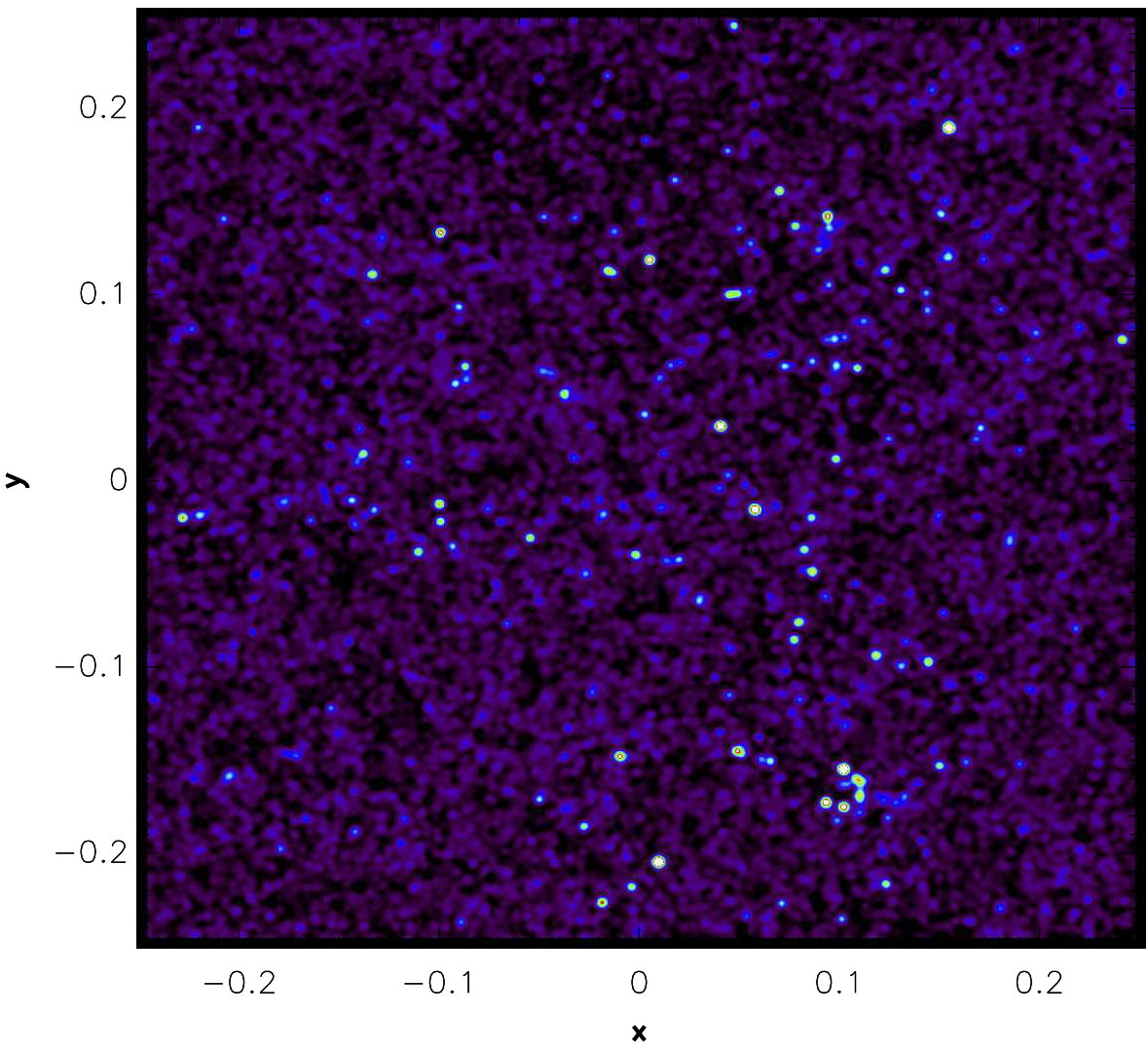}
\caption{Example mock image at 2.4~mm of the CO(3-2) flux of galaxies in and around a simulated $z=1.74$ proto-cluster, including homogeneous background noise and field galaxies along the line-of-sight. This $0.5\times0.5$ degree image constitutes a single pointing with a single-dish 50~m sub-mm telescope} \label{fig:protoclustersim}
\end{centering} 
\end{figure}

\subsubsection{The way forward}

\medskip
In order to make a big leap forward in understanding the evolution and formation of the largest structures and galaxies, a sub-mm observatory optimized for surveys is needed, i.e. a highly multiplexed instrument on a telescope with a single dish of at least 50~m. To visualize what such a telescope can achieve, a mock CO(3-2) image of a simulated proto-cluster at $z=1.74$ is shown in Fig.~\ref{fig:protoclustersim} (homogeneous noise is added), observed in a single pointing at 2.4~mm where the angular resolution would be of order of 12~arcsec for such a telescope, and up to 6$\times$ better at higher frequencies. This mock image is derived from a light-cone constructed out of a semi-analytical galaxy formation model \citep{vanKampen2005} in which a cluster simulation (using the same model set-up) was inserted at $z=1.74$ (a few hundred cluster galaxies were added in this way, of which around 50 are sufficiently bright to be detectable at the depth of this particular mock image). 

\medskip
We will target galaxies in already confirmed galaxy clusters (with known spectroscopic redshifts) as well as candidate clusters. In addition, we expect to discover new proto-clusters, especially dust obscured ones with large molecular gas reservoirs contained in their individual members, in the large surveys planned for AtLAST. At the time of conducting this survey with a large single dish telescope, a sample of several to a few ten thousand (proto-)galaxy clusters from $z=1-10$ will exist coming from future surveys and missions such as LSST and Euclid, the latter one with a dedicated study to discover galaxy clusters at high redshift \citep{Euclid2011}. Presently we already have a few thousand known candidate (proto-)galaxy clusters from Planck \citep{Aghanim2015,Ade2016} and Hyper Suprime-Cam \citep{Higuchi2019,Toshikawa2018}.

\section{Technical Justification}\label{sec:technical}

%\emph{Provide a brief summary and justification of the technical requirements for achieving the proposed science goals.}
%\emph{Will the proposed technical setup make AtLAST unique? and how?}
%\emph{How will the proposed setup make AtLAST complementary to other facilities?}

\medskip
To meet the observational requirements outlined here, we perceive the most salient instrumentation requirements to be the ability to achieve high mapping speeds over large areas, in multiple bands for continuum surveys, and with very wide bandwidths for emission line surveys. The current state of the art Transition Edge Sensor (TES) bolometers or Kinetic Inductance Detectors (KIDs) present the likeliest technological path to achieve this, as both of these are of high technical readiness level, have demonstrated background-limited performance in the mm/submm, and can be read out in large numbers (tens-hundreds of thousands as noted in \citealt{Klaassen2020}) through frequency multiplexing, allowing the construction of large imaging arrays and integral field units. In the following subsections we discuss the technical requirements for the science cases covered by this white paper.

%\begin{table*}
%	\centering
%    \rule{\textwidth}{0.4pt}\vspace{-6pt}
%    \caption{Frequencies, sensitivities and beam sizes for an AtLAST type of experiment. For more details refer to Di Mascolo et al.\ (in prep.), from which this table was duplicated for convenience. \medskip}
%	\begin{tabledata}{c^c^c^c^c^c} 
%
%    \header  band & ref.\ frequency & bandwidth &     beam &  sensitivity & survey noise \\
%    \header   --- &         [GHz] &       [GHz] & [arcsec] & [$\mathrm{\mu Jy~beam^{-1}~h^{1/2}}$] & [$\mathrm{\mu K_{cmb}-arcmin~h^{1/2}}$]\\

%    \row      2 &          42.0 &          24 &    35.34 &         6.60 &  2.40 \\
%    \row      3 &          91.5 &          51 &    16.22 &         6.46 &  1.27 \\
%    \row      4 &         151.0 &          62 &     9.83 &         7.14 &  1.21 \\
%    \row      5 &         217.5 &          69 &     6.82 &         9.22 &  1.86 \\
%    \row      6 &         288.5 &          73 &     5.14 &        11.91 &  3.71 \\
%    \row      7 &         350.0 &          50 &     4.24 &        23.59 & 12.26 \\
%    \row      8 &         403.0 &          38 &     3.68 &        39.98 & 34.70\\
%    \row      9 &         654.0 &         118 &     2.27 &        98.86 & $1.67\times10^3$\\
%    \row     10 &         845.5 &         119 &     1.76 &       162.51 & $3.70\times10^4$\\
% \end{tabledata}
%    \label{tab:freq_sens_beam}
%\end{table*}

\subsection{AtLAST as a sub-mm redshift machine}

\medskip
Because DSFGs are dust-obscured, their redshifts are best obtained in the mm/sub-mm wavebands instead of the classical optical or infrared parts of the spectrum.
As shown by e.g. \citet{Bakx2022}, a key requirement for such direct mm spectroscopy is a very wide frequency coverage, which is difficult to obtain with heterodyne receivers.
Several dedicated broad-bandwidth (but low spectral resolution) instruments were specifically designed for redshift searches on single-dish (sub-)mm telescopes, such as Z-Spec \citep{Naylor2003}, Zspectrometer \citep{Harris2007}, ZEUS \citep{Ferkinhoff2010}, the Redshift Search Receiver \citep{Erickson2007}, or DESHIMA \citep{Endo2019,Taniguchi2022}. These instruments have demonstrated the technical feasibility of innovative technology, but still resulted in only a few dozen new redshifts due to the limited sensitivity of the telescopes they were mounted on.
The ALMA Wideband Sensitivity Upgrade \citep[WSU;][]{Carpenter2023} will improve ALMA's capabilities for redshift determinations, especially in Band 2 \citep{Mroczkowski2019}. However, due to ALMA's very limited primary beam, redshift searches are done mostly on individual, known targets, or within very small fields.

\medskip
AtLAST promises to make a leap forward thanks to its unique combination of sensitivity, broad spectral bandwidth, wide field of view and multiplex spectroscopic capabilites. Based on the prototype instruments described above, the MKID-based IFUs will allow to cover frequency ranges of hundreds of GHz. This very broad bandwidth will be possible as the requirements on spectral resolution are rather low: $\sim$0.5\,GHz will be sufficient to avoid line smearing as the targets will be galaxies with line widths of several hundreds to $>$1000\,km\,s$^{-1}$. Rather than covering a single object line the prototype instruments mentioned above, the AtLAST IFUs will eventually cover the full focal plane of 2$^{\circ}$ diameter.

\medskip
The very broad spectral bandwidth is crucial for redshift determinations: for example instrument covering the atmospheric windows from 125 to 500 GHz would allow to use the brighter but more widely spaced FSLs rather than the fainter CO lines\footnote{furthermore, FSL would mostly circumvent the potential redshift degeneracies that follow the linear spacing of CO lines.} for redshift determinations (see Fig. 7 of the companion case study on CGM science by Lee et al.). An additional advantage of FSL over the CO ladder is that they cover a wider range of physical conditions in the gas clouds, covering not only the PDR's but also HII regions (see also Lee et al.); this will allow to obtain a more complete census of the sources in an unbiased wide-field redshift survey. Finally, very broad-band IFU studies will not only cover the emission lines, but will also allow to determine the slope of the continuum emission which is brighter at shorter wavelengths, providing an additional constraint on the redshifts.

%\begin{figure}[h]
%\begin{centering} 
%\includegraphics[width=0.48\textwidth]{fig/RedshiftSearchEfficiency\_individual.pdf}
%    \caption{This figure, taken from \citet{Bakx2022}, shows the redshift identification probability as a function of redshift between z = 0 and 8, and for smoothed redshift distributions based on two samples (i.e. HerBS; \citealt{Bakx2018,Bakx2020Erratum}, SPT; \citealt{Reuter2020}). Orange bars indicate the fraction of sources for which one would detect a single spectral line, while blue bars indicate the fraction where two or more spectral lines are detected. Orange with hatched blue fill indicates cases where one can identify the redshift robustly with even a single spectral line, while blue with hatched orange fill indicates the situation where redshift degeneracies remain down to a $5 \sigma$ uncertainty in $z_{\rm phot}$. This graph can be generated easily using the github tool found at \url{https://github.com/tjlcbakx/redshift-search-graphs}.}
    \label{fig:RSG}
%\end{centering}
%\end{figure}

\medskip
There exists the possibility for redshift degeneracies in a redshift survey targeting CO lines \citep{Bakx2022}. To assess this, a tool is available (\url{https://github.com/tjlcbakx/redshift-search-graphs}) to graph the ability of an instrument to derive the redshift of a galaxy in each of the redshift bins %(Figure~\ref{fig:RSG})
using solely the observational bandwidth of a receiver. It can be used to determine the fraction of sources in a redshift regime that will have no lines, one line, or multiple robust lines. Doing this exercise for the 2-3 mm windows shows that that a wide band IFU covering these windows would be highly efficient at $z>2$.  

\medskip
For the line intensity mapping method we need a broad spectral coverage to map the [CII] power spectrum, whereas we require a spectral resolution  $\Delta v=3000 km/s$ ($R=100$) to measure the 3D power spectrum. This can be achieved either with a dedicated low-resolution spectrometer, or by binning higher-resolution spectra.

\subsection{Surveying proto-cluster galaxies}

\medskip
To significantly increase the number of spectroscopically confirmed (proto-)clusters galaxies requires a fast survey sub-mm telescope of at least 50-m. Such a size guarantees the unambiguous identification of cluster galaxies due to the relatively high angular resolution. To get spectroscopic redshifts of several hundred to thousand member galaxies per cluster, a multiplex instrument with up to several thousand elements per field of view is indispensable. One option would be a heterodyne instrument with a wide field of view and extremely large spectral bandwidth, however the costs would be exorbitant, so this will not be feasible. Thus, we opt for for the MKID bolometer technology which should provide integral field unit spectroscopic capabilities.

\medskip
To guarantee spectroscopic redshifts from $z=1-10$ and a complete study of the most prominent lines emitted from the cold ISM such as multiple CO, the two [CI], the [CII], H$_{2}$O and HCN lines, the spectrometer should have an unprecedented bandwidth from 70 to 700~GHz. E.g., the brightest expected line emitted in the far-infrared is [CII] at 158$\mu$m. An instrument with spectral coverage from 180 to 345~GHz could thus follow the early stages of cluster evolution from $z=4.5-10$, whereas extending to higher frequencies ($\sim$700~GHz) would even allow us to map the peak of the star-formation and black hole activity of the Universe at $z=2$ \citep{Madau2014} with the same line. Furthermore, such a set-up guarantees the detection of several CO lines and the so-called CO SLED (spectral line energy distribution, e.g. \citealt{Dannerbauer2009, Daddi2015}) can be established. This enables us to securely determine physical properties such as the gas density, excitation temperature and even molecular gas mass. In addition, with both [CI] lines the total cold gas mass can be measured as well \citep{Papadopoulos2004,Tomassetti2014}. Getting a complete picture of the cold ISM supported by a large sample size is indispensable to study in a statistical way if environment plays a role by measuring parameters such as star-formation efficiency, molecular gas fraction and excitation.

\medskip
We need a fast enough survey telescope that not only allows the study of confirmed galaxy clusters (with known spectroscopic redshifts) and candidate clusters but in addition will yield an unbiased survey (negelecting the impact of cosmic variance) of a significant area of the sky (to beat cosmic variance) within a reasonable time span. Therefore, to conduct the survey and achieve the science goals following technical requirements should be fulfilled:
\newline
- a bolometer based on millimeter KID technology with many thousands of elements per field of view, which has spectroscopic properties similar to multi-object spectroscopy in the optical and near-infrared
\newline
- a large bandwidth from 70 to 700~GHz to obtain spectroscopic redshifts, 
\newline
- a field-of-view of $\sim$1~square degree to cover the typical size of (proto-)galaxy clusters,
\newline
- a spectral resolution of $500-1000$~km/s to detect cluster galaxies and determine their redshifts (preferably from two or more lines), 
\newline
-a survey speed of at least 15~arcmin$^2$ per minute to obtain a statistically significant sample of several thousand galaxy clusters.

%\subsection{Survey strategy and detector requirements}\label{sec:test:det}
%\neededit{...}

\section{Summary and conclusions}

\medskip
In this paper we outline several high-z science cases for AtLAST, a future 50-m submillimeter telescope in the Atacama dessert, focusing on the overall high-z galaxy population as well on galaxies in proto-clusters. Two companion AtLAST case studies focus on emission line probes of the CGM of galaxies (Lee et al., in prep.) and on probing the ICM using the full SZ spectrum (Di Mascolo et al., in prep.).

\medskip
AtLAST will have high angular resolution, a large collecting area and large focal plane, and therefore a high survey speed \citep[see e.g.][]{Klaassen2020, Ramasawmy2022, Mroczkowski2023, Mroczkowski2024}. This means AtLAST can cover large areas of the sky for high-$z$ galaxy surveys. A single pointing will be of order 200 time larger than the LMT, for example. Also, the photometric confusion noise at 350~$\mu$m will be four orders of magnitude lower than that of existing 6-m sub-mm telescopes.
These mayor improvements on existing facilities, combined with an excellent instrument suite, allows for a large leap forward in the active research field of early galaxy formation and evolution, and notably the study of Dusty Star-Forming Galaxies (DSFGs).

\medskip
One or more large, homogeneous surveys (continuum or spectral line) of DSFGs will yield classical statistical properties such as the auto-correlation function, the (photometric) redshift distribution, number counts, and so forth. It will also yield a high-$z$ counterpart to existing large galaxy samples at low and intermediate redshifts, and make mayor contributions to the understanding of how the evolution of DSFGs varies as a function the environment (voids, filaments, groups, clusters). Additionally, such surveys provide a rich catalogue of sources to follow-up with ALMA, JWST, and ELT.

\medskip
Using a large multi-chroic camera allows for a comprehensive multi-band imaging survey, uniquely mapping large parts of the sky to specifically target high-$z$ galaxies and map their distribution, where several of the bands can be observed simultaneously, allowing for accurate spectral slope determinations and photometric redshifts (especially in combination with complementary data available at other wavebands), for example. A worked example for a 1000 deg$^2$ continuum survey in 1000 hours was presented, which for AtLAST will have a 3$\sigma$ sensitivity limit of 570 $\mu$Jy at 350 $\mu$m (at this limit, 82\% of the Cosmic Infrared Background at 350$\mu$m will be resolved into individual sources) and 324 $\mu$Jy at 450 $\mu$m. We showed that the multi-wavelength approach significantly increases 
the accuracy one would achieve while deriving L$_{IR}$ and M$_{dust}$ as a function of redshift for different available bands, especially if data for three or more bands is available.

\medskip
In addition to a very wide continuum survey, we also considered a deep line survey with AtLAST using mostly the low-frequency part of the spectrum, which includes [CII] all the way to $z\sim$8. For a deep 1 deg$^2$ line survey in 3000 hours, in a 400km/s channel (R=750), we can go down to 3$\sigma$ sensitivity limits of 2330 $\mu$Jy at 350 $\mu$m and 27 $\mu$Jy at 3~mm. The model of \cite{Bethermin2017} was used predict peak flux densities for CO and various fine structure lines, and found that about 90\% and 50\% of our galaxies at $z<5$ and $z>5$, respectively, will have multiple line detections, which is crucial for redshift determination.
Combining this with multiple band continuum detections will allow us to obtain a wealth of information for each of these galaxies besides the redshift: gas content, cooling budget, star formation rate, dust mass, and dust temperature.
Therefore, such deep, wide-bandwidth spectral line surveys with AtLAST will be crucial for mapping the population of ``normal'' star-forming galaxies and their gas content across the cosmic history.  
Also note that especially the [CII] line will be very promising for this purpose as it is one the brightest FIR lines out to high-$z$.

\medskip
Besides studying the overall galaxy population at high redshifts, we can also use wide and deep surveys of these galaxies to extract cosmological parameters, especially the growth rate, f$\sigma_8$, and the Hubble constant, by measuring galaxy-galaxy clustering (including their monopole and quadrupole moments) and fit cosmological models to these data. At $z\sim 2.5$ this becomes difficult in the traditional optical/infrared bands, but high-redshift galaxy spectroscopic survey with AtLAST would move the successful strategy of the optical/infrared surveys to the mm/sub-mm part of the spectrum and combine this with the new possibility of measuring strong emission lines from the FIR regime at $z\gtrsim3$. Testing this using simulated dark matter halos at $z=3$ showed that we can measure the Hubble parameter at 0.7\% precision and the growth rate at 7.3\%.

\medskip
Another strong AtLAST science case revolves around  Line-intensity mapping (LIM), which provides an alternative method for studying the large-scale structure of the Universe across cosmic time: it measures 2- and 3-dimensional power spectra from spectral cubes. At sub-mm wavelengths this bridges the optical surveys at $z\leq1$ and the upcoming radio surveys of the 21-cm HI emission line in the Epoch of Reionisation ($z\geq6$). The [CII] and/or CO lines remain bright for $1<z<6$, making them ideal LIM tracers. 
Current effort are limited by small surveys areas, which AtLAST can resolve as it has a much larger collecting area, large focal plane, and superior site and dish quality. We demonstrated the power of LIM using a CO line to constrain the cosmic expansion history $H(z)$, i.e the Hubble expansion rate as a function of redshift. This is significantly improved for $z>3$ when one adds CO LIM to information from Supernovae (SN), galaxy surveys and the Lyman-$\alpha$ forest.

\medskip
Finally, we also presented the science case for mapping several thousand galaxy (proto)clusters at $z=1-10$ with AtLAST, producing a high-redshift counterpart to local large surveys of rich clusters like the well-studied Abell catalogue. 
The main aims of such a large survey of distant clusters are the formation and evolution of cluster galaxies over cosmic time and the impact of environment on the formation and evolution (possibly environmental) of these galaxies. To make a big leap forward in this emerging research field, we would need a large-format, wide-band, direct-detection spectrometer (based on MKID technology, for example), covering a wide field of $\sim$1~square degree and a frequency coverage from 70 to 700~GHz (which could be split over two instruments). 

\medskip
In concluding, we have shown that AtLAST is able to yield significant progress is a range of research topics focusing on the distant Universe and cosmology, notably the overall high-z galaxy population, the ones located in proto-clusters, and the measurement of several cosmological parameters to help constrain cosmological models. This is made possible because of the high angular resolution that a 50-meter aperture brings, its wide spectral coverage with moderately high spectral resolution, and an excellent sensitivity that can be reached over large patches of the mm/sub-mm sky, which is unprecedented.

\section*{Competing interests}
No competing interests were disclosed.
%All financial, personal, or professional competing interests for any of the authors that could be construed to unduly influence the content of the article must be disclosed and will be displayed alongside the article. If there are no relevant competing interests to declare, please add the following: 'No competing interests were disclosed'.

\section*{Grant information}
This project has received funding from the European Union’s Horizon 2020 research and innovation programme under grant agreement No.\ 951815 (AtLAST). 

F.M.M. acknowledges the UCM Mar{\'i}a Zambrano programme of the Spanish Ministry of Universities funded by the Next Generation European Union and is also partly supported by grant RTI2018-096188-B-I00 funded by the Spanish Ministry of Science and Innovation/State Agency of Research MCIN/AEI/10.13039/501100011033.

M.L. acknowledges support from the European Union’s Horizon Europe research and innovation programme under the Marie Sk\l odowska-Curie grant agreement No 101107795.

L.D.M.\ is supported by the ERC-StG ``ClustersXCosmo'' grant agreement 716762. L.D.M.\ further acknowledges financial contribution from the agreement ASI-INAF n.2017-14-H.0. This work has been supported by the French government, through the UCA\textsuperscript{J.E.D.I.} Investments in the Future project managed by the National Research Agency (ANR) with the reference number ANR-15-IDEX-01. 

M. R. is supported by the NWO Veni project "\textit{Under the lens}" (VI.Veni.202.225).

S.W. acknowledges support by the Research Council of Norway through the EMISSA project (project number 286853) and the Centres of Excellence scheme, project number 262622 (``Rosseland Centre for Solar Physics''). 

H.D. acknowledges financial support from the Agencia Estatal de Investigación del Ministerio de Ciencia e Innovación (AEI-MCINN) under grant (La evolución de los cúmulos de galaxias desde el amanecer hasta el mediodía cósmico) with reference (PID2019-105776GB-I00/DOI:10.13039/501100011033)  and del Ministerio de Ciencia, Innovación y Universidades (MCIU/AEI) under grant (Construcción de cúmulos de galaxias en formación a través de la formación estelar ocurecida por el polvo) and the European Regional Development Fund (ERDF) with reference (PID2022-143243NB-I00/DOI:10.13039/501100011033).

\section*{Acknowledgements}

\medskip

% \section*{References}

\begingroup
\small
\bibliographystyle{apj_mod}
\setlength{\parskip}{0pt}
\setlength{\bibsep}{0pt}
\bibliography{high-z}

\begin{thebibliography}{135}
\expandafter\ifx\csname natexlab\endcsname\relax\def\natexlab#1{#1}\fi

\bibitem[{{Abell} {et~al.}(1989){Abell}, {Corwin}, \& {Olowin}}]{ACO1989}
{Abell},~G.~O., {Corwin},Harold~G.,~J., \& {Olowin},~R.~P. 1989,
  \href{https://ui.adsabs.harvard.edu/abs/1989ApJS...70....1A}{\apjs},
  \href{http://doi.org/10.1086/191333}{\color{magenta}{70, 1}}

\bibitem[{{Alberts} \& {Noble}(2022)}]{AlbertsNoble2022}
{Alberts},~S., \& {Noble},~A. 2022,
  \href{https://ui.adsabs.harvard.edu/abs/2022Univ....8..554A}{Universe},
  \href{http://doi.org/10.3390/universe8110554}{\color{magenta}{8, 554}}

\bibitem[{{Alcock} \& {Paczynski}(1979)}]{Alcock1979}
{Alcock},~C., \& {Paczynski},~B. 1979,
  \href{https://ui.adsabs.harvard.edu/abs/1979Natur.281..358A}{\nat},
  \href{http://doi.org/10.1038/281358a0}{\color{magenta}{281, 358}}

\bibitem[{{Algera} {et~al.}(2023){Algera}, {Inami}, {Oesch}, {Sommovigo},
  {Bouwens}, {Topping}, {Schouws}, {Stefanon}, {Stark}, {Aravena}, {Barrufet},
  {da Cunha}, {Dayal}, {Endsley}, {Ferrara}, {Fudamoto}, {Gonzalez},
  {Graziani}, {Hodge}, {Hygate}, {de Looze}, {Nanayakkara}, {Schneider}, \&
  {van der Werf}}]{Algera2023}
{Algera},~H. S.~B., {Inami},~H., {Oesch},~P.~A., {et~al.} 2023,
  \href{https://ui.adsabs.harvard.edu/abs/2023MNRAS.518.6142A}{\mnras},
  \href{http://doi.org/10.1093/mnras/stac3195}{\color{magenta}{518, 6142}}

\bibitem[{{Amvrosiadis} {et~al.}(2019){Amvrosiadis}, {Valiante},
  {Gonzalez-Nuevo}, {Maddox}, {Negrello}, {Eales}, {Dunne}, {Wang}, {van
  Kampen}, {De Zotti}, {Smith}, {Andreani}, {Greenslade}, {Tai-An}, \&
  {Micha{\l}owski}}]{Amvrosiadis2019}
{Amvrosiadis},~A., {Valiante},~E., {Gonzalez-Nuevo},~J., {et~al.} 2019,
  \href{https://ui.adsabs.harvard.edu/abs/2019MNRAS.483.4649A}{\mnras},
  \href{http://doi.org/10.1093/mnras/sty3013}{\color{magenta}{483, 4649}}

\bibitem[{{Bakx} \& {Dannerbauer}(2022)}]{Bakx2022}
{Bakx},~T. J.~L.~C., \& {Dannerbauer},~H. 2022,
  \href{https://ui.adsabs.harvard.edu/abs/2022MNRAS.515..678B}{\mnras},
  \href{http://doi.org/10.1093/mnras/stac1306}{\color{magenta}{515, 678}}

\bibitem[{{Barrufet} {et~al.}(2023){Barrufet}, {Oesch}, {Bouwens}, {Inami},
  {Sommovigo}, {Algera}, {da Cunha}, {Aravena}, {Dayal}, {Ferrara}, {Fudamoto},
  {Gonzalez}, {Graziani}, {Hygate}, {de Looze}, {Nanayakkara}, {Pallottini},
  {Schneider}, {Stefanon}, {Topping}, \& {van der Werf}}]{Barrufet2023}
{Barrufet},~L., {Oesch},~P.~A., {Bouwens},~R., {et~al.} 2023,
  \href{https://ui.adsabs.harvard.edu/abs/2023MNRAS.522.3926B}{\mnras},
  \href{http://doi.org/10.1093/mnras/stad1259}{\color{magenta}{522, 3926}}

\bibitem[{{Bautista} {et~al.}(2021){Bautista}, {Paviot}, {Vargas Maga{\~n}a},
  {de la Torre}, {Fromenteau}, {Gil-Mar{\'\i}n}, {Ross}, {Burtin}, {Dawson},
  {Hou}, {Kneib}, {de Mattia}, {Percival}, {Rossi}, {Tojeiro}, {Zhao}, {Zhao},
  {Alam}, {Brownstein}, {Chapman}, {Choi}, {Chuang}, {Escoffier}, {de la
  Macorra}, {du Mas des Bourboux}, {Mohammad}, {Moon}, {M{\"u}ller},
  {Nadathur}, {Newman}, {Schneider}, {Seo}, \& {Wang}}]{Bautista2021}
{Bautista},~J.~E., {Paviot},~R., {Vargas Maga{\~n}a},~M., {et~al.} 2021,
  \href{https://ui.adsabs.harvard.edu/abs/2021MNRAS.500..736B}{\mnras},
  \href{http://doi.org/10.1093/mnras/staa2800}{\color{magenta}{500, 736}}

\bibitem[{{Behroozi} {et~al.}(2013){Behroozi}, {Wechsler}, \&
  {Wu}}]{Behroozi2013}
{Behroozi},~P.~S., {Wechsler},~R.~H., \& {Wu},~H.-Y. 2013,
  \href{https://ui.adsabs.harvard.edu/abs/2013ApJ...762..109B}{\apj},
  \href{http://doi.org/10.1088/0004-637X/762/2/109}{\color{magenta}{762, 109}}

\bibitem[{{Bernal} {et~al.}(2019){Bernal}, {Breysse}, \& {Kovetz}}]{Bernal2019}
{Bernal},~J.~L., {Breysse},~P.~C., \& {Kovetz},~E.~D. 2019,
  \href{https://ui.adsabs.harvard.edu/abs/2019PhRvL.123y1301B}{\prl},
  \href{http://doi.org/10.1103/PhysRevLett.123.251301}{\color{magenta}{123,
  251301}}

\bibitem[{{B{\'e}thermin} {et~al.}(2015){B{\'e}thermin}, {De Breuck},
  {Sargent}, \& {Daddi}}]{Bethermin2015}
{B{\'e}thermin},~M., {De Breuck},~C., {Sargent},~M., \& {Daddi},~E. 2015,
  \href{https://ui.adsabs.harvard.edu/abs/2015A&A...576L...9B}{\aap},
  \href{http://doi.org/10.1051/0004-6361/201525718}{\color{magenta}{576, L9}}

\bibitem[{{B{\'e}thermin} {et~al.}(2017){B{\'e}thermin}, {Wu}, {Lagache},
  {Davidzon}, {Ponthieu}, {Cousin}, {Wang}, {Dor{\'e}}, {Daddi}, \&
  {Lapi}}]{Bethermin2017}
{B{\'e}thermin},~M., {Wu},~H.-Y., {Lagache},~G., {et~al.} 2017,
  \href{https://ui.adsabs.harvard.edu/abs/2017A&A...607A..89B}{\aap},
  \href{http://doi.org/10.1051/0004-6361/201730866}{\color{magenta}{607, A89}}

\bibitem[{{B{\'e}thermin} {et~al.}(2020){B{\'e}thermin}, {Fudamoto}, {Ginolfi},
  {Loiacono}, {Khusanova}, {Capak}, {Cassata}, {Faisst}, {Le F{\`e}vre},
  {Schaerer}, {Silverman}, {Yan}, {Amorin}, {Bardelli}, {Boquien}, {Cimatti},
  {Davidzon}, {Dessauges-Zavadsky}, {Fujimoto}, {Gruppioni}, {Hathi}, {Ibar},
  {Jones}, {Koekemoer}, {Lagache}, {Lemaux}, {Moreau}, {Oesch}, {Pozzi},
  {Riechers}, {Talia}, {Toft}, {Vallini}, {Vergani}, {Zamorani}, \&
  {Zucca}}]{Bethermin2020}
{B{\'e}thermin},~M., {Fudamoto},~Y., {Ginolfi},~M., {et~al.} 2020,
  \href{https://ui.adsabs.harvard.edu/abs/2020A&A...643A...2B}{\aap},
  \href{http://doi.org/10.1051/0004-6361/202037649}{\color{magenta}{643, A2}}

\bibitem[{{B{\'e}thermin} {et~al.}(2022){B{\'e}thermin}, {Gkogkou}, {Van
  Cuyck}, {Lagache}, {Beelen}, {Aravena}, {Benoit}, {Bounmy}, {Calvo},
  {Catalano}, {de Batz de Trenquelleon}, {De Breuck}, {Fasano}, {Ferrara},
  {Goupy}, {Hoarau}, {Horellou}, {Hu}, {Julia}, {Knudsen}, {Lambert},
  {Macias-Perez}, {Marpaud}, {Monfardini}, {Pallottini}, {Ponthieu}, {Roehlly},
  {Vallini}, {Walter}, \& {Weiss}}]{Bethermin2022}
{B{\'e}thermin},~M., {Gkogkou},~A., {Van Cuyck},~M., {et~al.} 2022,
  \href{https://ui.adsabs.harvard.edu/abs/2022A&A...667A.156B}{\aap},
  \href{http://doi.org/10.1051/0004-6361/202243888}{\color{magenta}{667, A156}}

\bibitem[{{Blain} {et~al.}(2002){Blain}, {Smail}, {Ivison}, {Kneib}, \&
  {Frayer}}]{Blain2002}
{Blain},~A.~W., {Smail},~I., {Ivison},~R.~J., {Kneib},~J.~P., \&
  {Frayer},~D.~T. 2002,
  \href{https://ui.adsabs.harvard.edu/abs/2002PhR...369..111B}{\physrep},
  \href{http://doi.org/10.1016/S0370-1573(02)00134-5}{\color{magenta}{369,
  111}}

\bibitem[{{Boogaard} {et~al.}(2023){Boogaard}, {Decarli}, {Walter}, {Wei{\ss}},
  {Popping}, {Neri}, {Aravena}, {Riechers}, {Ellis}, {Carilli}, {Cox}, \&
  {Pety}}]{Boogaard2023}
{Boogaard},~L.~A., {Decarli},~R., {Walter},~F., {et~al.} 2023,
  \href{https://ui.adsabs.harvard.edu/abs/2023ApJ...945..111B}{\apj},
  \href{http://doi.org/10.3847/1538-4357/acb4f0}{\color{magenta}{945, 111}}

\bibitem[{{Bothwell} {et~al.}(2016){Bothwell}, {Maiolino}, {Peng}, {Cicone},
  {Griffith}, \& {Wagg}}]{Bothwell2016}
{Bothwell},~M.~S., {Maiolino},~R., {Peng},~Y., {et~al.} 2016,
  \href{https://ui.adsabs.harvard.edu/abs/2016MNRAS.455.1156B}{\mnras},
  \href{http://doi.org/10.1093/mnras/stv2121}{\color{magenta}{455, 1156}}

\bibitem[{{Bouwens} {et~al.}(2022){Bouwens}, {Smit}, {Schouws}, {Stefanon},
  {Bowler}, {Endsley}, {Gonzalez}, {Inami}, {Stark}, {Oesch}, {Hodge},
  {Aravena}, {da Cunha}, {Dayal}, {de Looze}, {Ferrara}, {Fudamoto},
  {Graziani}, {Li}, {Nanayakkara}, {Pallottini}, {Schneider}, {Sommovigo},
  {Topping}, {van der Werf}, {Algera}, {Barrufet}, {Hygate}, {Labb{\'e}},
  {Riechers}, \& {Witstok}}]{Bouwens2022}
{Bouwens},~R.~J., {Smit},~R., {Schouws},~S., {et~al.} 2022,
  \href{https://ui.adsabs.harvard.edu/abs/2022ApJ...931..160B}{\apj},
  \href{http://doi.org/10.3847/1538-4357/ac5a4a}{\color{magenta}{931, 160}}

\bibitem[{{Carilli} \& {Walter}(2013)}]{CarilliWalter2013}
{Carilli},~C.~L., \& {Walter},~F. 2013,
  \href{https://ui.adsabs.harvard.edu/abs/2013ARA&A..51..105C}{\araa},
  \href{http://doi.org/10.1146/annurev-astro-082812-140953}{\color{magenta}{51,
  105}}

\bibitem[{{Carpenter} {et~al.}(2023){Carpenter}, {Brogan}, {Iono}, \&
  {Mroczkowski}}]{Carpenter2023}
{Carpenter},~J., {Brogan},~C., {Iono},~D., \& {Mroczkowski},~T. 2023,
  \href{https://ui.adsabs.harvard.edu/abs/2023pcsf.conf..304C}{in Physics and
  Chemistry of Star Formation: The Dynamical ISM Across Time and Spatial
  Scales}\href{http://doi.org/10.48550/arXiv.2211.00195, }{,
  \color{magenta}{304}}

\bibitem[{{Carraro} {et~al.}(2020){Carraro}, {Rodighiero}, {Cassata}, {Brusa},
  {Shankar}, {Baronchelli}, {Daddi}, {Delvecchio}, {Franceschini}, {Griffiths},
  {Gruppioni}, {L{\'o}pez-Navas}, {Mancini}, {Marchesi}, {Negrello}, {Puglisi},
  {Sani}, \& {Suh}}]{Carraro2020}
{Carraro},~R., {Rodighiero},~G., {Cassata},~P., {et~al.} 2020,
  \href{https://ui.adsabs.harvard.edu/abs/2020A&A...642A..65C}{\aap},
  \href{http://doi.org/10.1051/0004-6361/201936649}{\color{magenta}{642, A65}}

\bibitem[{{Casey}(2016)}]{Casey2016}
{Casey},~C.~M. 2016,
  \href{https://ui.adsabs.harvard.edu/abs/2016ApJ...824...36C}{\apj},
  \href{http://doi.org/10.3847/0004-637X/824/1/36}{\color{magenta}{824, 36}}

\bibitem[{{Casey} {et~al.}(2014){Casey}, {Narayanan}, \& {Cooray}}]{Casey2014}
{Casey},~C.~M., {Narayanan},~D., \& {Cooray},~A. 2014,
  \href{https://ui.adsabs.harvard.edu/abs/2014PhR...541...45C}{\physrep},
  \href{http://doi.org/10.1016/j.physrep.2014.02.009}{\color{magenta}{541, 45}}

\bibitem[{{Casey} {et~al.}(2021){Casey}, {Zavala}, {Manning}, {Aravena},
  {B{\'e}thermin}, {Caputi}, {Champagne}, {Clements}, {Drew}, {Finkelstein},
  {Fujimoto}, {Hayward}, {Dekel}, {Kokorev}, {Lagos}, {Long}, {Magdis}, {Man},
  {Mitsuhashi}, {Popping}, {Spilker}, {Staguhn}, {Talia}, {Toft}, {Treister},
  {Weaver}, \& {Yun}}]{Casey2021}
{Casey},~C.~M., {Zavala},~J.~A., {Manning},~S.~M., {et~al.} 2021,
  \href{https://ui.adsabs.harvard.edu/abs/2021ApJ...923..215C}{\apj},
  \href{http://doi.org/10.3847/1538-4357/ac2eb4}{\color{magenta}{923, 215}}

\bibitem[{{Chapman} {et~al.}(2005){Chapman}, {Blain}, {Smail}, \&
  {Ivison}}]{Chapman2005}
{Chapman},~S.~C., {Blain},~A.~W., {Smail},~I., \& {Ivison},~R.~J. 2005,
  \href{https://ui.adsabs.harvard.edu/abs/2005ApJ...622..772C}{\apj},
  \href{http://doi.org/10.1086/428082}{\color{magenta}{622, 772}}

\bibitem[{{Chen} {et~al.}(2022){Chen}, {Liao}, {Smail}, {Swinbank}, {Ao},
  {Bunker}, {Chapman}, {Hatsukade}, {Ivison}, {Lee}, {Serjeant}, {Umehata},
  {Wang}, \& {Zhao}}]{Chen2022}
{Chen},~C.-C., {Liao},~C.-L., {Smail},~I., {et~al.} 2022,
  \href{https://ui.adsabs.harvard.edu/abs/2022ApJ...929..159C}{\apj},
  \href{http://doi.org/10.3847/1538-4357/ac61df}{\color{magenta}{929, 159}}

\bibitem[{{Chen} {et~al.}(2023){Chen}, {Ivison}, {Zwaan}, {Klitsch},
  {P{\'e}roux}, {Lovell}, {Lagos}, {Biggs}, \& {Bollo}}]{Jianhang2023}
{Chen},~J., {Ivison},~R.~J., {Zwaan},~M.~A., {et~al.} 2023,
  \href{https://ui.adsabs.harvard.edu/abs/2023A&A...675L..10C}{\aap},
  \href{http://doi.org/10.1051/0004-6361/202347107}{\color{magenta}{675, L10}}

\bibitem[{{Chiang} {et~al.}(2017){Chiang}, {Overzier}, {Gebhardt}, \&
  {Henriques}}]{Chiang2017}
{Chiang},~Y.-K., {Overzier},~R.~A., {Gebhardt},~K., \& {Henriques},~B. 2017,
  \href{https://ui.adsabs.harvard.edu/abs/2017ApJ...844L..23C}{\apjl},
  \href{http://doi.org/10.3847/2041-8213/aa7e7b}{\color{magenta}{844, L23}}

\bibitem[{{Cleary} {et~al.}(2022){Cleary}, {Borowska}, {Breysse}, {Catha},
  {Chung}, {Church}, {Dickinson}, {Eriksen}, {Foss}, {Gundersen}, {Harper},
  {Harris}, {Hobbs}, {Ihle}, {Kim}, {Kocz}, {Lamb}, {Lunde}, {Padmanabhan},
  {Pearson}, {Philip}, {Powell}, {Rasmussen}, {Readhead}, {Rennie}, {Silva},
  {Stutzer}, {Uzgil}, {Watts}, {Wehus}, {Woody}, {Basoalto}, {Bond}, {Dunne},
  {Gaier}, {Hensley}, {Keating}, {Lawrence}, {Murray}, {Paladini}, {Reeves},
  {Viero}, {Wechsler}, \& {Comap Collaboration}}]{Cleary2022}
{Cleary},~K.~A., {Borowska},~J., {Breysse},~P.~C., {et~al.} 2022,
  \href{https://ui.adsabs.harvard.edu/abs/2022ApJ...933..182C}{\apj},
  \href{http://doi.org/10.3847/1538-4357/ac63cc}{\color{magenta}{933, 182}}

\bibitem[{{Cochrane} {et~al.}(2018){Cochrane}, {Best}, {Sobral}, {Smail},
  {Geach}, {Stott}, \& {Wake}}]{Cochrane2018}
{Cochrane},~R.~K., {Best},~P.~N., {Sobral},~D., {et~al.} 2018,
  \href{https://ui.adsabs.harvard.edu/abs/2018MNRAS.475.3730C}{\mnras},
  \href{http://doi.org/10.1093/mnras/stx3345}{\color{magenta}{475, 3730}}

\bibitem[{{Cole} {et~al.}(2005){Cole}, {Percival}, {Peacock}, {Norberg},
  {Baugh}, {Frenk}, {Baldry}, {Bland-Hawthorn}, {Bridges}, {Cannon}, {Colless},
  {Collins}, {Couch}, {Cross}, {Dalton}, {Eke}, {De Propris}, {Driver},
  {Efstathiou}, {Ellis}, {Glazebrook}, {Jackson}, {Jenkins}, {Lahav}, {Lewis},
  {Lumsden}, {Maddox}, {Madgwick}, {Peterson}, {Sutherland}, \&
  {Taylor}}]{Cole2005}
{Cole},~S., {Percival},~W.~J., {Peacock},~J.~A., {et~al.} 2005,
  \href{https://ui.adsabs.harvard.edu/abs/2005MNRAS.362..505C}{\mnras},
  \href{http://doi.org/10.1111/j.1365-2966.2005.09318.x}{\color{magenta}{362,
  505}}

\bibitem[{{Combes}(2018)}]{Combes2018}
{Combes},~F. 2018,
  \href{https://ui.adsabs.harvard.edu/abs/2018A&ARv..26....5C}{\aapr},
  \href{http://doi.org/10.1007/s00159-018-0110-4}{\color{magenta}{26, 5}}

\bibitem[{{CONCERTO Collaboration} {et~al.}(2020){CONCERTO Collaboration},
  {Ade}, {Aravena}, {Barria}, {Beelen}, {Benoit}, {B{\'e}thermin}, {Bounmy},
  {Bourrion}, {Bres}, {De Breuck}, {Calvo}, {Cao}, {Catalano}, {D{\'e}sert},
  {Dur{\'a}n}, {Fasano}, {Fenouillet}, {Garcia}, {Garde}, {Goupy}, {Groppi},
  {Hoarau}, {Lagache}, {Lambert}, {Leggeri}, {Levy-Bertrand},
  {Mac{\'\i}as-P{\'e}rez}, {Mani}, {Marpaud}, {Mauskopf}, {Monfardini},
  {Pisano}, {Ponthieu}, {Prieur}, {Roni}, {Roudier}, {Tourres}, \&
  {Tucker}}]{Ade2020}
{CONCERTO Collaboration}, {Ade},~P., {Aravena},~M., {et~al.} 2020,
  \href{https://ui.adsabs.harvard.edu/abs/2020A&A...642A..60C}{\aap},
  \href{http://doi.org/10.1051/0004-6361/202038456}{\color{magenta}{642, A60}}

\bibitem[{{Coogan} {et~al.}(2018){Coogan}, {Daddi}, {Sargent}, {Strazzullo},
  {Valentino}, {Gobat}, {Magdis}, {Bethermin}, {Pannella}, {Onodera}, {Liu},
  {Cimatti}, {Dannerbauer}, {Carollo}, {Renzini}, \& {Tremou}}]{Coogan2018}
{Coogan},~R.~T., {Daddi},~E., {Sargent},~M.~T., {et~al.} 2018,
  \href{https://ui.adsabs.harvard.edu/abs/2018MNRAS.479..703C}{\mnras},
  \href{http://doi.org/10.1093/mnras/sty1446}{\color{magenta}{479, 703}}

\bibitem[{{Cox} {et~al.}(2023){Cox}, {Neri}, {Berta}, {Ismail}, {Stanley},
  {Young}, {Jin}, {Bakx}, {Beelen}, {Dannerbauer}, {Krips}, {Lehnert}, {Omont},
  {Riechers}, {Baker}, {Bendo}, {Borsato}, {Buat}, {Butler}, {Chartab},
  {Cooray}, {Dye}, {Eales}, {Gavazzi}, {Hughes}, {Ivison}, {Jones},
  {Marchetti}, {Messias}, {Nanni}, {Negrello}, {Perez-Fournon}, {Serjeant},
  {Urquhart}, {Vlahakis}, {Wei{\ss}}, {van der Werf}, \& {Yang}}]{Cox2023}
{Cox},~P., {Neri},~R., {Berta},~S., {et~al.} 2023,
  \href{https://ui.adsabs.harvard.edu/abs/2023A&A...678A..26C}{\aap},
  \href{http://doi.org/10.1051/0004-6361/202346801}{\color{magenta}{678, A26}}

\bibitem[{{Cramer} {et~al.}(2023){Cramer}, {Noble}, {Massingill}, {Cairns},
  {Clements}, {Cooper}, {Demarco}, {Matharu}, {McDonald}, {Muzzin}, {Nantais},
  {Rudnick}, {{\"U}bler}, {van Kampen}, {Webb}, {Wilson}, \&
  {Yee}}]{Cramer2023}
{Cramer},~W.~J., {Noble},~A.~G., {Massingill},~K., {et~al.} 2023,
  \href{https://ui.adsabs.harvard.edu/abs/2023ApJ...944..213C}{\apj},
  \href{http://doi.org/10.3847/1538-4357/acae96}{\color{magenta}{944, 213}}

\bibitem[{{Crites} {et~al.}(2014){Crites}, {Bock}, {Bradford}, {Chang},
  {Cooray}, {Duband}, {Gong}, {Hailey-Dunsheath}, {Hunacek}, {Koch}, {Li},
  {O'Brient}, {Prouve}, {Shirokoff}, {Silva}, {Staniszewski}, {Uzgil}, \&
  {Zemcov}}]{Crites2014}
{Crites},~A.~T., {Bock},~J.~J., {Bradford},~C.~M., {et~al.} 2014,
  \href{https://ui.adsabs.harvard.edu/abs/2014SPIE.9153E..1WC}{in Millimeter,
  Submillimeter, and Far-Infrared Detectors and Instrumentation for Astronomy
  VII}\href{http://doi.org/10.1117/12.2057207, }{, \color{magenta}{91531W}}

\bibitem[{{Daddi} {et~al.}(2015){Daddi}, {Dannerbauer}, {Liu}, {Aravena},
  {Bournaud}, {Walter}, {Riechers}, {Magdis}, {Sargent}, {B{\'e}thermin},
  {Carilli}, {Cibinel}, {Dickinson}, {Elbaz}, {Gao}, {Gobat}, {Hodge}, \&
  {Krips}}]{Daddi2015}
{Daddi},~E., {Dannerbauer},~H., {Liu},~D., {et~al.} 2015,
  \href{https://ui.adsabs.harvard.edu/abs/2015A&A...577A..46D}{\aap},
  \href{http://doi.org/10.1051/0004-6361/201425043}{\color{magenta}{577, A46}}

\bibitem[{{Dannerbauer} {et~al.}(2009){Dannerbauer}, {Daddi}, {Riechers},
  {Walter}, {Carilli}, {Dickinson}, {Elbaz}, \& {Morrison}}]{Dannerbauer2009}
{Dannerbauer},~H., {Daddi},~E., {Riechers},~D.~A., {et~al.} 2009,
  \href{https://ui.adsabs.harvard.edu/abs/2009ApJ...698L.178D}{\apjl},
  \href{http://doi.org/10.1088/0004-637X/698/2/L178}{\color{magenta}{698,
  L178}}

\bibitem[{{Dannerbauer} {et~al.}(2014){Dannerbauer}, {Kurk}, {De Breuck},
  {Wylezalek}, {Santos}, {Koyama}, {Seymour}, {Tanaka}, {Hatch}, {Altieri},
  {Coia}, {Galametz}, {Kodama}, {Miley}, {R{\"o}ttgering}, {Sanchez-Portal},
  {Valtchanov}, {Venemans}, \& {Ziegler}}]{Dannerbauer2014}
{Dannerbauer},~H., {Kurk},~J.~D., {De Breuck},~C., {et~al.} 2014,
  \href{https://ui.adsabs.harvard.edu/abs/2014A&A...570A..55D}{\aap},
  \href{http://doi.org/10.1051/0004-6361/201423771}{\color{magenta}{570, A55}}

\bibitem[{{Dannerbauer} {et~al.}(2017){Dannerbauer}, {Lehnert}, {Emonts},
  {Ziegler}, {Altieri}, {De Breuck}, {Hatch}, {Kodama}, {Koyama}, {Kurk},
  {Matiz}, {Miley}, {Narayanan}, {Norris}, {Overzier}, {R{\"o}ttgering},
  {Sargent}, {Seymour}, {Tanaka}, {Valtchanov}, \&
  {Wylezalek}}]{Dannerbauer2017}
{Dannerbauer},~H., {Lehnert},~M.~D., {Emonts},~B., {et~al.} 2017,
  \href{https://ui.adsabs.harvard.edu/abs/2017A&A...608A..48D}{\aap},
  \href{http://doi.org/10.1051/0004-6361/201730449}{\color{magenta}{608, A48}}

\bibitem[{{Dayal} {et~al.}(2022){Dayal}, {Ferrara}, {Sommovigo}, {Bouwens},
  {Oesch}, {Smit}, {Gonzalez}, {Schouws}, {Stefanon}, {Kobayashi}, {Bremer},
  {Algera}, {Aravena}, {Bowler}, {da Cunha}, {Fudamoto}, {Graziani}, {Hodge},
  {Inami}, {De Looze}, {Pallottini}, {Riechers}, {Schneider}, {Stark}, \&
  {Endsley}}]{Dayal2022}
{Dayal},~P., {Ferrara},~A., {Sommovigo},~L., {et~al.} 2022,
  \href{https://ui.adsabs.harvard.edu/abs/2022MNRAS.512..989D}{\mnras},
  \href{http://doi.org/10.1093/mnras/stac537}{\color{magenta}{512, 989}}

\bibitem[{{De Looze} {et~al.}(2014){De Looze}, {Cormier}, {Lebouteiller},
  {Madden}, {Baes}, {Bendo}, {Boquien}, {Boselli}, {Clements}, {Cortese},
  {Cooray}, {Galametz}, {Galliano}, {Graci{\'a}-Carpio}, {Isaak}, {Karczewski},
  {Parkin}, {Pellegrini}, {R{\'e}my-Ruyer}, {Spinoglio}, {Smith}, \&
  {Sturm}}]{Delooze2014}
{De Looze},~I., {Cormier},~D., {Lebouteiller},~V., {et~al.} 2014,
  \href{https://ui.adsabs.harvard.edu/abs/2014A&A...568A..62D}{\aap},
  \href{http://doi.org/10.1051/0004-6361/201322489}{\color{magenta}{568, A62}}

\bibitem[{{DES Collaboration} {et~al.}(2024){DES Collaboration}, {Abbott},
  {Adamow}, {Aguena}, {Allam}, {Alves}, {Amon}, {Andrade-Oliveira}, {Asorey},
  {Avila}, {Bacon}, {Bechtol}, {Bernstein}, {Bertin}, {Blazek}, {Bocquet},
  {Brooks}, {Burke}, {Camacho}, {Carnero Rosell}, {Carollo}, {Carretero},
  {Castander}, {Cawthon}, {Chan}, {Chang}, {Conselice}, {Costanzi}, {Crocce},
  {da Costa}, {Pereira}, {Davis}, {De Vicente}, {Deiosso}, {Desai}, {Diehl},
  {Dodelson}, {Doux}, {Drlica-Wagner}, {Elvin-Poole}, {Everett}, {Ferrero},
  {Fert{\'e}}, {Flaugher}, {Fosalba}, {Frieman}, {Garc{\'\i}a-Bellido},
  {Gaztanaga}, {Giannini}, {Gruendl}, {Gutierrez}, {Hartley}, {Hinton},
  {Hollowood}, {Honscheid}, {Huterer}, {James}, {Kent}, {Kuehn}, {Lahav},
  {Lee}, {Lidman}, {Lin}, {Marshall}, {Martini}, {Mena-Fern{\'a}ndez},
  {Menanteau}, {Miquel}, {Mohr}, {Myles}, {Nichol}, {Ogando}, {Palmese},
  {Percival}, {Pieres}, {Plazas Malag{\'o}n}, {Porredon}, {Prat},
  {Rodr{\'\i}guez-Monroy}, {Romer}, {Roodman}, {Rosenfeld}, {Ross}, {Rykoff},
  {Sako}, {Samuroff}, {S{\'a}nchez}, {Sanchez}, {Sanchez Cid}, {Santiago},
  {Schubnell}, {Sevilla-Noarbe}, {Sheldon}, {Smith}, {Suchyta}, {Swanson},
  {Tarle}, {Thomas}, {To}, {Toribio San Cipriano}, {Troxel}, {Tucker},
  {Tucker}, {Walker}, {Weaverdyck}, {Weller}, {Wiseman}, \& {Yanny}}]{DES2024}
{DES Collaboration}, {Abbott},~T.~M.~C., {Adamow},~M., {et~al.} 2024,
  \href{https://ui.adsabs.harvard.edu/abs/2024arXiv240210696D}{arXiv e-prints},
  \href{http://doi.org/10.48550/arXiv.2402.10696}{\color{magenta}{arXiv:2402.10696}}

\bibitem[{{DESI Collaboration} {et~al.}(2016){DESI Collaboration}, {Aghamousa},
  {Aguilar}, {Ahlen}, {Alam}, {Allen}, {Allende Prieto}, {Annis}, {Bailey},
  {Balland}, {Ballester}, {Baltay}, {Beaufore}, {Bebek}, {Beers}, {Bell},
  {Bernal}, {Besuner}, {Beutler}, {Blake}, {Bleuler}, {Blomqvist}, {Blum},
  {Bolton}, {Briceno}, {Brooks}, {Brownstein}, {Buckley-Geer}, {Burden},
  {Burtin}, {Busca}, {Cahn}, {Cai}, {Cardiel-Sas}, {Carlberg}, {Carton},
  {Casas}, {Castander}, {Cervantes-Cota}, {Claybaugh}, {Close}, {Coker},
  {Cole}, {Comparat}, {Cooper}, {Cousinou}, {Crocce}, {Cuby}, {Cunningham},
  {Davis}, {Dawson}, {de la Macorra}, {De Vicente}, {Delubac}, {Derwent},
  {Dey}, {Dhungana}, {Ding}, {Doel}, {Duan}, {Ealet}, {Edelstein},
  {Eftekharzadeh}, {Eisenstein}, {Elliott}, {Escoffier}, {Evatt}, {Fagrelius},
  {Fan}, {Fanning}, {Farahi}, {Farihi}, {Favole}, {Feng}, {Fernandez},
  {Findlay}, {Finkbeiner}, {Fitzpatrick}, {Flaugher}, {Flender}, {Font-Ribera},
  {Forero-Romero}, {Fosalba}, {Frenk}, {Fumagalli}, {Gaensicke}, {Gallo},
  {Garcia-Bellido}, {Gaztanaga}, {Pietro Gentile Fusillo}, {Gerard},
  {Gershkovich}, {Giannantonio}, {Gillet}, {Gonzalez-de-Rivera},
  {Gonzalez-Perez}, {Gott}, {Graur}, {Gutierrez}, {Guy}, {Habib}, {Heetderks},
  {Heetderks}, {Heitmann}, {Hellwing}, {Herrera}, {Ho}, {Holland}, {Honscheid},
  {Huff}, {Hutchinson}, {Huterer}, {Hwang}, {Illa Laguna}, {Ishikawa},
  {Jacobs}, {Jeffrey}, {Jelinsky}, {Jennings}, {Jiang}, {Jimenez}, {Johnson},
  {Joyce}, {Jullo}, {Juneau}, {Kama}, {Karcher}, {Karkar}, {Kehoe}, {Kennamer},
  {Kent}, {Kilbinger}, {Kim}, {Kirkby}, {Kisner}, {Kitanidis}, {Kneib},
  {Koposov}, {Kovacs}, {Koyama}, {Kremin}, {Kron}, {Kronig}, {Kueter-Young},
  {Lacey}, {Lafever}, {Lahav}, {Lambert}, {Lampton}, {Landriau}, {Lang},
  {Lauer}, {Le Goff}, {Le Guillou}, {Le Van Suu}, {Lee}, {Lee}, {Leitner},
  {Lesser}, {Levi}, {L'Huillier}, {Li}, {Liang}, {Lin}, {Linder}, {Loebman},
  {Luki{\'c}}, {Ma}, {MacCrann}, {Magneville}, {Makarem}, {Manera}, {Manser},
  {Marshall}, {Martini}, {Massey}, {Matheson}, {McCauley}, {McDonald},
  {McGreer}, {Meisner}, {Metcalfe}, {Miller}, {Miquel}, {Moustakas}, {Myers},
  {Naik}, {Newman}, {Nichol}, {Nicola}, {Nicolati da Costa}, {Nie}, {Niz},
  {Norberg}, {Nord}, {Norman}, {Nugent}, {O'Brien}, {Oh}, {Olsen}, {Padilla},
  {Padmanabhan}, {Padmanabhan}, {Palanque-Delabrouille}, {Palmese},
  {Pappalardo}, {P{\^a}ris}, {Park}, {Patej}, {Peacock}, {Peiris}, {Peng},
  {Percival}, {Perruchot}, {Pieri}, {Pogge}, {Pollack}, {Poppett}, {Prada},
  {Prakash}, {Probst}, {Rabinowitz}, {Raichoor}, {Ree}, {Refregier}, {Regal},
  {Reid}, {Reil}, {Rezaie}, {Rockosi}, {Roe}, {Ronayette}, {Roodman}, {Ross},
  {Ross}, {Rossi}, {Rozo}, {Ruhlmann-Kleider}, {Rykoff}, {Sabiu}, {Samushia},
  {Sanchez}, {Sanchez}, {Schlegel}, {Schneider}, {Schubnell}, {Secroun},
  {Seljak}, {Seo}, {Serrano}, {Shafieloo}, {Shan}, {Sharples}, {Sholl},
  {Shourt}, {Silber}, {Silva}, {Sirk}, {Slosar}, {Smith}, {Smoot}, {Som},
  {Song}, {Sprayberry}, {Staten}, {Stefanik}, {Tarle}, {Sien Tie}, {Tinker},
  {Tojeiro}, {Valdes}, {Valenzuela}, {Valluri}, {Vargas-Magana}, {Verde},
  {Walker}, {Wang}, {Wang}, {Weaver}, {Weaverdyck}, {Wechsler}, {Weinberg},
  {White}, {Yang}, {Yeche}, {Zhang}, {Zhao}, {Zheng}, {Zhou}, {Zhou}, {Zhu},
  {Zou}, \& {Zu}}]{DESI2016}
{DESI Collaboration}, {Aghamousa},~A., {Aguilar},~J., {et~al.} 2016,
  \href{https://ui.adsabs.harvard.edu/abs/2016arXiv161100036D}{arXiv e-prints},
  \href{http://doi.org/10.48550/arXiv.1611.00036}{\color{magenta}{arXiv:1611.00036}}

\bibitem[{{Di Cesare} {et~al.}(2023){Di Cesare}, {Graziani}, {Schneider},
  {Ginolfi}, {Venditti}, {Santini}, \& {Hunt}}]{DiCesare2023}
{Di Cesare},~C., {Graziani},~L., {Schneider},~R., {et~al.} 2023,
  \href{https://ui.adsabs.harvard.edu/abs/2023MNRAS.519.4632D}{\mnras},
  \href{http://doi.org/10.1093/mnras/stac3702}{\color{magenta}{519, 4632}}

\bibitem[{{Di Valentino} {et~al.}(2021){Di Valentino}, {Mena}, {Pan},
  {Visinelli}, {Yang}, {Melchiorri}, {Mota}, {Riess}, \&
  {Silk}}]{DiValentino2021}
{Di Valentino},~E., {Mena},~O., {Pan},~S., {et~al.} 2021,
  \href{https://ui.adsabs.harvard.edu/abs/2021CQGra..38o3001D}{Classical and
  Quantum Gravity},
  \href{http://doi.org/10.1088/1361-6382/ac086d}{\color{magenta}{38, 153001}}

\bibitem[{{Dole} {et~al.}(2006){Dole}, {Lagache}, {Puget}, {Caputi},
  {Fern{\'a}ndez-Conde}, {Le Floc'h}, {Papovich}, {P{\'e}rez-Gonz{\'a}lez},
  {Rieke}, \& {Blaylock}}]{Dole06}
{Dole},~H., {Lagache},~G., {Puget},~J.~L., {et~al.} 2006,
  \href{https://ui.adsabs.harvard.edu/abs/2006A&A...451..417D}{\aap},
  \href{http://doi.org/10.1051/0004-6361:20054446}{\color{magenta}{451, 417}}

\bibitem[{{Draine} \& {Li}(2007)}]{DraineLi2007}
{Draine},~B.~T., \& {Li},~A. 2007,
  \href{https://ui.adsabs.harvard.edu/abs/2007ApJ...657..810D}{\apj},
  \href{http://doi.org/10.1086/511055}{\color{magenta}{657, 810}}

\bibitem[{{Drew} \& {Casey}(2022)}]{DrewCasey2022}
{Drew},~P.~M., \& {Casey},~C.~M. 2022,
  \href{https://ui.adsabs.harvard.edu/abs/2022ApJ...930..142D}{\apj},
  \href{http://doi.org/10.3847/1538-4357/ac6270}{\color{magenta}{930, 142}}

\bibitem[{{Dunlop} {et~al.}(2017){Dunlop}, {McLure}, {Biggs}, {Geach},
  {Micha{\l}owski}, {Ivison}, {Rujopakarn}, {van Kampen}, {Kirkpatrick},
  {Pope}, {Scott}, {Swinbank}, {Targett}, {Aretxaga}, {Austermann}, {Best},
  {Bruce}, {Chapin}, {Charlot}, {Cirasuolo}, {Coppin}, {Ellis}, {Finkelstein},
  {Hayward}, {Hughes}, {Ibar}, {Jagannathan}, {Khochfar}, {Koprowski},
  {Narayanan}, {Nyland}, {Papovich}, {Peacock}, {Rieke}, {Robertson},
  {Vernstrom}, {Werf}, {Wilson}, \& {Yun}}]{Dunlop2017}
{Dunlop},~J.~S., {McLure},~R.~J., {Biggs},~A.~D., {et~al.} 2017,
  \href{https://ui.adsabs.harvard.edu/abs/2017MNRAS.466..861D}{\mnras},
  \href{http://doi.org/10.1093/mnras/stw3088}{\color{magenta}{466, 861}}

\bibitem[{{Eales} {et~al.}(2010){Eales}, {Dunne}, {Clements}, {Cooray}, {De
  Zotti}, {Dye}, {Ivison}, {Jarvis}, {Lagache}, {Maddox}, {Negrello},
  {Serjeant}, {Thompson}, {Van Kampen}, {Amblard}, {Andreani}, {Baes},
  {Beelen}, {Bendo}, {Benford}, {Bertoldi}, {Bock}, {Bonfield}, {Boselli},
  {Bridge}, {Buat}, {Burgarella}, {Carlberg}, {Cava}, {Chanial}, {Charlot},
  {Christopher}, {Coles}, {Cortese}, {Dariush}, {da Cunha}, {Dalton}, {Danese},
  {Dannerbauer}, {Driver}, {Dunlop}, {Fan}, {Farrah}, {Frayer}, {Frenk},
  {Geach}, {Gardner}, {Gomez}, {Gonz{\'a}lez-Nuevo}, {Gonz{\'a}lez-Solares},
  {Griffin}, {Hardcastle}, {Hatziminaoglou}, {Herranz}, {Hughes}, {Ibar},
  {Jeong}, {Lacey}, {Lapi}, {Lawrence}, {Lee}, {Leeuw}, {Liske},
  {L{\'o}pez-Caniego}, {M{\"u}ller}, {Nandra}, {Panuzzo}, {Papageorgiou},
  {Patanchon}, {Peacock}, {Pearson}, {Phillipps}, {Pohlen}, {Popescu},
  {Rawlings}, {Rigby}, {Rigopoulou}, {Robotham}, {Rodighiero}, {Sansom},
  {Schulz}, {Scott}, {Smith}, {Sibthorpe}, {Smail}, {Stevens}, {Sutherland},
  {Takeuchi}, {Tedds}, {Temi}, {Tuffs}, {Trichas}, {Vaccari}, {Valtchanov},
  {van der Werf}, {Verma}, {Vieria}, {Vlahakis}, \& {White}}]{Eales2010}
{Eales},~S., {Dunne},~L., {Clements},~D., {et~al.} 2010,
  \href{https://ui.adsabs.harvard.edu/abs/2010PASP..122..499E}{\pasp},
  \href{http://doi.org/10.1086/653086}{\color{magenta}{122, 499}}

\bibitem[{{Eisenstein} {et~al.}(2005){Eisenstein}, {Zehavi}, {Hogg},
  {Scoccimarro}, {Blanton}, {Nichol}, {Scranton}, {Seo}, {Tegmark}, {Zheng},
  {Anderson}, {Annis}, {Bahcall}, {Brinkmann}, {Burles}, {Castander},
  {Connolly}, {Csabai}, {Doi}, {Fukugita}, {Frieman}, {Glazebrook}, {Gunn},
  {Hendry}, {Hennessy}, {Ivezi{\'c}}, {Kent}, {Knapp}, {Lin}, {Loh}, {Lupton},
  {Margon}, {McKay}, {Meiksin}, {Munn}, {Pope}, {Richmond}, {Schlegel},
  {Schneider}, {Shimasaku}, {Stoughton}, {Strauss}, {SubbaRao}, {Szalay},
  {Szapudi}, {Tucker}, {Yanny}, \& {York}}]{Eisenstein2005}
{Eisenstein},~D.~J., {Zehavi},~I., {Hogg},~D.~W., {et~al.} 2005,
  \href{https://ui.adsabs.harvard.edu/abs/2005ApJ...633..560E}{\apj},
  \href{http://doi.org/10.1086/466512}{\color{magenta}{633, 560}}

\bibitem[{{Endo} {et~al.}(2019){Endo}, {Karatsu}, {Tamura}, {Oshima},
  {Taniguchi}, {Takekoshi}, {Asayama}, {Bakx}, {Bosma}, {Bueno}, {Chin},
  {Fujii}, {Fujita}, {Huiting}, {Ikarashi}, {Ishida}, {Ishii}, {Kawabe},
  {Klapwijk}, {Kohno}, {Kouchi}, {Llombart}, {Maekawa}, {Murugesan},
  {Nakatsubo}, {Naruse}, {Ohtawara}, {Pascual Laguna}, {Suzuki}, {Suzuki},
  {Thoen}, {Tsukagoshi}, {Ueda}, {de Visser}, {van der Werf}, {Yates},
  {Yoshimura}, {Yurduseven}, \& {Baselmans}}]{Endo2019}
{Endo},~A., {Karatsu},~K., {Tamura},~Y., {et~al.} 2019,
  \href{https://ui.adsabs.harvard.edu/abs/2019NatAs...3..989E}{Nature
  Astronomy},
  \href{http://doi.org/10.1038/s41550-019-0850-8}{\color{magenta}{3, 989}}

\bibitem[{{Erickson} {et~al.}(2007){Erickson}, {Narayanan}, {Goeller}, \&
  {Grosslein}}]{Erickson2007}
{Erickson},~N., {Narayanan},~G., {Goeller},~R., \& {Grosslein},~R. 2007,
  \href{https://ui.adsabs.harvard.edu/abs/2007ASPC..375...71E}{in From
  Z-Machines to ALMA: (Sub)Millimeter Spectroscopy of Galaxies}, 71

\bibitem[{{Everett} {et~al.}(2020){Everett}, {Zhang}, {Crawford}, {Vieira},
  {Aravena}, {Archipley}, {Austermann}, {Benson}, {Bleem}, {Carlstrom},
  {Chang}, {Chapman}, {Crites}, {de Haan}, {Dobbs}, {George}, {Halverson},
  {Harrington}, {Holder}, {Holzapfel}, {Hrubes}, {Knox}, {Lee}, {Luong-Van},
  {Mangian}, {Marrone}, {McMahon}, {Meyer}, {Mocanu}, {Mohr}, {Natoli},
  {Padin}, {Pryke}, {Reichardt}, {Reuter}, {Ruhl}, {Sayre}, {Schaffer},
  {Shirokoff}, {Spilker}, {Stalder}, {Staniszewski}, {Stark}, {Story},
  {Switzer}, {Vanderlinde}, {Wei{\ss}}, \& {Williamson}}]{Everett2020}
{Everett},~W.~B., {Zhang},~L., {Crawford},~T.~M., {et~al.} 2020,
  \href{https://ui.adsabs.harvard.edu/abs/2020ApJ...900...55E}{\apj},
  \href{http://doi.org/10.3847/1538-4357/ab9df7}{\color{magenta}{900, 55}}

\bibitem[{{Faisst} {et~al.}(2020){Faisst}, {Schaerer}, {Lemaux}, {Oesch},
  {Fudamoto}, {Cassata}, {B{\'e}thermin}, {Capak}, {Le F{\`e}vre}, {Silverman},
  {Yan}, {Ginolfi}, {Koekemoer}, {Morselli}, {Amor{\'\i}n}, {Bardelli},
  {Boquien}, {Brammer}, {Cimatti}, {Dessauges-Zavadsky}, {Fujimoto},
  {Gruppioni}, {Hathi}, {Hemmati}, {Ibar}, {Jones}, {Khusanova}, {Loiacono},
  {Pozzi}, {Talia}, {Tasca}, {Riechers}, {Rodighiero}, {Romano}, {Scoville},
  {Toft}, {Vallini}, {Vergani}, {Zamorani}, \& {Zucca}}]{Faisst2020}
{Faisst},~A.~L., {Schaerer},~D., {Lemaux},~B.~C., {et~al.} 2020,
  \href{https://ui.adsabs.harvard.edu/abs/2020ApJS..247...61F}{\apjs},
  \href{http://doi.org/10.3847/1538-4365/ab7ccd}{\color{magenta}{247, 61}}

\bibitem[{{Ferkinhoff} {et~al.}(2010){Ferkinhoff}, {Nikola}, {Parshley},
  {Stacey}, {Irwin}, {Cho}, \& {Halpern}}]{Ferkinhoff2010}
{Ferkinhoff},~C., {Nikola},~T., {Parshley},~S.~C., {et~al.} 2010,
  \href{https://ui.adsabs.harvard.edu/abs/2010SPIE.7741E..0YF}{in Millimeter,
  Submillimeter, and Far-Infrared Detectors and Instrumentation for Astronomy
  V}\href{http://doi.org/10.1117/12.857018, }{, \color{magenta}{77410Y}}

\bibitem[{{Fujimoto} {et~al.}(2023){Fujimoto}, {Kohno}, {Ouchi}, {Oguri},
  {Kokorev}, {Brammer}, {Sun}, {Gonzalez-Lopez}, {Bauer}, {Caminha},
  {Hatsukade}, {Richard}, {Smail}, {Tsujita}, {Ueda}, {Uematsu}, {Zitrin},
  {Coe}, {Kneib}, {Postman}, {Umetsu}, {Lagos}, {Popping}, {Ao}, {Bradley},
  {Caputi}, {Dessauges-Zavadsky}, {Egami}, {Espada}, {Ivison}, {Jauzac},
  {Knudsen}, {Koekemoer}, {Magdis}, {Mahler}, {Munoz Arancibia}, {Rawle},
  {Shimasaku}, {Toft}, {Umehata}, {Valentino}, {Wang}, \&
  {Wang}}]{Fujimoto2023}
{Fujimoto},~S., {Kohno},~K., {Ouchi},~M., {et~al.} 2023,
  \href{https://ui.adsabs.harvard.edu/abs/2023arXiv230301658F}{arXiv e-prints},
  \href{http://doi.org/10.48550/arXiv.2303.01658}{\color{magenta}{arXiv:2303.01658}}

\bibitem[{{Geach} {et~al.}(2017){Geach}, {Dunlop}, {Halpern}, {Smail}, {van der
  Werf}, {Alexander}, {Almaini}, {Aretxaga}, {Arumugam}, {Asboth}, {Banerji},
  {Beanlands}, {Best}, {Blain}, {Birkinshaw}, {Chapin}, {Chapman}, {Chen},
  {Chrysostomou}, {Clarke}, {Clements}, {Conselice}, {Coppin}, {Cowley},
  {Danielson}, {Eales}, {Edge}, {Farrah}, {Gibb}, {Harrison}, {Hine}, {Hughes},
  {Ivison}, {Jarvis}, {Jenness}, {Jones}, {Karim}, {Koprowski}, {Knudsen},
  {Lacey}, {Mackenzie}, {Marsden}, {McAlpine}, {McMahon}, {Meijerink},
  {Micha{\l}owski}, {Oliver}, {Page}, {Peacock}, {Rigopoulou}, {Robson},
  {Roseboom}, {Rotermund}, {Scott}, {Serjeant}, {Simpson}, {Simpson}, {Smith},
  {Spaans}, {Stanley}, {Stevens}, {Swinbank}, {Targett}, {Thomson}, {Valiante},
  {Wake}, {Webb}, {Willott}, {Zavala}, \& {Zemcov}}]{Geach2017}
{Geach},~J.~E., {Dunlop},~J.~S., {Halpern},~M., {et~al.} 2017,
  \href{https://ui.adsabs.harvard.edu/abs/2017MNRAS.465.1789G}{\mnras},
  \href{http://doi.org/10.1093/mnras/stw2721}{\color{magenta}{465, 1789}}

\bibitem[{{Graci{\'a}-Carpio} {et~al.}(2011){Graci{\'a}-Carpio}, {Sturm},
  {Hailey-Dunsheath}, {Fischer}, {Contursi}, {Poglitsch}, {Genzel},
  {Gonz{\'a}lez-Alfonso}, {Sternberg}, {Verma}, {Christopher}, {Davies},
  {Feuchtgruber}, {de Jong}, {Lutz}, \& {Tacconi}}]{GraciaCarpio2011}
{Graci{\'a}-Carpio},~J., {Sturm},~E., {Hailey-Dunsheath},~S., {et~al.} 2011,
  \href{https://ui.adsabs.harvard.edu/abs/2011ApJ...728L...7G}{\apjl},
  \href{http://doi.org/10.1088/2041-8205/728/1/L7}{\color{magenta}{728, L7}}

\bibitem[{{Gralla} {et~al.}(2020){Gralla}, {Marriage}, {Addison}, {Baker},
  {Bond}, {Crichton}, {Datta}, {Devlin}, {Dunkley}, {D{\"u}nner}, {Fowler},
  {Gallardo}, {Hall}, {Halpern}, {Hasselfield}, {Hilton}, {Hincks},
  {Huffenberger}, {Hughes}, {Kosowsky}, {L{\'o}pez-Caraballo}, {Louis},
  {Marsden}, {Moodley}, {Niemack}, {Page}, {Partridge}, {Rivera}, {Sievers},
  {Staggs}, {Su}, {Swetz}, \& {Wollack}}]{Gralla2020}
{Gralla},~M.~B., {Marriage},~T.~A., {Addison},~G., {et~al.} 2020,
  \href{https://ui.adsabs.harvard.edu/abs/2020ApJ...893..104G}{\apj},
  \href{http://doi.org/10.3847/1538-4357/ab7915}{\color{magenta}{893, 104}}

\bibitem[{{Harris} {et~al.}(2007){Harris}, {Baker}, {Jewell}, {Rauch}, {Zonak},
  {O'Neil}, {Shelton}, {Norrod}, {Ray}, \& {Watts}}]{Harris2007}
{Harris},~A.~I., {Baker},~A.~J., {Jewell},~P.~R., {et~al.} 2007,
  \href{https://ui.adsabs.harvard.edu/abs/2007ASPC..375...82H}{in From
  Z-Machines to ALMA: (Sub)Millimeter Spectroscopy of Galaxies}, 82

\bibitem[{{Hayashi} {et~al.}(2017){Hayashi}, {Kodama}, {Kohno}, {Yamaguchi},
  {Tadaki}, {Hatsukade}, {Koyama}, {Shimakawa}, {Tamura}, \&
  {Suzuki}}]{Hayashi2017}
{Hayashi},~M., {Kodama},~T., {Kohno},~K., {et~al.} 2017,
  \href{https://ui.adsabs.harvard.edu/abs/2017ApJ...841L..21H}{\apjl},
  \href{http://doi.org/10.3847/2041-8213/aa71ad}{\color{magenta}{841, L21}}

\bibitem[{{Hayashi} {et~al.}(2018){Hayashi}, {Tadaki}, {Kodama}, {Kohno},
  {Yamaguchi}, {Hatsukade}, {Koyama}, {Shimakawa}, {Tamura}, \&
  {Suzuki}}]{Hayashi2018}
{Hayashi},~M., {Tadaki},~K.-i., {Kodama},~T., {et~al.} 2018,
  \href{https://ui.adsabs.harvard.edu/abs/2018ApJ...856..118H}{\apj},
  \href{http://doi.org/10.3847/1538-4357/aab3e7}{\color{magenta}{856, 118}}

\bibitem[{{Higuchi} {et~al.}(2019){Higuchi}, {Ouchi}, {Ono}, {Shibuya},
  {Toshikawa}, {Harikane}, {Kojima}, {Chiang}, {Egami}, {Kashikawa},
  {Overzier}, {Konno}, {Inoue}, {Hasegawa}, {Fujimoto}, {Goto}, {Ishikawa},
  {Ito}, {Komiyama}, \& {Tanaka}}]{Higuchi2019}
{Higuchi},~R., {Ouchi},~M., {Ono},~Y., {et~al.} 2019,
  \href{https://ui.adsabs.harvard.edu/abs/2019ApJ...879...28H}{\apj},
  \href{http://doi.org/10.3847/1538-4357/ab2192}{\color{magenta}{879, 28}}

\bibitem[{{Hirashita} \& {Il'in}(2022)}]{Hirashita2022}
{Hirashita},~H., \& {Il'in},~V.~B. 2022,
  \href{https://ui.adsabs.harvard.edu/abs/2022MNRAS.509.5771H}{\mnras},
  \href{http://doi.org/10.1093/mnras/stab3455}{\color{magenta}{509, 5771}}

\bibitem[{{Inami} {et~al.}(2022){Inami}, {Algera}, {Schouws}, {Sommovigo},
  {Bouwens}, {Smit}, {Stefanon}, {Bowler}, {Endsley}, {Ferrara}, {Oesch},
  {Stark}, {Aravena}, {Barrufet}, {da Cunha}, {Dayal}, {De Looze}, {Fudamoto},
  {Gonzalez}, {Graziani}, {Hodge}, {Hygate}, {Nanayakkara}, {Pallottini},
  {Riechers}, {Schneider}, {Topping}, \& {van der Werf}}]{Inami2022}
{Inami},~H., {Algera},~H. S.~B., {Schouws},~S., {et~al.} 2022,
  \href{https://ui.adsabs.harvard.edu/abs/2022MNRAS.515.3126I}{\mnras},
  \href{http://doi.org/10.1093/mnras/stac1779}{\color{magenta}{515, 3126}}

\bibitem[{{Ivison} {et~al.}(2013){Ivison}, {Swinbank}, {Smail}, {Harris},
  {Bussmann}, {Cooray}, {Cox}, {Fu}, {Kov{\'a}cs}, {Krips}, {Narayanan},
  {Negrello}, {Neri}, {Pe{\~n}arrubia}, {Richard}, {Riechers}, {Rowlands},
  {Staguhn}, {Targett}, {Amber}, {Baker}, {Bourne}, {Bertoldi}, {Bremer},
  {Calanog}, {Clements}, {Dannerbauer}, {Dariush}, {De Zotti}, {Dunne},
  {Eales}, {Farrah}, {Fleuren}, {Franceschini}, {Geach}, {George}, {Helly},
  {Hopwood}, {Ibar}, {Jarvis}, {Kneib}, {Maddox}, {Omont}, {Scott}, {Serjeant},
  {Smith}, {Thompson}, {Valiante}, {Valtchanov}, {Vieira}, \& {van der
  Werf}}]{Ivison2013}
{Ivison},~R.~J., {Swinbank},~A.~M., {Smail},~I., {et~al.} 2013,
  \href{https://ui.adsabs.harvard.edu/abs/2013ApJ...772..137I}{\apj},
  \href{http://doi.org/10.1088/0004-637X/772/2/137}{\color{magenta}{772, 137}}

\bibitem[{{Jin} {et~al.}(2021){Jin}, {Dannerbauer}, {Emonts}, {Serra}, {Lagos},
  {Thomson}, {Bassini}, {Lehnert}, {Allison}, {Champagne}, {Inderm{\"u}hle},
  {Norris}, {Seymour}, {Shimakawa}, {Casey}, {De Breuck}, {Drouart}, {Hatch},
  {Kodama}, {Koyama}, {Macgregor}, {Miley}, {Overzier},
  {P{\'e}rez-Mart{\'\i}nez}, {Rodr{\'\i}guez-Espinosa}, {R{\"o}ttgering},
  {S{\'a}nchez Portal}, \& {Ziegler}}]{Jin2021}
{Jin},~S., {Dannerbauer},~H., {Emonts},~B., {et~al.} 2021,
  \href{https://ui.adsabs.harvard.edu/abs/2021A&A...652A..11J}{\aap},
  \href{http://doi.org/10.1051/0004-6361/202040232}{\color{magenta}{652, A11}}

\bibitem[{{Johnston} {et~al.}(2021){Johnston}, {Joachimi}, {Norberg},
  {Hoekstra}, {Eriksen}, {Fortuna}, {Manzoni}, {Serrano}, {Siudek},
  {Tortorelli}, {Asorey}, {Cabayol}, {Carretero}, {Casas}, {Castander},
  {Crocce}, {Fernandez}, {Garc{\'\i}a-Bellido}, {Gaztanaga}, {Hildebrandt},
  {Miquel}, {Navarro-Girones}, {Padilla}, {Sanchez}, {Sevilla-Noarbe}, \&
  {Tallada-Cresp{\'\i}}}]{Johnston2021}
{Johnston},~H., {Joachimi},~B., {Norberg},~P., {et~al.} 2021,
  \href{https://ui.adsabs.harvard.edu/abs/2021A&A...646A.147J}{\aap},
  \href{http://doi.org/10.1051/0004-6361/202039682}{\color{magenta}{646, A147}}

\bibitem[{{Kaiser}(1987)}]{Kaiser1987}
{Kaiser},~N. 1987,
  \href{https://ui.adsabs.harvard.edu/abs/1987MNRAS.227....1K}{\mnras},
  \href{http://doi.org/10.1093/mnras/227.1.1}{\color{magenta}{227, 1}}

\bibitem[{{Karoumpis} {et~al.}(2022){Karoumpis}, {Magnelli},
  {Romano-D{\'\i}az}, {Haslbauer}, \& {Bertoldi}}]{Karoumpis2022}
{Karoumpis},~C., {Magnelli},~B., {Romano-D{\'\i}az},~E., {Haslbauer},~M., \&
  {Bertoldi},~F. 2022,
  \href{https://ui.adsabs.harvard.edu/abs/2022A&A...659A..12K}{\aap},
  \href{http://doi.org/10.1051/0004-6361/202141293}{\color{magenta}{659, A12}}

\bibitem[{{Klaassen} {et~al.}(2020){Klaassen}, {Mroczkowski}, {Cicone},
  {Hatziminaoglou}, {Sartori}, {De Breuck}, {Bryan}, {Dicker}, {Duran},
  {Groppi}, {Kaercher}, {Kawabe}, {Kohno}, \& {Geach}}]{Klaassen2020}
{Klaassen},~P.~D., {Mroczkowski},~T.~K., {Cicone},~C., {et~al.} 2020,
  \href{https://ui.adsabs.harvard.edu/abs/2020SPIE11445E..2FK}{in Ground-based
  and Airborne Telescopes VIII}\href{http://doi.org/10.1117/12.2561315, }{,
  \color{magenta}{114452F}}

\bibitem[{{Klitsch} {et~al.}(2019){Klitsch}, {P{\'e}roux}, {Zwaan}, {Smail},
  {Nelson}, {Popping}, {Chen}, {Diemer}, {Ivison}, {Allison}, {Muller},
  {Swinbank}, {Hamanowicz}, {Biggs}, \& {Dutta}}]{Klitsch2019}
{Klitsch},~A., {P{\'e}roux},~C., {Zwaan},~M.~A., {et~al.} 2019,
  \href{https://ui.adsabs.harvard.edu/abs/2019MNRAS.490.1220K}{\mnras},
  \href{http://doi.org/10.1093/mnras/stz2660}{\color{magenta}{490, 1220}}

\bibitem[{{Kravtsov} \& {Borgani}(2012)}]{Kravtsov2012}
{Kravtsov},~A.~V., \& {Borgani},~S. 2012,
  \href{https://ui.adsabs.harvard.edu/abs/2012ARA&A..50..353K}{\araa},
  \href{http://doi.org/10.1146/annurev-astro-081811-125502}{\color{magenta}{50,
  353}}

\bibitem[{{Lagache} {et~al.}(2018){Lagache}, {Cousin}, \&
  {Chatzikos}}]{Lagache2018}
{Lagache},~G., {Cousin},~M., \& {Chatzikos},~M. 2018,
  \href{https://ui.adsabs.harvard.edu/abs/2018A&A...609A.130L}{\aap},
  \href{http://doi.org/10.1051/0004-6361/201732019}{\color{magenta}{609, A130}}

\bibitem[{{Lagos} {et~al.}(2020){Lagos}, {da Cunha}, {Robotham}, {Obreschkow},
  {Valentino}, {Fujimoto}, {Magdis}, \& {Tobar}}]{Lagos2020}
{Lagos},~C. d.~P., {da Cunha},~E., {Robotham},~A. S.~G., {et~al.} 2020,
  \href{https://ui.adsabs.harvard.edu/abs/2020MNRAS.499.1948L}{\mnras},
  \href{http://doi.org/10.1093/mnras/staa2861}{\color{magenta}{499, 1948}}

\bibitem[{{Lamb} {et~al.}(2022){Lamb}, {Cleary}, {Woody}, {Catha}, {Chung},
  {Gundersen}, {Harper}, {Harris}, {Hobbs}, {Ihle}, {Kocz}, {Pearson},
  {Philip}, {Powell}, {Basoalto}, {Bond}, {Borowska}, {Breysse}, {Church},
  {Dickinson}, {Dunne}, {Eriksen}, {Foss}, {Gaier}, {Kim}, {Lawrence}, {Lunde},
  {Padmanabhan}, {Rasmussen}, {Readhead}, {Reeves}, {Rennie}, {Stutzer},
  {Viero}, {Watts}, {Wehus}, \& {Comap Collaboration}}]{Lamb2022}
{Lamb},~J.~W., {Cleary},~K.~A., {Woody},~D.~P., {et~al.} 2022,
  \href{https://ui.adsabs.harvard.edu/abs/2022ApJ...933..183L}{\apj},
  \href{http://doi.org/10.3847/1538-4357/ac63c6}{\color{magenta}{933, 183}}

\bibitem[{{Lammers} {et~al.}(2022){Lammers}, {Hill}, {Lim}, {Scott},
  {Ca{\~n}ameras}, \& {Dole}}]{Lammers2022}
{Lammers},~C., {Hill},~R., {Lim},~S., {et~al.} 2022,
  \href{https://ui.adsabs.harvard.edu/abs/2022MNRAS.514.5004L}{\mnras},
  \href{http://doi.org/10.1093/mnras/stac1555}{\color{magenta}{514, 5004}}

\bibitem[{{Laureijs} {et~al.}(2011){Laureijs}, {Amiaux}, {Arduini},
  {Augu{\`e}res}, {Brinchmann}, {Cole}, {Cropper}, {Dabin}, {Duvet}, {Ealet},
  {Garilli}, {Gondoin}, {Guzzo}, {Hoar}, {Hoekstra}, {Holmes}, {Kitching},
  {Maciaszek}, {Mellier}, {Pasian}, {Percival}, {Rhodes}, {Saavedra Criado},
  {Sauvage}, {Scaramella}, {Valenziano}, {Warren}, {Bender}, {Castander},
  {Cimatti}, {Le F{\`e}vre}, {Kurki-Suonio}, {Levi}, {Lilje}, {Meylan},
  {Nichol}, {Pedersen}, {Popa}, {Rebolo Lopez}, {Rix}, {Rottgering},
  {Zeilinger}, {Grupp}, {Hudelot}, {Massey}, {Meneghetti}, {Miller}, {Paltani},
  {Paulin-Henriksson}, {Pires}, {Saxton}, {Schrabback}, {Seidel}, {Walsh},
  {Aghanim}, {Amendola}, {Bartlett}, {Baccigalupi}, {Beaulieu}, {Benabed},
  {Cuby}, {Elbaz}, {Fosalba}, {Gavazzi}, {Helmi}, {Hook}, {Irwin}, {Kneib},
  {Kunz}, {Mannucci}, {Moscardini}, {Tao}, {Teyssier}, {Weller}, {Zamorani},
  {Zapatero Osorio}, {Boulade}, {Foumond}, {Di Giorgio}, {Guttridge}, {James},
  {Kemp}, {Martignac}, {Spencer}, {Walton}, {Bl{\"u}mchen}, {Bonoli},
  {Bortoletto}, {Cerna}, {Corcione}, {Fabron}, {Jahnke}, {Ligori}, {Madrid},
  {Martin}, {Morgante}, {Pamplona}, {Prieto}, {Riva}, {Toledo}, {Trifoglio},
  {Zerbi}, {Abdalla}, {Douspis}, {Grenet}, {Borgani}, {Bouwens}, {Courbin},
  {Delouis}, {Dubath}, {Fontana}, {Frailis}, {Grazian}, {Koppenh{\"o}fer},
  {Mansutti}, {Melchior}, {Mignoli}, {Mohr}, {Neissner}, {Noddle}, {Poncet},
  {Scodeggio}, {Serrano}, {Shane}, {Starck}, {Surace}, {Taylor},
  {Verdoes-Kleijn}, {Vuerli}, {Williams}, {Zacchei}, {Altieri}, {Escudero
  Sanz}, {Kohley}, {Oosterbroek}, {Astier}, {Bacon}, {Bardelli}, {Baugh},
  {Bellagamba}, {Benoist}, {Bianchi}, {Biviano}, {Branchini}, {Carbone},
  {Cardone}, {Clements}, {Colombi}, {Conselice}, {Cresci}, {Deacon}, {Dunlop},
  {Fedeli}, {Fontanot}, {Franzetti}, {Giocoli}, {Garcia-Bellido}, {Gow},
  {Heavens}, {Hewett}, {Heymans}, {Holland}, {Huang}, {Ilbert}, {Joachimi},
  {Jennins}, {Kerins}, {Kiessling}, {Kirk}, {Kotak}, {Krause}, {Lahav}, {van
  Leeuwen}, {Lesgourgues}, {Lombardi}, {Magliocchetti}, {Maguire}, {Majerotto},
  {Maoli}, {Marulli}, {Maurogordato}, {McCracken}, {McLure}, {Melchiorri},
  {Merson}, {Moresco}, {Nonino}, {Norberg}, {Peacock}, {Pello}, {Penny},
  {Pettorino}, {Di Porto}, {Pozzetti}, {Quercellini}, {Radovich}, {Rassat},
  {Roche}, {Ronayette}, {Rossetti}, {Sartoris}, {Schneider}, {Semboloni},
  {Serjeant}, {Simpson}, {Skordis}, {Smadja}, {Smartt}, {Spano}, {Spiro},
  {Sullivan}, {Tilquin}, {Trotta}, {Verde}, {Wang}, {Williger}, {Zhao},
  {Zoubian}, \& {Zucca}}]{Euclid2011}
{Laureijs},~R., {Amiaux},~J., {Arduini},~S., {et~al.} 2011,
  \href{https://ui.adsabs.harvard.edu/abs/2011arXiv1110.3193L}{arXiv e-prints},
  \href{http://doi.org/10.48550/arXiv.1110.3193}{\color{magenta}{arXiv:1110.3193}}

\bibitem[{{Li} {et~al.}(2018){Li}, {Bradford}, {Crites}, {Hunacek}, {Wei},
  {Cheng}, {Chang}, \& {Bock}}]{Li2018}
{Li},~C.-T., {Bradford},~C.~M., {Crites},~A., {et~al.} 2018,
  \href{https://ui.adsabs.harvard.edu/abs/2018SPIE10708E..3FL}{in Millimeter,
  Submillimeter, and Far-Infrared Detectors and Instrumentation for Astronomy
  IX}\href{http://doi.org/10.1117/12.2311415, }{, \color{magenta}{107083F}}

\bibitem[{{Lim} {et~al.}(2020){Lim}, {Chen}, {Smail}, {Wang}, {Tee}, {Lin},
  {Scott}, {Toba}, {Chang}, {Ao}, {Babul}, {Bunker}, {Chapman}, {Clements},
  {Conselice}, {Gao}, {Greve}, {Ho}, {Hong}, {Hwang}, {Koprowski},
  {Micha{\l}owski}, {Shim}, {Shu}, \& {Simpson}}]{Lim2020}
{Lim},~C.-F., {Chen},~C.-C., {Smail},~I., {et~al.} 2020,
  \href{https://ui.adsabs.harvard.edu/abs/2020ApJ...895..104L}{\apj},
  \href{http://doi.org/10.3847/1538-4357/ab8eaf}{\color{magenta}{895, 104}}

\bibitem[{{Lovell} {et~al.}(2018){Lovell}, {Thomas}, \& {Wilkins}}]{Lovell2018}
{Lovell},~C.~C., {Thomas},~P.~A., \& {Wilkins},~S.~M. 2018,
  \href{https://ui.adsabs.harvard.edu/abs/2018MNRAS.474.4612L}{\mnras},
  \href{http://doi.org/10.1093/mnras/stx3090}{\color{magenta}{474, 4612}}

\bibitem[{{Lutz} {et~al.}(2001){Lutz}, {Dunlop}, {Almaini}, {Andreani},
  {Blain}, {Efstathiou}, {Fox}, {Genzel}, {Hasinger}, {Hughes}, {Ivison},
  {Lawrence}, {Mann}, {Oliver}, {Peacock}, {Rigopoulou}, {Rowan-Robinson},
  {Scott}, {Serjeant}, \& {Tacconi}}]{Lutz2001}
{Lutz},~D., {Dunlop},~J.~S., {Almaini},~O., {et~al.} 2001,
  \href{https://ui.adsabs.harvard.edu/abs/2001A&A...378...70L}{\aap},
  \href{http://doi.org/10.1051/0004-6361:20011120}{\color{magenta}{378, 70}}

\bibitem[{{Madau} \& {Dickinson}(2014)}]{Madau2014}
{Madau},~P., \& {Dickinson},~M. 2014,
  \href{https://ui.adsabs.harvard.edu/abs/2014ARA&A..52..415M}{\araa},
  \href{http://doi.org/10.1146/annurev-astro-081811-125615}{\color{magenta}{52,
  415}}

\bibitem[{{Madhavacheril} {et~al.}(2023){Madhavacheril}, {Qu}, {Sherwin},
  {MacCrann}, {Li}, {Abril-Cabezas}, {Ade}, {Aiola}, {Alford}, {Amiri},
  {Amodeo}, {An}, {Atkins}, {Austermann}, {Battaglia}, {Battistelli}, {Beall},
  {Bean}, {Beringue}, {Bhandarkar}, {Biermann}, {Bolliet}, {Bond}, {Cai},
  {Calabrese}, {Calafut}, {Capalbo}, {Carrero}, {Challinor}, {Chesmore}, {Cho},
  {Choi}, {Clark}, {C{\'o}rdova Rosado}, {Cothard}, {Coughlin}, {Coulton},
  {Crowley}, {Dalal}, {Darwish}, {Devlin}, {Dicker}, {Doze}, {Duell}, {Duff},
  {Duivenvoorden}, {Dunkley}, {D{\"u}nner}, {Fanfani}, {Fankhanel}, {Farren},
  {Ferraro}, {Freundt}, {Fuzia}, {Gallardo}, {Garrido}, {Givans}, {Gluscevic},
  {Golec}, {Guan}, {Hall}, {Halpern}, {Han}, {Harrison}, {Hasselfield},
  {Healy}, {Henderson}, {Hensley}, {Herv{\'\i}as-Caimapo}, {Hill}, {Hilton},
  {Hilton}, {Hincks}, {Hlo{\v{z}}ek}, {Ho}, {Huber}, {Hubmayr}, {Huffenberger},
  {Hughes}, {Irwin}, {Isopi}, {Jense}, {Keller}, {Kim}, {Knowles}, {Koopman},
  {Kosowsky}, {Kramer}, {Kusiak}, {La Posta}, {Lague}, {Lakey}, {Lee}, {Li},
  {Limon}, {Lokken}, {Louis}, {Lungu}, {MacInnis}, {Maldonado}, {Maldonado},
  {Mallaby-Kay}, {Marques}, {McMahon}, {Mehta}, {Menanteau}, {Moodley},
  {Morris}, {Mroczkowski}, {Naess}, {Namikawa}, {Nati}, {Newburgh}, {Nicola},
  {Niemack}, {Nolta}, {Orlowski-Scherer}, {Page}, {Pandey}, {Partridge},
  {Prince}, {Puddu}, {Radiconi}, {Robertson}, {Rojas}, {Sakuma}, {Salatino},
  {Schaan}, {Schmitt}, {Sehgal}, {Shaikh}, {Sierra}, {Sievers}, {Sif{\'o}n},
  {Simon}, {Sonka}, {Spergel}, {Staggs}, {Storer}, {Switzer}, {Tampier},
  {Thornton}, {Trac}, {Treu}, {Tucker}, {Ulluom}, {Vale}, {Van Engelen}, {Van
  Lanen}, {van Marrewijk}, {Vargas}, {Vavagiakis}, {Wagoner}, {Wang}, {Wenzl},
  {Wollack}, {Xu}, {Zago}, \& {Zhang}}]{Madhavacheril2023}
{Madhavacheril},~M.~S., {Qu},~F.~J., {Sherwin},~B.~D., {et~al.} 2023,
  \href{https://ui.adsabs.harvard.edu/abs/2023arXiv230405203M}{arXiv e-prints},
  \href{http://doi.org/10.48550/arXiv.2304.05203}{\color{magenta}{arXiv:2304.05203}}

\bibitem[{{Monfardini} {et~al.}(2022){Monfardini}, {Beelen}, {Benoit},
  {Bounmy}, {Calvo}, {Catalano}, {Goupy}, {Lagache}, {Ade}, {Barria},
  {B{\'e}thermin}, {Bourrion}, {Bres}, {Breuck}, {D{\'e}sert}, {Duvauchelle},
  {Fasano}, {Fenouillet}, {Garcia}, {Garde}, {Hoarau}, {Hu}, {Lambert},
  {Levy-Bertrand}, {Lundgren}, {Macias-Perez}, {Marpaud}, {Pisano}, {Ponthieu},
  {Prieur}, {Roni}, {Roudier}, {Tourres}, {Tucker}, {Cantzler}, {Caro}, {Diaz},
  {Dur{\'a}n}, {Montenegro}, {Navarro}, {Olguin}, {Palma}, {Parra}, \&
  {Santana}}]{Monfardini2022}
{Monfardini},~A., {Beelen},~A., {Benoit},~A., {et~al.} 2022,
  \href{https://ui.adsabs.harvard.edu/abs/2022JLTP..209..751M}{Journal of Low
  Temperature Physics},
  \href{http://doi.org/10.1007/s10909-022-02690-3}{\color{magenta}{209, 751}}

\bibitem[{{Mountrichas} \& {Shankar}(2023)}]{Mountrichas2023}
{Mountrichas},~G., \& {Shankar},~F. 2023,
  \href{https://ui.adsabs.harvard.edu/abs/2023MNRAS.518.2088M}{\mnras},
  \href{http://doi.org/10.1093/mnras/stac3211}{\color{magenta}{518, 2088}}

\bibitem[{{Mroczkowski} {et~al.}(2019){Mroczkowski}, {De Breuck}, {Kemper},
  {Phillips}, {Fuller}, {Beltr{\'a}n}, {Laing}, {Marconi}, {Testi}, {Yagoubov},
  {George}, \& {McGenn}}]{Mroczkowski2019}
{Mroczkowski},~T., {De Breuck},~C., {Kemper},~C., {et~al.} 2019,
  \href{https://ui.adsabs.harvard.edu/abs/2019arXiv190509064M}{arXiv e-prints},
  \href{http://doi.org/10.48550/arXiv.1905.09064}{\color{magenta}{arXiv:1905.09064}}

\bibitem[{{Mroczkowski} {et~al.}(2023){Mroczkowski}, {Cicone}, {Reichert},
  {Gallardo}, {Kaercher}, {Hills}, {Bok}, {Dahl}, {Dubois-dit-Bonclaude},
  {Kiselev}, {Timpe}, {Zimmerer}, {Dicker}, {Macintosh}, {Klaassen}, \&
  {Niemack}}]{Mroczkowski2023}
{Mroczkowski},~T., {Cicone},~C., {Reichert},~M., {et~al.} 2023,
  \href{https://ui.adsabs.harvard.edu/abs/2023arXiv230810952M}{in 2023 XXXVth
  General Assembly and Scientific Symposium of the International Union of Radio
  Science (URSI
  GASS)}\href{http://doi.org/10.23919/URSIGASS57860.2023.10265372, }{,
  \color{magenta}{1--4}}

\bibitem[{Mroczkowski {et~al.}(2024)Mroczkowski, Gallardo, Timpe, Kiselev,
  Groh, Kaercher, Reichert, Cicone, Puddu, dit Bonclaude, Bok, Dahl, Macintosh,
  Dicker, Viole, Sartori, Venegas, Zeyringer, Niemack, Poppi, Olguin,
  Hatziminaoglou, Breuck, Klaassen, Montenegro-Montes, \&
  Zimmerer}]{Mroczkowski2024}
Mroczkowski,~T., Gallardo,~P.~A., Timpe,~M., {et~al.} 2024,
  \href{https://ui.adsabs.harvard.edu/abs/2024arXiv240218645M}{arXiv e-prints},
  \href{http://doi.org/10.48550/arXiv.2402.18645}{\color{magenta}{arXiv:2402.18645}}

\bibitem[{{Muldrew} {et~al.}(2015){Muldrew}, {Hatch}, \& {Cooke}}]{Muldrew2015}
{Muldrew},~S.~I., {Hatch},~N.~A., \& {Cooke},~E.~A. 2015,
  \href{https://ui.adsabs.harvard.edu/abs/2015MNRAS.452.2528M}{\mnras},
  \href{http://doi.org/10.1093/mnras/stv1449}{\color{magenta}{452, 2528}}

\bibitem[{{Naylor} {et~al.}(2003){Naylor}, {Ade}, {Bock}, {Bradford},
  {Dragovan}, {Duband}, {Earle}, {Glenn}, {Matsuhara}, {Nguyen}, {Yun}, \&
  {Zmuidzinas}}]{Naylor2003}
{Naylor},~B.~J., {Ade},~P. A.~R., {Bock},~J.~J., {et~al.} 2003,
  \href{https://ui.adsabs.harvard.edu/abs/2003SPIE.4855..239N}{in Millimeter
  and Submillimeter Detectors for
  Astronomy}\href{http://doi.org/10.1117/12.459419, }{,
  \color{magenta}{239--248}}

\bibitem[{{Nishimichi} {et~al.}(2019){Nishimichi}, {Takada}, {Takahashi},
  {Osato}, {Shirasaki}, {Oogi}, {Miyatake}, {Oguri}, {Murata}, {Kobayashi}, \&
  {Yoshida}}]{Nishimichi2019}
{Nishimichi},~T., {Takada},~M., {Takahashi},~R., {et~al.} 2019,
  \href{https://ui.adsabs.harvard.edu/abs/2019ApJ...884...29N}{\apj},
  \href{http://doi.org/10.3847/1538-4357/ab3719}{\color{magenta}{884, 29}}

\bibitem[{{Noble} {et~al.}(2017){Noble}, {McDonald}, {Muzzin}, {Nantais},
  {Rudnick}, {van Kampen}, {Webb}, {Wilson}, {Yee}, {Boone}, {Cooper},
  {DeGroot}, {Delahaye}, {Demarco}, {Foltz}, {Hayden}, {Lidman},
  {Manilla-Robles}, \& {Perlmutter}}]{Noble2017}
{Noble},~A.~G., {McDonald},~M., {Muzzin},~A., {et~al.} 2017,
  \href{https://ui.adsabs.harvard.edu/abs/2017ApJ...842L..21N}{\apjl},
  \href{http://doi.org/10.3847/2041-8213/aa77f3}{\color{magenta}{842, L21}}

\bibitem[{{Noble} {et~al.}(2019){Noble}, {Muzzin}, {McDonald}, {Rudnick},
  {Matharu}, {Cooper}, {Demarco}, {Lidman}, {Nantais}, {van Kampen}, {Webb},
  {Wilson}, \& {Yee}}]{Noble2019}
{Noble},~A.~G., {Muzzin},~A., {McDonald},~M., {et~al.} 2019,
  \href{https://ui.adsabs.harvard.edu/abs/2019ApJ...870...56N}{\apj},
  \href{http://doi.org/10.3847/1538-4357/aaf1c6}{\color{magenta}{870, 56}}

\bibitem[{{Norberg} {et~al.}(2009){Norberg}, {Baugh}, {Gazta{\~n}aga}, \&
  {Croton}}]{Norberg2009}
{Norberg},~P., {Baugh},~C.~M., {Gazta{\~n}aga},~E., \& {Croton},~D.~J. 2009,
  \href{https://ui.adsabs.harvard.edu/abs/2009MNRAS.396...19N}{\mnras},
  \href{http://doi.org/10.1111/j.1365-2966.2009.14389.x}{\color{magenta}{396,
  19}}

\bibitem[{{Oliver} {et~al.}(2012){Oliver}, {Bock}, {Altieri}, {Amblard},
  {Arumugam}, {Aussel}, {Babbedge}, {Beelen}, {B{\'e}thermin}, {Blain},
  {Boselli}, {Bridge}, {Brisbin}, {Buat}, {Burgarella},
  {Castro-Rodr{\'\i}guez}, {Cava}, {Chanial}, {Cirasuolo}, {Clements},
  {Conley}, {Conversi}, {Cooray}, {Dowell}, {Dubois}, {Dwek}, {Dye}, {Eales},
  {Elbaz}, {Farrah}, {Feltre}, {Ferrero}, {Fiolet}, {Fox}, {Franceschini},
  {Gear}, {Giovannoli}, {Glenn}, {Gong}, {Gonz{\'a}lez Solares}, {Griffin},
  {Halpern}, {Harwit}, {Hatziminaoglou}, {Heinis}, {Hurley}, {Hwang}, {Hyde},
  {Ibar}, {Ilbert}, {Isaak}, {Ivison}, {Lagache}, {Le Floc'h}, {Levenson},
  {Faro}, {Lu}, {Madden}, {Maffei}, {Magdis}, {Mainetti}, {Marchetti},
  {Marsden}, {Marshall}, {Mortier}, {Nguyen}, {O'Halloran}, {Omont}, {Page},
  {Panuzzo}, {Papageorgiou}, {Patel}, {Pearson}, {P{\'e}rez-Fournon}, {Pohlen},
  {Rawlings}, {Raymond}, {Rigopoulou}, {Riguccini}, {Rizzo}, {Rodighiero},
  {Roseboom}, {Rowan-Robinson}, {S{\'a}nchez Portal}, {Schulz}, {Scott},
  {Seymour}, {Shupe}, {Smith}, {Stevens}, {Symeonidis}, {Trichas}, {Tugwell},
  {Vaccari}, {Valtchanov}, {Vieira}, {Viero}, {Vigroux}, {Wang}, {Ward},
  {Wardlow}, {Wright}, {Xu}, \& {Zemcov}}]{Oliver2012}
{Oliver},~S.~J., {Bock},~J., {Altieri},~B., {et~al.} 2012,
  \href{https://ui.adsabs.harvard.edu/abs/2012MNRAS.424.1614O}{\mnras},
  \href{http://doi.org/10.1111/j.1365-2966.2012.20912.x}{\color{magenta}{424,
  1614}}

\bibitem[{{Overzier}(2016)}]{Overzier2016}
{Overzier},~R.~A. 2016,
  \href{https://ui.adsabs.harvard.edu/abs/2016A&ARv..24...14O}{\aapr},
  \href{http://doi.org/10.1007/s00159-016-0100-3}{\color{magenta}{24, 14}}

\bibitem[{{Papadopoulos} {et~al.}(2004){Papadopoulos}, {Thi}, \&
  {Viti}}]{Papadopoulos2004}
{Papadopoulos},~P.~P., {Thi},~W.~F., \& {Viti},~S. 2004,
  \href{https://ui.adsabs.harvard.edu/abs/2004MNRAS.351..147P}{\mnras},
  \href{http://doi.org/10.1111/j.1365-2966.2004.07762.x}{\color{magenta}{351,
  147}}

\bibitem[{{Peacock} {et~al.}(2001){Peacock}, {Cole}, {Norberg}, {Baugh},
  {Bland-Hawthorn}, {Bridges}, {Cannon}, {Colless}, {Collins}, {Couch},
  {Dalton}, {Deeley}, {De Propris}, {Driver}, {Efstathiou}, {Ellis}, {Frenk},
  {Glazebrook}, {Jackson}, {Lahav}, {Lewis}, {Lumsden}, {Maddox}, {Percival},
  {Peterson}, {Price}, {Sutherland}, \& {Taylor}}]{Peacock2001}
{Peacock},~J.~A., {Cole},~S., {Norberg},~P., {et~al.} 2001,
  \href{https://ui.adsabs.harvard.edu/abs/2001Natur.410..169P}{\nat},
  \href{http://doi.org/10.1038/35065528}{\color{magenta}{410, 169}}

\bibitem[{{Planck Collaboration} {et~al.}(2015){Planck Collaboration},
  {Aghanim}, {Altieri}, {Arnaud}, {Ashdown}, {Aumont}, {Baccigalupi}, {Banday},
  {Barreiro}, {Bartolo}, {Battaner}, {Beelen}, {Benabed}, {Benoit-L{\'e}vy},
  {Bernard}, {Bersanelli}, {Bethermin}, {Bielewicz}, {Bonavera}, {Bond},
  {Borrill}, {Bouchet}, {Boulanger}, {Burigana}, {Calabrese}, {Canameras},
  {Cardoso}, {Catalano}, {Chamballu}, {Chary}, {Chiang}, {Christensen},
  {Clements}, {Colombi}, {Couchot}, {Crill}, {Curto}, {Danese}, {Dassas},
  {Davies}, {Davis}, {de Bernardis}, {de Rosa}, {de Zotti}, {Delabrouille},
  {Diego}, {Dole}, {Donzelli}, {Dor{\'e}}, {Douspis}, {Ducout}, {Dupac},
  {Efstathiou}, {Elsner}, {En{\ss}lin}, {Falgarone}, {Flores-Cacho}, {Forni},
  {Frailis}, {Fraisse}, {Franceschi}, {Frejsel}, {Frye}, {Galeotta}, {Galli},
  {Ganga}, {Giard}, {Gjerl{\o}w}, {Gonz{\'a}lez-Nuevo}, {G{\'o}rski},
  {Gregorio}, {Gruppuso}, {Gu{\'e}ry}, {Hansen}, {Hanson}, {Harrison}, {Helou},
  {Hern{\'a}ndez-Monteagudo}, {Hildebrandt}, {Hivon}, {Hobson}, {Holmes},
  {Hovest}, {Huffenberger}, {Hurier}, {Jaffe}, {Jaffe}, {Keih{\"a}nen},
  {Keskitalo}, {Kisner}, {Kneissl}, {Knoche}, {Kunz}, {Kurki-Suonio},
  {Lagache}, {Lamarre}, {Lasenby}, {Lattanzi}, {Lawrence}, {Le Floc'h},
  {Leonardi}, {Levrier}, {Liguori}, {Lilje}, {Linden-V{\o}rnle},
  {L{\'o}pez-Caniego}, {Lubin}, {Mac{\'\i}as-P{\'e}rez}, {MacKenzie}, {Maffei},
  {Mandolesi}, {Maris}, {Martin}, {Martinache}, {Mart{\'\i}nez-Gonz{\'a}lez},
  {Masi}, {Matarrese}, {Mazzotta}, {Melchiorri}, {Mennella}, {Migliaccio},
  {Moneti}, {Montier}, {Morgante}, {Mortlock}, {Munshi}, {Murphy}, {Natoli},
  {Negrello}, {Nesvadba}, {Novikov}, {Novikov}, {Omont}, {Pagano}, {Pajot},
  {Pasian}, {Perdereau}, {Perotto}, {Perrotta}, {Pettorino}, {Piacentini},
  {Piat}, {Plaszczynski}, {Pointecouteau}, {Polenta}, {Popa}, {Pratt},
  {Prunet}, {Puget}, {Rachen}, {Reach}, {Reinecke}, {Remazeilles}, {Renault},
  {Ristorcelli}, {Rocha}, {Roudier}, {Rusholme}, {Sandri}, {Santos}, {Savini},
  {Scott}, {Spencer}, {Stolyarov}, {Sunyaev}, {Sutton}, {Sygnet}, {Tauber},
  {Terenzi}, {Toffolatti}, {Tomasi}, {Tristram}, {Tucci}, {Umana},
  {Valenziano}, {Valiviita}, {Valtchanov}, {Van Tent}, {Vieira}, {Vielva},
  {Wade}, {Wandelt}, {Wehus}, {Welikala}, {Zacchei}, \& {Zonca}}]{Aghanim2015}
{Planck Collaboration}, {Aghanim},~N., {Altieri},~B., {et~al.} 2015,
  \href{https://ui.adsabs.harvard.edu/abs/2015A&A...582A..30P}{\aap},
  \href{http://doi.org/10.1051/0004-6361/201424790}{\color{magenta}{582, A30}}

\bibitem[{{Planck Collaboration} {et~al.}(2016){Planck Collaboration}, {Ade},
  {Aghanim}, {Arnaud}, {Aumont}, {Baccigalupi}, {Banday}, {Barreiro},
  {Bartolo}, {Battaner}, {Benabed}, {Benoit-L{\'e}vy}, {Bernard}, {Bersanelli},
  {Bielewicz}, {Bonaldi}, {Bonavera}, {Bond}, {Borrill}, {Bouchet},
  {Boulanger}, {Burigana}, {Butler}, {Calabrese}, {Catalano}, {Chiang},
  {Christensen}, {Clements}, {Colombo}, {Couchot}, {Coulais}, {Crill}, {Curto},
  {Cuttaia}, {Danese}, {Davies}, {Davis}, {de Bernardis}, {de Rosa}, {de
  Zotti}, {Delabrouille}, {Dickinson}, {Diego}, {Dole}, {Dor{\'e}}, {Douspis},
  {Ducout}, {Dupac}, {Elsner}, {En{\ss}lin}, {Eriksen}, {Falgarone}, {Finelli},
  {Flores-Cacho}, {Frailis}, {Fraisse}, {Franceschi}, {Galeotta}, {Galli},
  {Ganga}, {Giard}, {Giraud-H{\'e}raud}, {Gjerl{\o}w}, {Gonz{\'a}lez-Nuevo},
  {G{\'o}rski}, {Gregorio}, {Gruppuso}, {Gudmundsson}, {Hansen}, {Harrison},
  {Helou}, {Hern{\'a}ndez-Monteagudo}, {Herranz}, {Hildebrandt}, {Hivon},
  {Hobson}, {Hornstrup}, {Hovest}, {Huffenberger}, {Hurier}, {Jaffe}, {Jaffe},
  {Keih{\"a}nen}, {Keskitalo}, {Kisner}, {Kneissl}, {Knoche}, {Kunz},
  {Kurki-Suonio}, {Lagache}, {Lamarre}, {Lasenby}, {Lattanzi}, {Lawrence},
  {Leonardi}, {Levrier}, {Liguori}, {Lilje}, {Linden-V{\o}rnle},
  {L{\'o}pez-Caniego}, {Lubin}, {Mac{\'\i}as-P{\'e}rez}, {Maffei}, {Maggio},
  {Maino}, {Mandolesi}, {Mangilli}, {Maris}, {Martin},
  {Mart{\'\i}nez-Gonz{\'a}lez}, {Masi}, {Matarrese}, {Melchiorri}, {Mennella},
  {Migliaccio}, {Mitra}, {Miville-Desch{\^e}nes}, {Moneti}, {Montier},
  {Morgante}, {Mortlock}, {Munshi}, {Murphy}, {Nati}, {Natoli}, {Nesvadba},
  {Noviello}, {Novikov}, {Novikov}, {Oxborrow}, {Pagano}, {Pajot}, {Paoletti},
  {Partridge}, {Pasian}, {Pearson}, {Perdereau}, {Perotto}, {Pettorino},
  {Piacentini}, {Piat}, {Plaszczynski}, {Pointecouteau}, {Polenta}, {Pratt},
  {Prunet}, {Puget}, {Rachen}, {Reinecke}, {Remazeilles}, {Renault}, {Renzi},
  {Ristorcelli}, {Rocha}, {Rosset}, {Rossetti}, {Roudier},
  {Rubi{\~n}o-Mart{\'\i}n}, {Rusholme}, {Sandri}, {Santos}, {Savelainen},
  {Savini}, {Scott}, {Spencer}, {Stolyarov}, {Stompor}, {Sudiwala}, {Sunyaev},
  {Suur-Uski}, {Sygnet}, {Tauber}, {Terenzi}, {Toffolatti}, {Tomasi},
  {Tristram}, {Tucci}, {T{\"u}rler}, {Umana}, {Valenziano}, {Valiviita}, {Van
  Tent}, {Vielva}, {Villa}, {Wade}, {Wandelt}, {Wehus}, {Welikala}, {Yvon},
  {Zacchei}, \& {Zonca}}]{Ade2016}
{Planck Collaboration}, {Ade},~P.~A.~R., {Aghanim},~N., {et~al.} 2016,
  \href{https://ui.adsabs.harvard.edu/abs/2016A&A...596A.100P}{\aap},
  \href{http://doi.org/10.1051/0004-6361/201527206}{\color{magenta}{596, A100}}

\bibitem[{{Popping} {et~al.}(2016){Popping}, {van Kampen}, {Decarli}, {Spaans},
  {Somerville}, \& {Trager}}]{Popping2016}
{Popping},~G., {van Kampen},~E., {Decarli},~R., {et~al.} 2016,
  \href{https://ui.adsabs.harvard.edu/abs/2016MNRAS.461...93P}{\mnras},
  \href{http://doi.org/10.1093/mnras/stw1323}{\color{magenta}{461, 93}}

\bibitem[{{Ramasawmy} {et~al.}(2022){Ramasawmy}, {Klaassen}, {Cicone},
  {Mroczkowski}, {Chen}, {Cornish}, {da Cunha}, {Hatziminaoglou}, {Johnstone},
  {Liu}, {Perrott}, {Schimek}, {Stanke}, \& {Wedemeyer}}]{Ramasawmy2022}
{Ramasawmy},~J., {Klaassen},~P.~D., {Cicone},~C., {et~al.} 2022,
  \href{https://ui.adsabs.harvard.edu/abs/2022SPIE12190E..07R}{in Millimeter,
  Submillimeter, and Far-Infrared Detectors and Instrumentation for Astronomy
  XI}\href{http://doi.org/10.1117/12.2627505, }{, \color{magenta}{1219007}}

\bibitem[{{Reuter} {et~al.}(2020){Reuter}, {Vieira}, {Spilker}, {Weiss},
  {Aravena}, {Archipley}, {B{\'e}thermin}, {Chapman}, {De Breuck}, {Dong},
  {Everett}, {Fu}, {Greve}, {Hayward}, {Hill}, {Hezaveh}, {Jarugula}, {Litke},
  {Malkan}, {Marrone}, {Narayanan}, {Phadke}, {Stark}, \&
  {Strandet}}]{Reuter2020}
{Reuter},~C., {Vieira},~J.~D., {Spilker},~J.~S., {et~al.} 2020,
  \href{https://ui.adsabs.harvard.edu/abs/2020ApJ...902...78R}{\apj},
  \href{http://doi.org/10.3847/1538-4357/abb599}{\color{magenta}{902, 78}}

\bibitem[{{Rudnick} {et~al.}(2017){Rudnick}, {Hodge}, {Walter}, {Momcheva},
  {Tran}, {Papovich}, {da Cunha}, {Decarli}, {Saintonge}, {Willmer}, {Lotz}, \&
  {Lentati}}]{Rudnick2017}
{Rudnick},~G., {Hodge},~J., {Walter},~F., {et~al.} 2017,
  \href{https://ui.adsabs.harvard.edu/abs/2017ApJ...849...27R}{\apj},
  \href{http://doi.org/10.3847/1538-4357/aa87b2}{\color{magenta}{849, 27}}

\bibitem[{{Salim} \& {Narayanan}(2020)}]{SalimNarayanan2020}
{Salim},~S., \& {Narayanan},~D. 2020,
  \href{https://ui.adsabs.harvard.edu/abs/2020ARA&A..58..529S}{\araa},
  \href{http://doi.org/10.1146/annurev-astro-032620-021933}{\color{magenta}{58,
  529}}

\bibitem[{{Schimek} {et~al.}(2024){Schimek}, {Decataldo}, {Shen}, {Cicone},
  {Baumschlager}, {van Kampen}, {Klaassen}, {Madau}, {Di Mascolo}, {Mayer},
  {Montoya Arroyave}, {Mroczkowski}, \& {Warraich}}]{Schimek24}
{Schimek},~A., {Decataldo},~D., {Shen},~S., {et~al.} 2024,
  \href{https://ui.adsabs.harvard.edu/abs/2024A&A...682A..98S}{\aap},
  \href{http://doi.org/10.1051/0004-6361/202346945}{\color{magenta}{682, A98}}

\bibitem[{{Shirley} {et~al.}(2021){Shirley}, {Duncan}, {Campos Varillas},
  {Hurley}, {Ma{\l}ek}, {Roehlly}, {Smith}, {Aussel}, {Bakx}, {Buat},
  {Burgarella}, {Christopher}, {Duivenvoorden}, {Eales}, {Efstathiou},
  {Gonz{\'a}lez Solares}, {Griffin}, {Jarvis}, {Faro}, {Marchetti}, {McCheyne},
  {Papadopoulos}, {Penner}, {Pons}, {Prescott}, {Rigby}, {Rottgering},
  {Saxena}, {Scudder}, {Vaccari}, {Wang}, \& {Oliver}}]{Shirley2021}
{Shirley},~R., {Duncan},~K., {Campos Varillas},~M.~C., {et~al.} 2021,
  \href{https://ui.adsabs.harvard.edu/abs/2021MNRAS.507..129S}{\mnras},
  \href{http://doi.org/10.1093/mnras/stab1526}{\color{magenta}{507, 129}}

\bibitem[{{Silva} {et~al.}(2021){Silva}, {Kovetz}, {Keating}, {Dizgah},
  {Bethermin}, {Breysse}, {Karkare}, {Bernal}, \& {Delabrouille}}]{Silva2021}
{Silva},~M.~B., {Kovetz},~E.~D., {Keating},~G.~K., {et~al.} 2021,
  \href{https://ui.adsabs.harvard.edu/abs/2021ExA....51.1593S}{Experimental
  Astronomy},
  \href{http://doi.org/10.1007/s10686-021-09755-3}{\color{magenta}{51, 1593}}

\bibitem[{{Sommovigo} {et~al.}(2022){Sommovigo}, {Ferrara}, {Pallottini},
  {Dayal}, {Bouwens}, {Smit}, {da Cunha}, {De Looze}, {Bowler}, {Hodge},
  {Inami}, {Oesch}, {Endsley}, {Gonzalez}, {Schouws}, {Stark}, {Stefanon},
  {Aravena}, {Graziani}, {Riechers}, {Schneider}, {van der Werf}, {Algera},
  {Barrufet}, {Fudamoto}, {Hygate}, {Labb{\'e}}, {Li}, {Nanayakkara}, \&
  {Topping}}]{Sommovigo2022}
{Sommovigo},~L., {Ferrara},~A., {Pallottini},~A., {et~al.} 2022,
  \href{https://ui.adsabs.harvard.edu/abs/2022MNRAS.513.3122S}{\mnras},
  \href{http://doi.org/10.1093/mnras/stac302}{\color{magenta}{513, 3122}}

\bibitem[{{Stach} {et~al.}(2017){Stach}, {Swinbank}, {Smail}, {Hilton},
  {Simpson}, \& {Cooke}}]{Stach2017}
{Stach},~S.~M., {Swinbank},~A.~M., {Smail},~I., {et~al.} 2017,
  \href{https://ui.adsabs.harvard.edu/abs/2017ApJ...849..154S}{\apj},
  \href{http://doi.org/10.3847/1538-4357/aa93f6}{\color{magenta}{849, 154}}

\bibitem[{{Stach} {et~al.}(2021){Stach}, {Smail}, {Amvrosiadis}, {Swinbank},
  {Dudzevi{\v{c}}i{\={u}}t{\.{e}}}, {Geach}, {Almaini}, {Birkin}, {Chen},
  {Conselice}, {Cooke}, {Coppin}, {Dunlop}, {Farrah}, {Ikarashi}, {Ivison}, \&
  {Wardlow}}]{Stach2021}
{Stach},~S.~M., {Smail},~I., {Amvrosiadis},~A., {et~al.} 2021,
  \href{https://ui.adsabs.harvard.edu/abs/2021MNRAS.504..172S}{\mnras},
  \href{http://doi.org/10.1093/mnras/stab714}{\color{magenta}{504, 172}}

\bibitem[{{Tacconi} {et~al.}(2020){Tacconi}, {Genzel}, \&
  {Sternberg}}]{Tacconi2020}
{Tacconi},~L.~J., {Genzel},~R., \& {Sternberg},~A. 2020,
  \href{https://ui.adsabs.harvard.edu/abs/2020ARA&A..58..157T}{\araa},
  \href{http://doi.org/10.1146/annurev-astro-082812-141034}{\color{magenta}{58,
  157}}

\bibitem[{{Tacconi} {et~al.}(2018){Tacconi}, {Genzel}, {Saintonge}, {Combes},
  {Garc{\'\i}a-Burillo}, {Neri}, {Bolatto}, {Contini}, {F{\"o}rster Schreiber},
  {Lilly}, {Lutz}, {Wuyts}, {Accurso}, {Boissier}, {Boone}, {Bouch{\'e}},
  {Bournaud}, {Burkert}, {Carollo}, {Cooper}, {Cox}, {Feruglio}, {Freundlich},
  {Herrera-Camus}, {Juneau}, {Lippa}, {Naab}, {Renzini}, {Salome}, {Sternberg},
  {Tadaki}, {{\"U}bler}, {Walter}, {Weiner}, \& {Weiss}}]{Tacconi2018}
{Tacconi},~L.~J., {Genzel},~R., {Saintonge},~A., {et~al.} 2018,
  \href{https://ui.adsabs.harvard.edu/abs/2018ApJ...853..179T}{\apj},
  \href{http://doi.org/10.3847/1538-4357/aaa4b4}{\color{magenta}{853, 179}}

\bibitem[{{Tadaki} {et~al.}(2019){Tadaki}, {Kodama}, {Hayashi}, {Shimakawa},
  {Koyama}, {Lee}, {Tanaka}, {Hatsukade}, {Iono}, {Kohno}, {Matsuda}, {Suzuki},
  {Tamura}, {Toshikawa}, \& {Umehata}}]{Tadaki2019}
{Tadaki},~K.-i., {Kodama},~T., {Hayashi},~M., {et~al.} 2019,
  \href{https://ui.adsabs.harvard.edu/abs/2019PASJ...71...40T}{\pasj},
  \href{http://doi.org/10.1093/pasj/psz005}{\color{magenta}{71, 40}}

\bibitem[{{Takada} {et~al.}(2014){Takada}, {Ellis}, {Chiba}, {Greene},
  {Aihara}, {Arimoto}, {Bundy}, {Cohen}, {Dor{\'e}}, {Graves}, {Gunn},
  {Heckman}, {Hirata}, {Ho}, {Kneib}, {Le F{\`e}vre}, {Lin}, {More},
  {Murayama}, {Nagao}, {Ouchi}, {Seiffert}, {Silverman}, {Sodr{\'e}},
  {Spergel}, {Strauss}, {Sugai}, {Suto}, {Takami}, \& {Wyse}}]{Takada2014}
{Takada},~M., {Ellis},~R.~S., {Chiba},~M., {et~al.} 2014,
  \href{https://ui.adsabs.harvard.edu/abs/2014PASJ...66R...1T}{\pasj},
  \href{http://doi.org/10.1093/pasj/pst019}{\color{magenta}{66, R1}}

\bibitem[{{Taniguchi} {et~al.}(2022){Taniguchi}, {Bakx}, {Baselmans},
  {Huiting}, {Karatsu}, {Llombart}, {Rybak}, {Takekoshi}, {Tamura}, {Akamatsu},
  {Brackenhoff}, {Bueno}, {Buijtendorp}, {Dabironezare}, {Doing}, {Fujii},
  {Fujita}, {Gouwerok}, {H{\"a}hnle}, {Ishida}, {Ishii}, {Kawabe}, {Kitayama},
  {Kohno}, {Kouchi}, {Maekawa}, {Matsuda}, {Murugesan}, {Nakatsubo}, {Oshima},
  {Pascual Laguna}, {Thoen}, {van der Werf}, {Yates}, \&
  {Endo}}]{Taniguchi2022}
{Taniguchi},~A., {Bakx},~T. J.~L.~C., {Baselmans},~J. J.~A., {et~al.} 2022,
  \href{https://ui.adsabs.harvard.edu/abs/2022JLTP..209..278T}{Journal of Low
  Temperature Physics},
  \href{http://doi.org/10.1007/s10909-022-02888-5}{\color{magenta}{209, 278}}

\bibitem[{{Tomassetti} {et~al.}(2014){Tomassetti}, {Porciani}, {Romano-Diaz},
  {Ludlow}, \& {Papadopoulos}}]{Tomassetti2014}
{Tomassetti},~M., {Porciani},~C., {Romano-Diaz},~E., {Ludlow},~A.~D., \&
  {Papadopoulos},~P.~P. 2014,
  \href{https://ui.adsabs.harvard.edu/abs/2014MNRAS.445L.124T}{\mnras},
  \href{http://doi.org/10.1093/mnrasl/slu137}{\color{magenta}{445, L124}}

\bibitem[{{Toshikawa} {et~al.}(2018){Toshikawa}, {Uchiyama}, {Kashikawa},
  {Ouchi}, {Overzier}, {Ono}, {Harikane}, {Ishikawa}, {Kodama}, {Matsuda},
  {Lin}, {Onoue}, {Tanaka}, {Nagao}, {Akiyama}, {Komiyama}, {Goto}, \&
  {Lee}}]{Toshikawa2018}
{Toshikawa},~J., {Uchiyama},~H., {Kashikawa},~N., {et~al.} 2018,
  \href{https://ui.adsabs.harvard.edu/abs/2018PASJ...70S..12T}{\pasj},
  \href{http://doi.org/10.1093/pasj/psx102}{\color{magenta}{70, S12}}

\bibitem[{{Vakili} {et~al.}(2023){Vakili}, {Hoekstra}, {Bilicki}, {Fortuna},
  {Kuijken}, {Wright}, {Asgari}, {Brown}, {Dombrovskij}, {Erben}, {Giblin},
  {Heymans}, {Hildebrandt}, {Johnston}, {Joudaki}, \&
  {Kannawadi}}]{Mohammadjavad2023}
{Vakili},~M., {Hoekstra},~H., {Bilicki},~M., {et~al.} 2023,
  \href{https://ui.adsabs.harvard.edu/abs/2023A&A...675A.202V}{\aap},
  \href{http://doi.org/10.1051/0004-6361/202039293}{\color{magenta}{675, A202}}

\bibitem[{{van Kampen} {et~al.}(2005){van Kampen}, {Percival}, {Crawford},
  {Dunlop}, {Scott}, {Bevis}, {Oliver}, {Pearce}, {Kay}, {Gazta{\~n}aga},
  {Hughes}, \& {Aretxaga}}]{vanKampen2005}
{van Kampen},~E., {Percival},~W.~J., {Crawford},~M., {et~al.} 2005,
  \href{https://ui.adsabs.harvard.edu/abs/2005MNRAS.359..469V}{\mnras},
  \href{http://doi.org/10.1111/j.1365-2966.2005.08899.x}{\color{magenta}{359,
  469}}

\bibitem[{{van Kampen} {et~al.}(2023){van Kampen}, {Lacy}, {Farrah}, {Lagos},
  {Jarvis}, {Maraston}, {Nyland}, {Oliver}, {Surace}, \&
  {Thorne}}]{vanKampen2023}
{van Kampen},~E., {Lacy},~M., {Farrah},~D., {et~al.} 2023,
  \href{https://ui.adsabs.harvard.edu/abs/2023MNRAS.523..251V}{\mnras},
  \href{http://doi.org/10.1093/mnras/stad1466}{\color{magenta}{523, 251}}

\bibitem[{{Walter} {et~al.}(2016){Walter}, {Decarli}, {Aravena}, {Carilli},
  {Bouwens}, {da Cunha}, {Daddi}, {Ivison}, {Riechers}, {Smail}, {Swinbank},
  {Weiss}, {Anguita}, {Assef}, {Bacon}, {Bauer}, {Bell}, {Bertoldi}, {Chapman},
  {Colina}, {Cortes}, {Cox}, {Dickinson}, {Elbaz}, {G{\'o}nzalez-L{\'o}pez},
  {Ibar}, {Inami}, {Infante}, {Hodge}, {Karim}, {Le Fevre}, {Magnelli}, {Neri},
  {Oesch}, {Ota}, {Popping}, {Rix}, {Sargent}, {Sheth}, {van der Wel}, {van der
  Werf}, \& {Wagg}}]{Walter2016}
{Walter},~F., {Decarli},~R., {Aravena},~M., {et~al.} 2016,
  \href{https://ui.adsabs.harvard.edu/abs/2016ApJ...833...67W}{\apj},
  \href{http://doi.org/10.3847/1538-4357/833/1/67}{\color{magenta}{833, 67}}

\bibitem[{{Weiss} {et~al.}(2007){Weiss}, {Downes}, {Walter}, \&
  {Henkel}}]{Weiss2007}
{Weiss},~A., {Downes},~D., {Walter},~F., \& {Henkel},~C. 2007,
  \href{https://ui.adsabs.harvard.edu/abs/2007ASPC..375...25W}{in From
  Z-Machines to ALMA: (Sub)Millimeter Spectroscopy of Galaxies}, 25

\bibitem[{{Wei{\ss}} {et~al.}(2009){Wei{\ss}}, {Kov{\'a}cs}, {Coppin}, {Greve},
  {Walter}, {Smail}, {Dunlop}, {Knudsen}, {Alexander}, {Bertoldi}, {Brandt},
  {Chapman}, {Cox}, {Dannerbauer}, {De Breuck}, {Gawiser}, {Ivison}, {Lutz},
  {Menten}, {Koekemoer}, {Kreysa}, {Kurczynski}, {Rix}, {Schinnerer}, \& {van
  der Werf}}]{Weiss2009}
{Wei{\ss}},~A., {Kov{\'a}cs},~A., {Coppin},~K., {et~al.} 2009,
  \href{https://ui.adsabs.harvard.edu/abs/2009ApJ...707.1201W}{\apj},
  \href{http://doi.org/10.1088/0004-637X/707/2/1201}{\color{magenta}{707,
  1201}}

\bibitem[{{Wilkinson} {et~al.}(2017){Wilkinson}, {Almaini}, {Chen}, {Smail},
  {Arumugam}, {Blain}, {Chapin}, {Chapman}, {Conselice}, {Cowley}, {Dunlop},
  {Farrah}, {Geach}, {Hartley}, {Ivison}, {Maltby}, {Micha{\l}owski},
  {Mortlock}, {Scott}, {Simpson}, {Simpson}, {van der Werf}, \&
  {Wild}}]{Wilkinson2017}
{Wilkinson},~A., {Almaini},~O., {Chen},~C.-C., {et~al.} 2017,
  \href{https://ui.adsabs.harvard.edu/abs/2017MNRAS.464.1380W}{\mnras},
  \href{http://doi.org/10.1093/mnras/stw2405}{\color{magenta}{464, 1380}}

\bibitem[{{Williams} {et~al.}(2022){Williams}, {Alberts}, {Spilker}, {Noble},
  {Stefanon}, {Willmer}, {Bezanson}, {Narayanan}, \& {Whitaker}}]{Williams2022}
{Williams},~C.~C., {Alberts},~S., {Spilker},~J.~S., {et~al.} 2022,
  \href{https://ui.adsabs.harvard.edu/abs/2022ApJ...929...35W}{\apj},
  \href{http://doi.org/10.3847/1538-4357/ac58fa}{\color{magenta}{929, 35}}

\bibitem[{{Wolfire} {et~al.}(2022){Wolfire}, {Vallini}, \&
  {Chevance}}]{Wolfire2022}
{Wolfire},~M.~G., {Vallini},~L., \& {Chevance},~M. 2022,
  \href{https://ui.adsabs.harvard.edu/abs/2022ARA&A..60..247W}{\araa},
  \href{http://doi.org/10.1146/annurev-astro-052920-010254}{\color{magenta}{60,
  247}}

\bibitem[{{Yang} {et~al.}(2021){Yang}, {Somerville}, {Pullen}, {Popping},
  {Breysse}, \& {Maniyar}}]{Yang2021}
{Yang},~S., {Somerville},~R.~S., {Pullen},~A.~R., {et~al.} 2021,
  \href{https://ui.adsabs.harvard.edu/abs/2021ApJ...911..132Y}{\apj},
  \href{http://doi.org/10.3847/1538-4357/abec75}{\color{magenta}{911, 132}}

\bibitem[{{Yue} \& {Ferrara}(2019)}]{Yue2019}
{Yue},~B., \& {Ferrara},~A. 2019,
  \href{https://ui.adsabs.harvard.edu/abs/2019MNRAS.490.1928Y}{\mnras},
  \href{http://doi.org/10.1093/mnras/stz2728}{\color{magenta}{490, 1928}}

\bibitem[{{Zanella} {et~al.}(2018){Zanella}, {Daddi}, {Magdis}, {Diaz Santos},
  {Cormier}, {Liu}, {Cibinel}, {Gobat}, {Dickinson}, {Sargent}, {Popping},
  {Madden}, {Bethermin}, {Hughes}, {Valentino}, {Rujopakarn}, {Pannella},
  {Bournaud}, {Walter}, {Wang}, {Elbaz}, \& {Coogan}}]{Zanella2018}
{Zanella},~A., {Daddi},~E., {Magdis},~G., {et~al.} 2018,
  \href{https://ui.adsabs.harvard.edu/abs/2018MNRAS.481.1976Z}{\mnras},
  \href{http://doi.org/10.1093/mnras/sty2394}{\color{magenta}{481, 1976}}

\bibitem[{{Zhao} {et~al.}(2016){Zhao}, {Lu}, {Xu}, {Gao}, {Lord},
  {Charmandaris}, {Diaz-Santos}, {Evans}, {Howell}, {Petric}, {van der Werf},
  \& {Sanders}}]{Zhao2016}
{Zhao},~Y., {Lu},~N., {Xu},~C.~K., {et~al.} 2016,
  \href{https://ui.adsabs.harvard.edu/abs/2016ApJ...819...69Z}{\apj},
  \href{http://doi.org/10.3847/0004-637X/819/1/69}{\color{magenta}{819, 69}}

\end{thebibliography}
\endgroup

%\bibliographystyle{atlast}
%\setlength{\parskip}{0pt}
%\setlength{\bibsep}{0pt}
%\bibliography{high-$z$}

\end{document}